\documentclass[12pt]{iopart}

\usepackage{graphicx}
\usepackage{psfrag}
\usepackage{epstopdf}
\usepackage{color}
\usepackage{mathrsfs}
\usepackage{comment}
\usepackage{multirow}

\expandafter\let\csname equation*\endcsname\relax
\expandafter\let\csname endequation*\endcsname\relax
\usepackage{amsmath,amssymb,amsthm,wasysym}

\DeclareSymbolFontAlphabet{\mathrsfs}{rsfs}
\DeclareMathAlphabet{\mathcal}{OMS}{cmsy}{m}{n}

\def\p{\partial}
\def\non{\nonumber}

\def\ph{\varphi}

\def\t{\theta}

\def\del{\triangle}
\def\C{\tilde{C}}
\def\xx{\times}

\newcommand{\scri}{\mathrsfs{I}}
\newcommand{\be}{\begin{equation}}
\newcommand{\ee}{\end{equation}}

\usepackage{float}
\definecolor{gray}{rgb}{0.5,0.5,0.5}
\definecolor{cyan}{rgb}{0,0.9,0.9}
\definecolor{orange}{rgb}{0.9,0.5,0}
\definecolor{magenta}{rgb}{1,0,1}
\definecolor{purple}{rgb}{0.8,0.4,0.8}
\definecolor{darkgreen}{rgb}{0,.6,0}

\begin{document}

\title
    [Numerical solution of the 2+1 Teukolsky equation,
     application to late-time decays]
    {Numerical solution of the 2+1 Teukolsky equation on a
      hyperboloidal and horizon penetrating foliation of Kerr and 
      application to late-time decays}

\author{Enno Harms, Sebastiano Bernuzzi, and Bernd Br\"{u}gmann}
\address{Theoretical Physics Institute, University of Jena,
  07743 Jena, Germany}

\begin{abstract}
In this work we present a formulation of the Teukolsky equation for
generic spin perturbations on the hyperboloidal and horizon
penetrating foliation of Kerr recently proposed by R{\'a}cz and T{\'o}th.
An additional, spin-dependent rescaling of the field variable can be
used to achieve stable, long-term, and accurate time-domain evolutions
of generic spin perturbations.  
As an application (and a severe numerical test), we investigate
the late-time decays of electromagnetic and gravitational
perturbations at the horizon and
future null infinity by means of 2+1 evolutions.  
As initial data we consider four combinations of (non-)stationary and
(non-)compact-support initial data with a pure spin-weighted spherical
harmonic profile. We present an extensive study
of late time decays of axisymmetric perturbations. 
We verify the power-law decay rates predicted
analytically, 
together with a certain ``splitting''
behaviour of the power-law exponent.
We also present results for non-axisymmetric perturbations. In
particular, our approach allows to study the behaviour of the late
time decays of gravitational fields for nearly extremal and
extremal black holes. For rapid rotation we observe a very prolonged,
weakly damped, quasi-normal-mode phase. For extremal rotation the
field at future null infinity shows an oscillatory behaviour 
decaying as the inverse power of time, while at the horizon it is
amplified by several orders of magnitude over long
timescales. 
This behaviour can be understood in terms of the superradiance cavity
argument.  
\end{abstract}

 \maketitle

\section{Introduction}
\label{sec:intro}

The Teukolsky equation (TE)~\cite{Teukolsky:1972my,Teukolsky:1973ha} 
describes linear perturbations of a field $\Psi$ of spin $s$ around the
Kerr black hole solution.
Black hole perturbation theory~\cite{Chandrasekhar:1985kt} has
widespread applications in general/mathematical relativity and,
ultimately, in astrophysics. To give a few examples, the study of TE solutions
is related to fundamental topics like the stability of 
Kerr~\cite{Press:1973zz,Hartle:1974,Stewart:1975,Whiting:1989JMP....30.1301W},  
no-hair
theorems~\cite{Ori:1997GReGr..29..881O,Hod:1998dt,Barack:1999ma}, 
black hole quasi-normal-modes (QNMs)~\cite{Berti:2009kk}, 
and gravitational radiation from binary
systems within the post-Newtonian approximation~\cite{Sasaki:2003xr},
the test-mass~\cite{Hughes:2001jr,Drasco:2005kz,Glampedakis:2002ya,Glampedakis:2002cb},  
or the self-force (see e.g.~\cite{Dolan:2011dx} for a recent
work and~\cite{Barack:2009ux} for a review) approximations. 

The TE is separable if a Fourier transform in time is performed
simultaneously with the one in the angular azimuthal direction
(axisymmetric background). Exploiting this fact, most of the numerical
calculations have been perfomed in the frequency domain. 
Time-domain numerical solutions of the TE are mainly limited to the
scalar case ($s=0$), which is extensively investigated also in
multidimensional simulations, see
e.g.~\cite{Krivan:1996da,Scheel:2003vs,Lehner:2004cf,Dorband:2006gg,Zenginoglu:2010zm,Racz:2011qu,Jasiulek:2011ce}.   
The relevant case of gravitational perturbations ($s=-2$)
has been considered for the first time by Krivan et
al.~\cite{Krivan:1997hc}. There,  
compact-support initial data have been evolved with the 2+1
mode-decomposed homogeneous TE using a particular first order
reduction and a Lax-Wendroff second order scheme.
The same approach has been extended to fourth order
accuracy in the radial direction in~\cite{PazosAvalos:2004rp}, and
considered in comparison to an approximate 1+1 approach in~\cite{Nunez:2010ra}.
To date, the numerical scheme of~\cite{Krivan:1997hc} is, to our knowledge, the
only successful method to solve the TE in the time domain for a generic spin
field, and it has been applied in late-time decay 
studies of gravitational perturbations (at finite
radius)~\cite{Krivan:1997hc,Glampedakis:2001js,PazosAvalos:2004rp},
and in simulations of gravitational waves from a particle plunging into a rotating black
hole~\cite{Sundararajan:2007jg,Sundararajan:2008zm,Sundararajan:2010sr,Zenginoglu:2011zz}.

A long-standing problem in perturbation theory of black holes is the
study of the late-time decay of the perturbative field, the so-called  ``tail''.
Pioneering work by Price~\cite{Price:1971fb} pointed out that the scattering
of a scalar field off a non-rotating black hole is
characterized by a power-law late-time decay $\propto t^{-\mu}$,
the decay rate $\mu$ depending on the particular multipoles of the
radiation considered and on the initial data employed. 
The behaviour can be related to the asymptotic form
(large radii) of the black hole potential. Similar considerations hold
for the gravitational case described by the
Regge-Wheeler-Zerilli equations~\cite{Regge:1957td,Zerilli:1970se} which have
been studied in several works,
e.g.~\cite{Vishveshwara:1970zz,Press:1971wr,Davis:1972ud,Lousto:2005ip,Bernuzzi:2008rq,Zenginoglu:2008uc}. 
Working in null-cone coordinates, Gundlach et al.~\cite{Gundlach:1993tp} have
computed for the first time the decay rates at future null infinity
($\scri^+$) of scalar and electromagnetic perturbations of non-rotating black
holes. It was found that they are different from those at finite radii.  

The numerical work of Krivan et al.~\cite{Krivan:1996da,Krivan:1997hc}
considered 
for the first time the problem of late-time decay of fields on
Kerr. As expected, the rotation of the background causes mode coupling
(or mode mixing) between the polar modes, labelled by the integer
$l$. In this paper we will often refer to the projections of a field against 
these polar modes as the {\it projected modes}. 
Given an initial angular profile, specified by a particular
$l'$-multipole ({\it pure multipole} initial data), all the other
projected modes allowed by symmetry/parity arguments can be excited
during the evolution.   
Each projected mode decays with a specific rate, $\mu_l$, while the
main (unprojected) field is dominated by the slowest one. 
Both $s=0$ and $s=-2$ perturbations were studied
in~\cite{Krivan:1996da,Krivan:1997hc}, but numerically the decay rates
of the projected modes 
were studied only for the $s=0$ case.  

Tails of generic spin perturbations around rotating black holes
have been studied analytically in two series of works by
Hod~\cite{Hod:1999rx,Hod:1999ci,Hod:2000fh} and Barack and
Ori~\cite{Barack:1998bv,Barack:1998bw,Barack:1999ma,Barack:1999ya,Barack:1999st}.   
Power law tails were studied at the  
horizon, finite radius and null-infinity for compact support,
non-stationary initial data with angular dependence given by both pure
$l'$-mode and generic profiles. The decay rates at the horizon, finite
radius and null-infinity generically differ. Also, they depend on the 
spin-weight of the field.
Mode coupling can be understood in the expansion of the spin-weighted
spheroidal harmonics in terms of spin-weighted spherical harmonics.
Such an expansion emerges naturally in a time domain analysis, whereas
in the frequency domain the TE is
separable and the spin-weighted spheroidal harmonics (depending on the
frequency parameter) are eigenfunctions of the angular operator.
Another observation related to mode coupling is that the decay rates of
the field depend on the index $l'$ of the initial multipole; this fact
was referred to as the {\it memory effect} in~\cite{Burko:2002bt}.

Burko and Khanna~\cite{Burko:2002bt} have reported numerical
experiments in which the decay rates differ from the above
predictions. In particular their data, computed in ingoing
coordinates, supported a {\it simple picture} 
in which each projected mode decays with the Price law.
The late-time dynamics is then dominated by the smallest of the projected
modes which is excited (i.e.~allowed by symmetry and initial data.)
For $s=0$ and axisymmetric ($m=0$) perturbations the two predictions start
to differ for $l'\geq4$. 

Further theoretical and numerical
studies in the $s=0$ case~\cite{Poisson:2002jz,Scheel:2003vs,Gleiser:2007ti,Tiglio:2007jp,Burko:2007ju,Burko:2010zj}
helped to clarify that the decay rates depend upon
the slicing of the background and the initial data. Hence there is no
contradiction between the different predictions. In a recent work  
Burko and Khanna~\cite{Burko:2010zj} proposed the general formula
\be
\label{eq:BK_mul}
\mu_l =  \left\{  \begin{array}{l l }
    l'+l+3 \ , &  l\geq l' \\ 
    l' + l +1 \ , & l<l'   \\
  \end{array} \right.    
\ee
for decay rates of compact support, non-stationary, pure $l'$-multipole
initial data at $i^+$. 

Decay rates at $\scri^+$ for $s=-2$ perturbations of non-rotating black
holes were computed by Zengino{\u g}lu using hyperboloidal
foliations~\cite{Zenginoglu:2008uc}. The same technique allowed the first
calculation of $s=0$ perturbations of rotating black holes (Kerr) at
null infinity~\cite{Zenginoglu:2009hd}. 
Recently, R{\'a}cz and T{\'o}th~\cite{Racz:2011qu} (hereafter RT) performed a
detailed numerical analysis of $s=0$ decay rates on Kerr confirming
the formula in Eq.~\eqref{eq:BK_mul} for the decay rates of the
projected modes on the horizon and at finite radius (at
$\scri^+$ different formulas hold, see~\cite{Hod:1999rx,Zenginoglu:2009hd,Racz:2011qu}).
Their work is based on a particular choice of coordinates (see
also~\cite{Moncrief:2000,Fodor:2003yg,Fodor:2006ue}) that realizes a horizon
penetrating, hyperboloidal foliation of Kerr. To our knowledge this is
the only foliation of Kerr present in the literature that
smoothly connects the horizon with future null-infinity.
References~\cite{Racz:2011qu,Jasiulek:2011ce} pointed out an
intermediate ``splitting'' behaviour of the decay rates of scalar 
fields on Kerr at finite radius. The phenomenon was clarified
in~\cite{Zenginoglu:2012us} using a horizon penetrating, hyperboloidal
foliation constructed with transmitting layers
(see~\cite{Bernuzzi:2011aj} and references therein). 

Finally, Andersson and
Glampedakis~\cite{Andersson:1999wj,Glampedakis:2001js} have shown
that the late time decay of non-axisymmetric perturbations of rapidly
spinning black holes is not characterized by a Price-like power law but dominated
by the QNM contribution. In particular, for nearly extremal black holes,
the collective effect from several, slowly damped QNMs results in an
oscillatory and weakly exponentially damped signal.
In contrast, for extremal black holes the weakly damped QNMs combine to
give asymptotically an overall decay rate $\propto t^{-1}$.

In this paper we consider a novel formulation of the TE on the
hyperboloidal and horizon-penetrating foliation of RT, and perform
numerical experiments in the time domain studying the late time decay of generic
spin perturbations of Kerr.
The use of hyperboloidal and horizon penetrating foliations with
compactification of null infinity has two well-known properties, e.g.~\cite{Zenginoglu:2007jw,Zenginoglu:2010cq}:  
(i)~$\scri^+$ and the horizon are included in the computational domain,
thus allowing an unambiguous extraction of the radiation, 
(ii)~artificial boundary conditions are not needed for the solution of
the Cauchy problem.
The late time decay problem is a severe test for the robustness and accuracy of a
numerical scheme for the TE. Note also that, for instance, 
Price-like power law tails can not arise if Sommerfeld-type
artificial boundary conditions are employed~\cite{Dafermos:2004wt}.  
Here, we compute numerically, for the first time to our knowledge,
late-time decay rates for gravitational and electromagnetic
perturbations of Kerr both at the horizon and at future null infinity.

The paper is organized as follows. 
In Sec.~\ref{sec:TE_RT} the novel formulation of the TE on the RT foliation is
described. The key addition to the construction of RT is a $s$-dependent
rescaling of the master variable, resulting in equation
coefficients regular over the whole domain of integration. A
two-parameter family of (RT-like) 
coordinate transformations is also given.
In Sec.~\ref{sec:num} the numerical method employed is detailed.
In Sec.~\ref{sec:exp} we give an overview of the numerical experiments
performed, including convergence tests and a discussion of numerical difficulties.
In Sec.~\ref{sec:decaym0} results for power law decay rates are
presented in the axisymmetric case ($m=0$).
In Sec.~\ref{sec:decaym2} results for late time decays for a few
non-axisymmetric cases are  presented ($s=0,-2$). 
We conclude in Sec.~\ref{sec:conc}. 
In~\ref{app:coefs} the TE coefficients are stated
in RT coordinates. In~\ref{app:green} the analytic calculation of the
Green's function of~\cite{Hod:2000fh} is reviewed.

Geometric units ($c=G=1$) are employed.

\section{Teukolsky equation on the RT foliation}
\label{sec:TE_RT}

In this section we review the construction of the RT coordinates, present
a two-parameter generalization and derive a form of the TE with
coefficients regular at $\scri^+$ and the horizon for every spin field
value $s$.  
In the scalar case, $s=0$, the equation reduces to the one given in
RT, modulo an overall factor.

\subsection{The RT foliation}
\label{sec:TE_RT:coords}

\begin{figure}[t]
  \centering
  \includegraphics[width=0.49\textwidth]{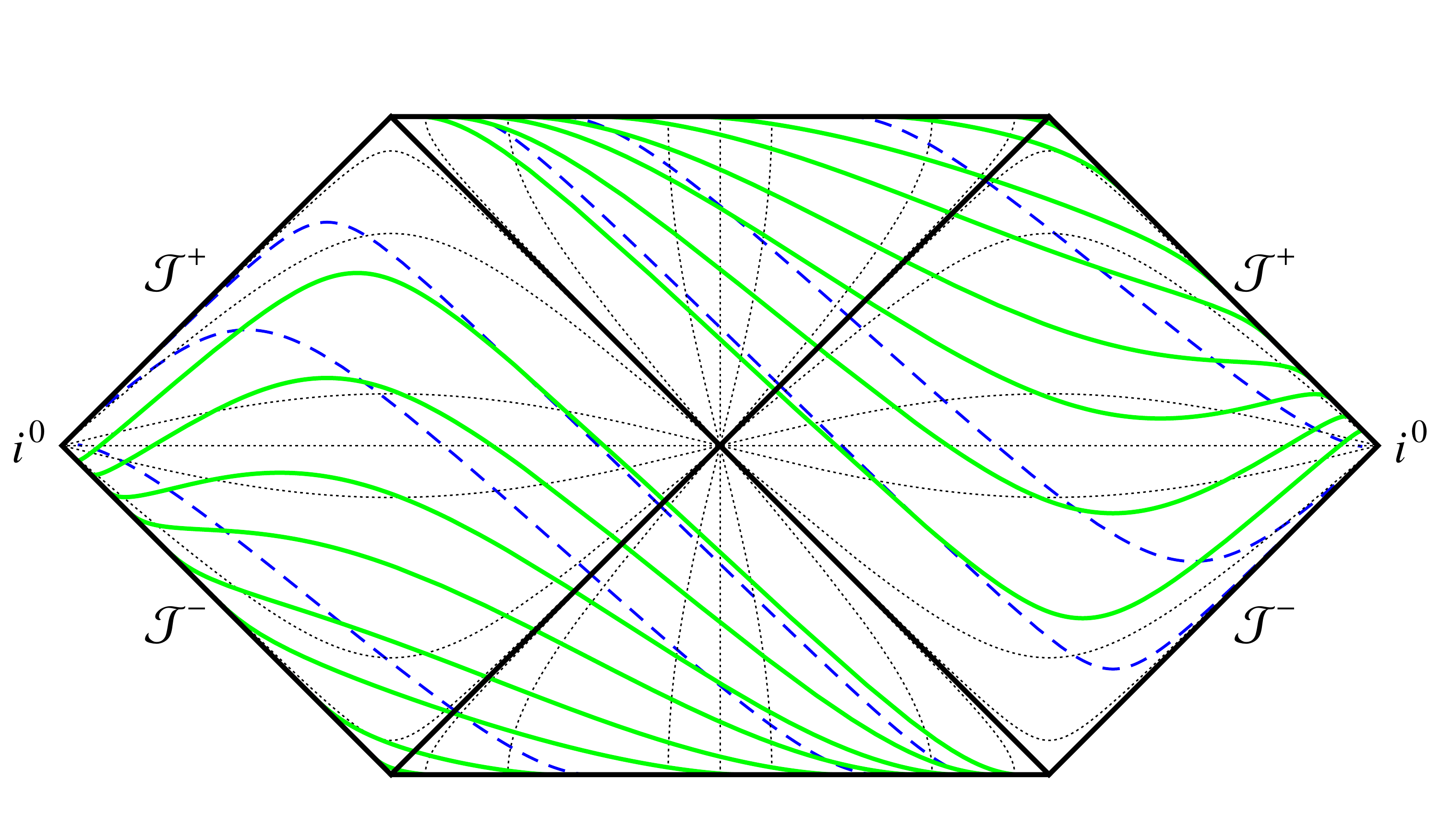} 
  \caption{Conformal diagram for the Schwarzschild spacetime ($a=0$). The
    $\tau$-slices (Kerr ingoing coordinates) are the dashed blue
    lines, the $T$-slices (RT coordinates) are the solid green lines,
    the $t$-slices (BL coordinates) are the dotted black lines.}  
  \label{fig:confdiag}
\end{figure}

R{\'a}cz and T{\'o}th~\cite{Racz:2011qu} have proposed a hyperboloidal foliation of
Kerr that penetrates the horizon and reaches future null infinity.
The foliation can be explicitly constructed from the Kerr solution in
Boyer-Lindquist (BL) coordinates $x^\mu=\{t,r,\theta,\phi\}$ by two successive
coordinate transformations.
To fix the notation, we recall the Kerr metric in BL coordinates 
\begin{align}
\label{eq:metric_BL} 
g_{\textrm{BL}}= -\left(1-\frac{2Mr}{\rho}\right) dt^2 - \frac{4 a
  M r}{\rho}\sin^2\t \,dt\,d\ph +
\frac{\rho}{\triangle}\, dr^2  +\rho \,d\t^2 \non \\ 
+\left(r^2+a^2+\frac{2Ma^2r\sin^2\t}{\rho}\right)\,\sin^2\t\,d\ph^2
\ ,
\end{align}
where $M$ and $a\,M$ are the mass and angular momentum of the black
hole,  $\rho = r^2 + a^2\cos\t^2$, and $\del =r^2 -2Mr + a^2
:=(r-r_+)(r-r_-)$. The two step procedure to construct the RT foliation
is the following. First, switch to ingoing-Kerr coordinates
$x^\mu=\{\tau,r,\theta,\varphi\}$ defined by 
\begin{eqnarray}
\label{eq:RT1}
\tau &=& t-r + \int dr \frac{a^2+r^2}{\del}\\
\label{eq:RT2}
\varphi &=& \phi  + a \int \frac{dr}{\del}  \ .
\end{eqnarray}
Second, introduce the coordinates $x^a=\{T,R,\theta,\varphi\}$ such that
\begin{eqnarray}
\label{eq:RT3}
\tau &=& T -4 M \log \left(\left|1-R^2\right|\right)+\frac{1+R^2}{1-R^2}\\
\label{eq:RT4}
r &=& \frac{2R}{1-R^2} \ .
\end{eqnarray}
The transformation in Eq.~\eqref{eq:RT3}-\eqref{eq:RT4} realizes a
hyperboloidal and horizon penetrating foliation with $\scri^+$
compactified at $R=1$~\cite{Moncrief:2000,Fodor:2003yg} and the $r_+$ horizon
located at  
\be
R_+ = \frac{2 \sqrt{2 M \sqrt{M^2-a^2}-a^2+2 M^2+1}-2}{2
   \left(\sqrt{M^2-a^2}+M\right)} \ .
\ee
Note that $R_+$ depends on $M$ and $a$. In Fig.~\ref{fig:confdiag}
(compare Fig.~1 of RT)  the two foliations of the spacetime are
shown in the non-rotating $a=0$ limit for simplicity. 

Following~\cite{Moncrief:2000,Fodor:2003yg,Racz:2011qu}
we can also write Eq.~\eqref{eq:RT3}-\eqref{eq:RT4} in the more general form
\begin{eqnarray}
\label{eq:RT5}
\tau &=& T + \sqrt{ \kappa^2 + (R/\Omega(R))^2} - 4\, M \log(2 \Omega(R)) \\
\label{eq:RT6}
r &=& \frac{R}{\Omega_S(R)} \ , 
\end{eqnarray}
where $\Omega_S(R)$ is a general {\it compress
  function}~\cite{Zenginoglu:2010cq} 
(eventually containing a parameter $S$ to fix the coordinate location of
$\scri^+$) and $\kappa$ is a parameter determining the coordinate speed
of the outgoing characteristic at null infinity (in RT $\kappa=S=1$).

\subsection{Scaling the TE master variable}
\label{sec:TE_RT:scaling}

The homogeneous TE in BL coordinates reads
\begin{align}
\label{eq:TE} 
&D^2 \p_{tt}\Psi = - \del^{-1}4aMr \p_{\ph t}\Psi 
+ \del^{-s}\p_r(\del^{s+1}\p_r\Psi)
+ \sin\t^{-1}\p_\t(\sin\t\p_\t\Psi) \non\\
&+ (\sin\t^{-2} - a^2\del^{-1})\p_{\ph \ph}\Psi
- 2s (\del^{-1}M(a^2-r^2) + (r+ia\cos\t))\p_t\Psi \non\\
&+ 2s (\del^{-1}a(r-M) + i \cot\t\sin\t^{-1})\p_{\ph}\Psi
- s (s\cot\t^2 - 1)\Psi \ ,
\end{align}
where $D^{2}=(r^2+a^2)\del^{-1}-a^2\sin\t^2$. 
The introduction of a conformal compactification like that of 
Eqs~\eqref{eq:RT5}-\eqref{eq:RT6} requires to rescale the field
variable like  
\be
\label{eq:rescale1}
\Psi = r^{-(2s+1)}\psi \ , 
\ee
in order to avoid the singularity of the physical metric at $\scri^+$ 
that is induced also in the wave
equation~\cite{Zenginoglu:2007jw,Racz:2011qu}.
However, once Eq.~\eqref{eq:TE} is expressed in RT coordinates that
include the horizon, some
coefficients have singularities on the horizon if $s\neq0$.
The simple rescaling 
\be
\label{eq:rescale2}
\Psi = r^{-(2s+1)} \left(\Delta\, r^{-2}\right)^{-s}\psi  =
\Delta^{-s} r^{-1} \psi\ , 
\ee
or 
\be
\label{eq:rescale3}
\Psi = \Delta(R)^{-s} \frac{R}{\Omega(R)}\psi \ ,  
\ee
cures the singularities. Note that the factor
$1/r^2$ in the term $\left(\Delta\, r^{-2}\right)^{-s}$ is essential 
since $\del\sim\,r^2$ for $r\to\infty$. The rescaling in
Eq.~\eqref{eq:rescale2} has been, in some form, widely used in the
literature about the TE since~\cite{Teukolsky:1973ha}, and thus should
be considered as the ``natural'' one. In connection with the
hyperboloidal transformation, it has been recently considered
in the frequency domain framework of~\cite{Zenginoglu:2011jz} (see
Eq.~(13) there). For time-domain studies it is employed here for the
first time together with the RT coordinates (Eq.~\ref{eq:rescale3}).
The resulting TE is of the form 
\begin{align}
\label{eq:TE_RT} 
 &C_{0} \psi 
+ C_{T} \partial_T \psi
+ C_{R} \partial_R\psi  
+ C_{\theta} \partial_{\theta}\psi 
+ C_{\varphi}  \partial_{\varphi}\psi 
+ C_{TT} \partial_{TT}\psi
+ C_{RR} \partial_{RR}\psi \non \\
& + C_{\theta\theta} \partial_{\theta\theta}\psi
+ C_{\varphi\varphi} \partial_{\varphi\varphi}\psi 
+ C_{R\varphi}  \partial_{R\varphi} \psi 
+ C_{TR} \partial_{TR} \psi 
+ C_{T\varphi} \partial_{T\varphi}\psi  = 0
\qquad , 
\end{align}
where the coefficients are given explicitly in~\ref{app:coefs}.
As one can check from direct inspection, the
coefficients are rational polynomials and they are regular over the
domain $R\in[R_+,1]$. This property holds also for the general
rescaling in Eqs.~\eqref{eq:RT5}-\eqref{eq:RT6}.
Note that the new variable $\psi$ is asymptotically constant, for example
for $s=-2$ it is related to the Weyl scalar $\Psi_4$ by $\psi = r \,
\Psi_4$, since $\Psi \sim r^4 \, \Psi_4$ asymptotically.
Note also that the equation in the time domain is
not separable in the standard way due to the term $\propto\p_{\varphi t}$. 

For numerical applications Eq.~\eqref{eq:TE_RT} can be decomposed in
2+1 form separating each Fourier $m$-mode in the azimuthal direction and
exploiting the axisymmetry of the background.  
The resulting 2+1 equation has the form
\begin{align}
 \label{eq:TE_RTm} 
 & \C_{0} \psi + \C_{T}  \partial_T \psi + \C_{\theta}
 \partial_{\theta}\psi  + \C_{R}  \partial_R\psi + \C_{TT}
 \partial_{TT}\psi + \C_{RR}  \partial_{RR}\psi +
 \C_{\theta\theta}  \partial_{\theta\theta}\psi  +  \C_{TR} 
 \partial_{TR} \psi  = 0 \ , 
\end{align}
with coefficients presented in~\ref{app:coefs}, and the index $m$ in
the field variable $\psi_m$ has been suppressed for brevity.

\subsection{The initial-boundary value problem}
\label{sec:TE_RT:IBVP}

\begin{figure}[t]
  \centering
  \includegraphics[width=0.9\textwidth]{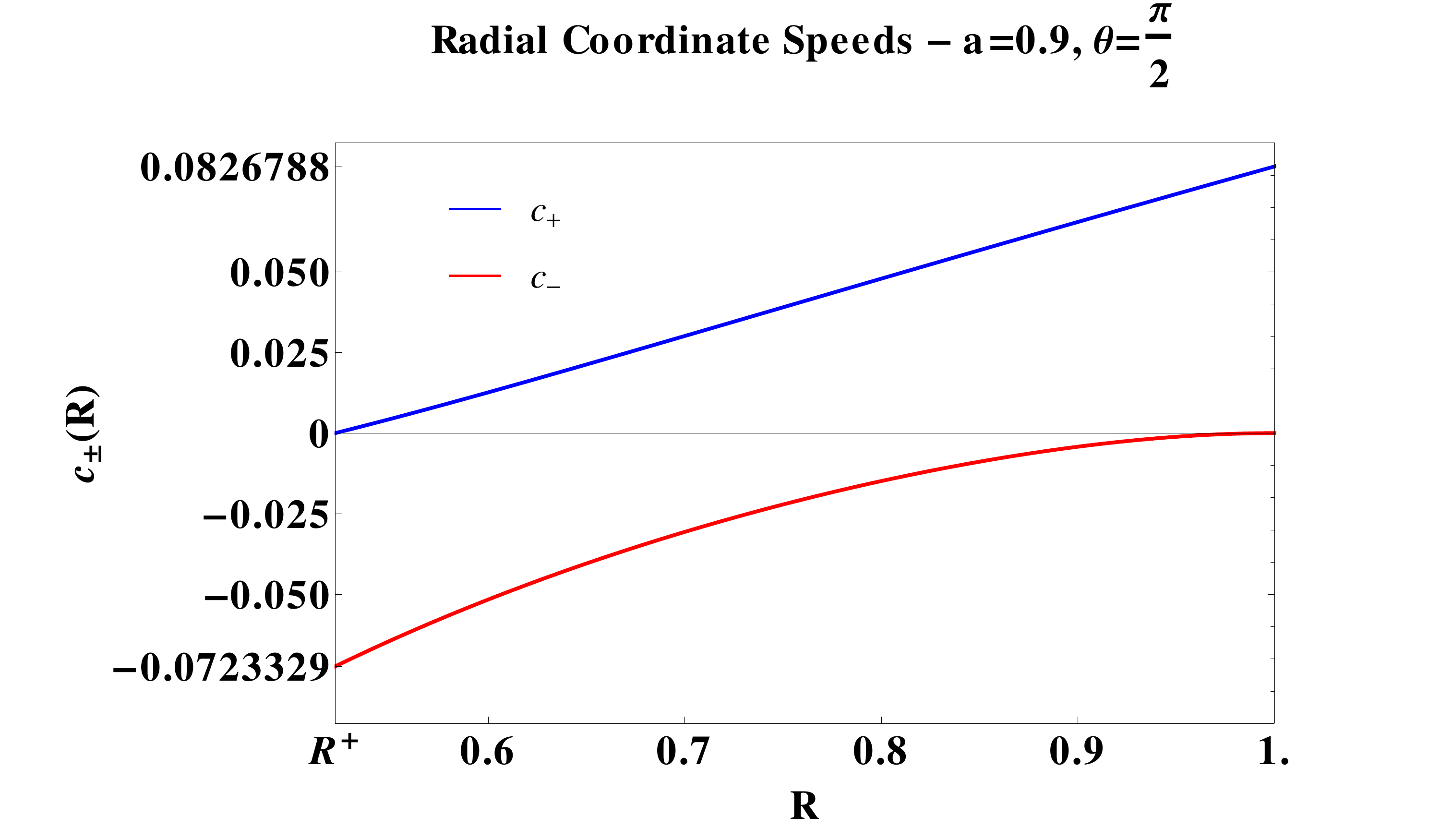}    \\
  \includegraphics[width=0.49\textwidth]{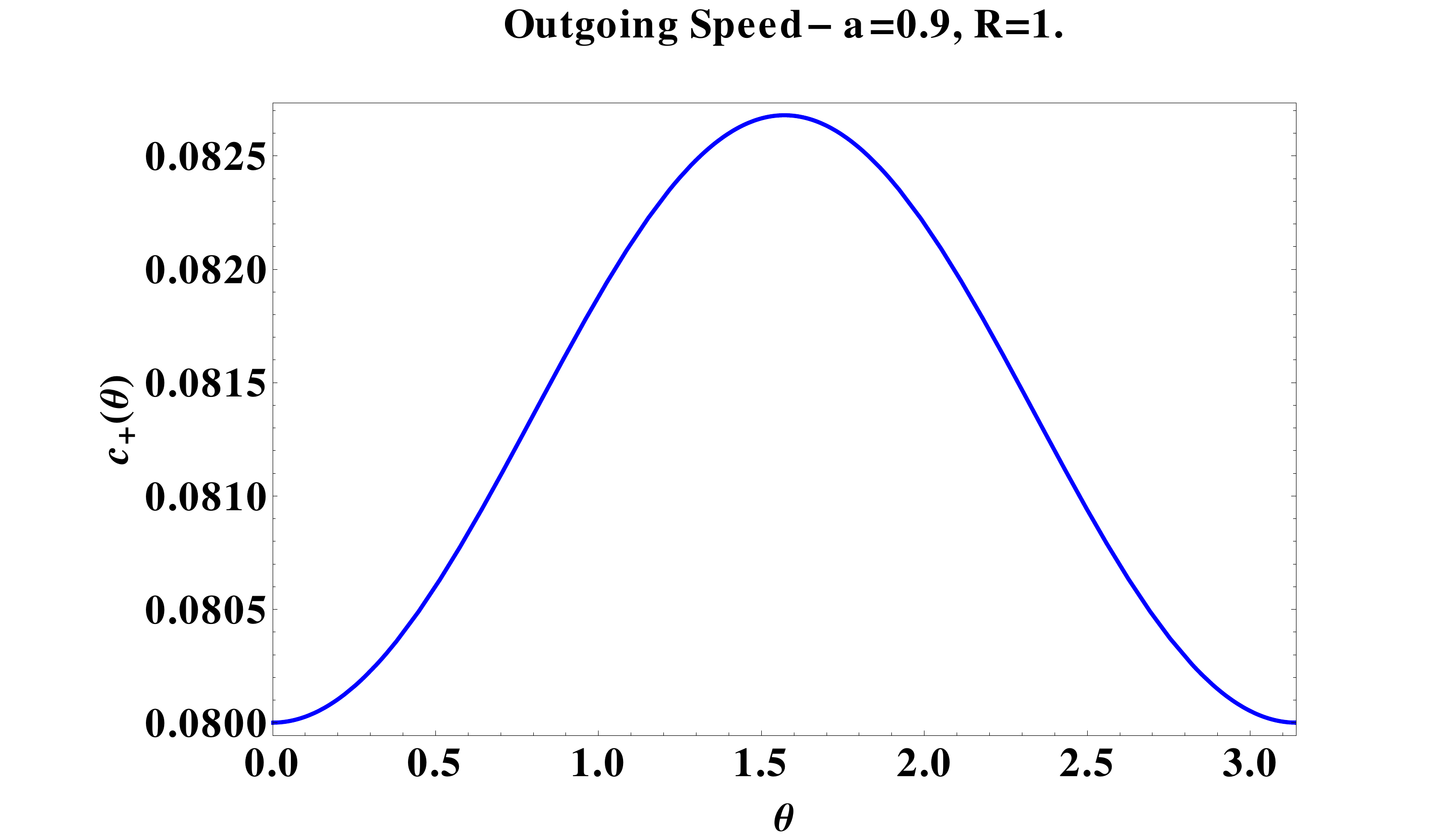}   
  \includegraphics[width=0.49\textwidth]{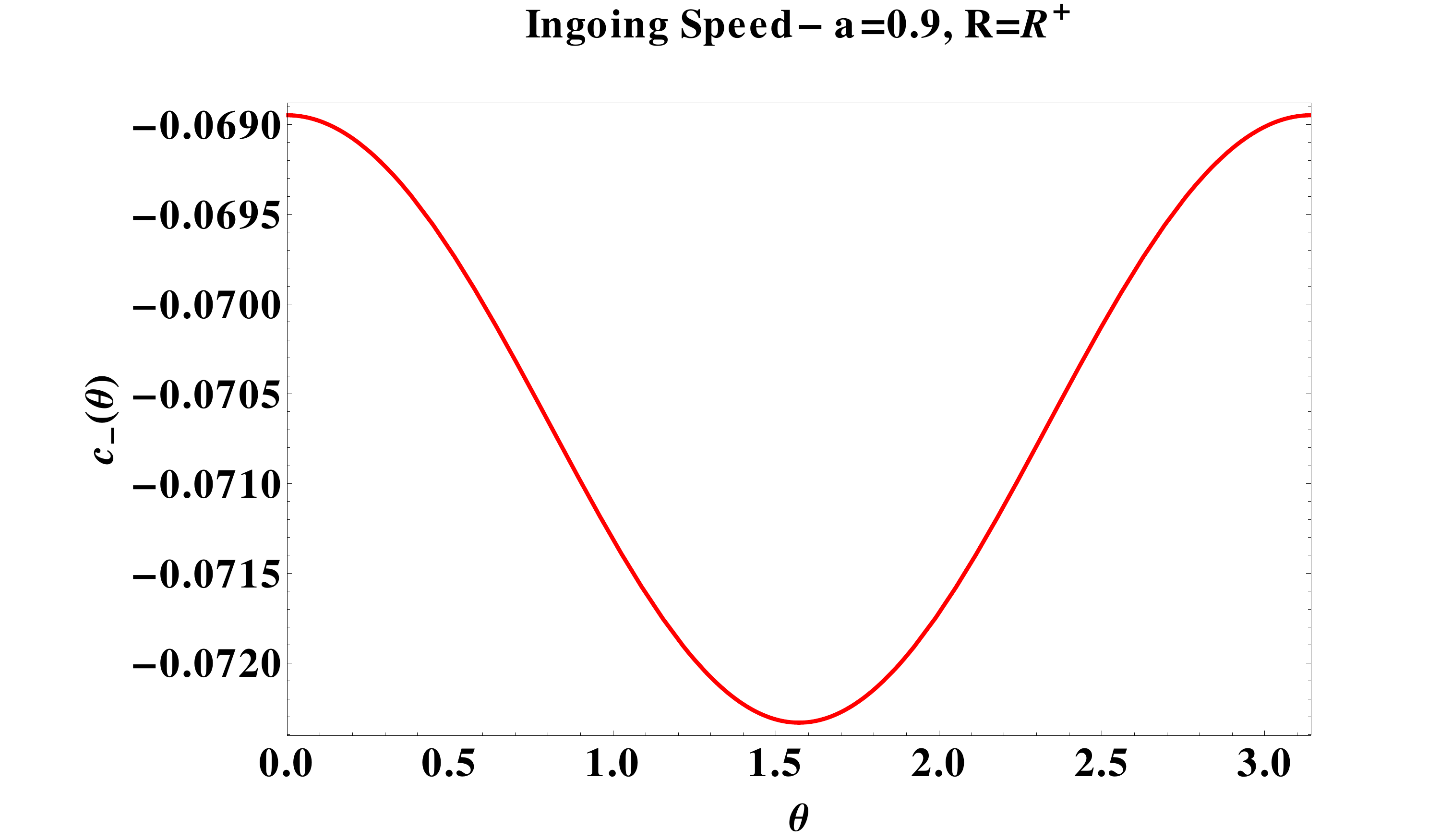}   
  \caption{Radial coordinate speeds for $M=1$ and
    $a=0.9$. Different panels show the functional dependences on
    $R$ and $\theta$, $c_{\pm}(R,\theta)$. Because the incoming 
    (outgoing) coordinate speed vanishes at the outer (inner) boundary
    artificial boundary conditions are not needed. Note the 
    weak dependence on $\theta$ for $a=0.9$.} 
  \label{fig:speeds}
\end{figure}

From a PDE point of view, Eq.~\eqref{eq:TE_RT} (as Eq.~\eqref{eq:TE})
is a 3+1 linear wave equation with variable coefficients in spherical
coordinates.  
The principal part of the equation is independent of the spin weight, hence
symmetric hyperbolicity and well-posedness of the Cauchy
problem follow from the scalar $s=0$ case 
when appropriate boundaries are considered~\footnote{ 
Numerical instabilities can still arise from unstable
discretization, stiff non-principal terms, or exponentially growing
continuum modes triggered by numerical noise. }.

An advantage of using a hyperboloidal foliation and compactification
is that no artificial timelike outer boundary condition is needed, 
e.g.~\cite{Zenginoglu:2010cq}. Furthermore, because the foliation is horizon
penetrating, also the inner boundary condition is not needed. 
Inspection of the equation at $R=1$ and $R=R_+$ gives that
both boundaries are ``outflow'' boundaries. Taking the limits 
$R\to R_+$ and $R\to1$ of Eq.~\eqref{eq:TE_RT} with $M=1$ (and
omitting the angular terms and the non-principal part for simplicity) 
one obtains 
\be
\partial_T \left( \tilde{C}_{TT}\partial_T \psi + \tilde{C}_{TR} \partial_R \psi \right) = 0
\ .
\ee
The speeds, $c_{R}(R,\theta)=\tilde{C}_{TR}/\tilde{C}_{TT}$, for $a=0$ are given by 
\begin{eqnarray}
  \label{eq:values_speeds_a0}
  c_{R,-}(R_+,\frac{\pi}{2}) 
  &\simeq -0.0460655 \ ,  \\ \nonumber
  c_{R,+}(1,\frac{\pi}{2}) 
  &= 0.08 \ .
\end{eqnarray}
Similarly, one obtains for $a=0.9$
\begin{eqnarray}
  \label{eq:values_speeds_a9}
  c_{R,-}(R_+,\frac{\pi}{2}) 
  & \simeq -0.0723329 \ ,\\ \nonumber
  c_{R,+}(1,\frac{\pi}{2}) 
  &\simeq 0.0826788 \ .
\end{eqnarray}
The speeds $c_{R,\pm}$ are the coordinate speeds of light in radial direction that can be in general calculated as 
\be
 c_{R,+} = -\beta^R + \sqrt{\gamma^{RR}} \, \alpha \ ,\ \ \ \  c_{R,-} = -
 \beta^R - \sqrt{\gamma^{RR}} \, \alpha \ , 
\ee
where $\alpha$ is the lapse, $\beta^R$ the radial component of the
shift vector, and $\gamma^{RR}$ the radial component of the spatial
$3$-metric in RT coordinates (see~\ref{app:coefs}.) 
The speeds are illustrated in Fig.~\ref{fig:speeds} for $M=1$ and
$a=0.9$, and agree with the above estimates from the limiting
equation.

\section{Numerical method}
\label{sec:num}

In this section we describe two different numerical strategies
implemented for the solution of the 2+1 TE. 

The numerical algorithms rely in both cases on a method-of-lines
approach for the time integration. Equation~\eqref{eq:TE_RTm} (or
Eq.~\eqref{eq:TE_RT}) is written in a first-order-in-time form, 
\be
\frac{d u}{dt} = R(u) \ ,
\ee
and evolved as an ODE system. Above $u$ is a certain vector of real
functions and $R(.)$ a discrete representation of the right-hand-side
of the system. 
A standard fourth order Runge-Kutta integrator is employed. The time
step is choosen according to a Courant-Friedrich-Lewy (CFL) condition of
type $\del t = C_{\rm CFL} \min(h_R,h_\theta)$, 
where $h_x$ is the minimum grid
spacing in direction $x$ and the factor $C_{\rm CFL}$ 
accounts for the maximum speed of the system. The spatial
discretization is perfomed in two different ways, and the specific
form of the first-order-in-time systems also differs. 

The {\it PS-Teukode} implements a fully first-order reduction of
Eq.~\eqref{eq:TE_RTm} with reduction 
variables~\unskip\footnote[5]{ 
  The $\psi$ variable is complex, the split into real and imaginary
  part is understood and omitted in the formulas for brevity.
}
$u=\{\psi,\ \p_T\psi,\ \p_R\psi,\ \p_\theta\psi\}$ and
pseudo-spectral (PS) representation of first derivatives. The radial
direction is covered by a Gauss-Lobatto grid and Chebychev polynomials
are employed for the PS derivatives. The angular direction
$\theta\in(0,\pi)$ is represented on a staggered equidistant grid
with {\it double covering} (i.e.~the function is extended to
$\theta\in(0,2\pi)$) in order to employ the Fourier basis for the
derivatives, e.g.~\cite{Bruegmann:2011zj}. 
Differently from~\cite{Bruegmann:2011zj}, here we extend the
function by imposing a given parity. Specifically, by inspection of
spin-weighted spherical harmonics, we assume the field $\psi$ has
parity $\pi = (-1)^{m+s}$ across the axis.  
We experimentally found that this prescription is important for
long-term stability. Note that this assumption is compatible with the
use of pure multipole initial data in the 3+1 equation 
and with generic initial data in the 2+1 equation.
The {\it PS-Teukode} additionally implements the following options: 
(i)~finite differencing derivatives in the radial direction (as
described below); 
(ii)~the 3+1 Eq.~\eqref{eq:TE_RT} by using the Fourier basis also in
the azimuthal direction;
(iii)~different floating-point arithmetics: double,
long-double and quadruple. Although quadruple precision requires
several changes in the code, and since it is not available in standard hardware
leads to a slow-down by a factor of $\sim50$, it turned out
to be essential for the investigation of $s=-2$ decay rates.

The {\it FD-Teukode} implements a second-order reduction in space of
Eq.~\eqref{eq:TE_RTm} with reduction variables $u=\{\psi,\ \p_T
\psi\}$, and finite difference (FD)  representation of the derivatives up
to sixth order accuracy. The stencils in the radial direction are
centered in the bulk of the domain and lop-sided at the
boundaries. The angular grid is staggered (the same as in the PS code)
and ghosts points are employed to implement the boundary conditions on
the axis. The ghosts points are filled according to the parity
condition $\pi = (-1)^{m+s}$. Artificial dissipation operators are
also implemented and employed for long term stability.

Stable, long-term evolutions are obtained with both codes. In this
paper we mainly restrict attention to the results obtained with the PS code
(in some cases used with FD in the radial direction as indicated below) since the late time
decays can be computed more accurately with this approach.

\section{Overview of the numerical experiments}
\label{sec:exp}

\begin{figure}[t]
  \centering
  \includegraphics[width=\textwidth]{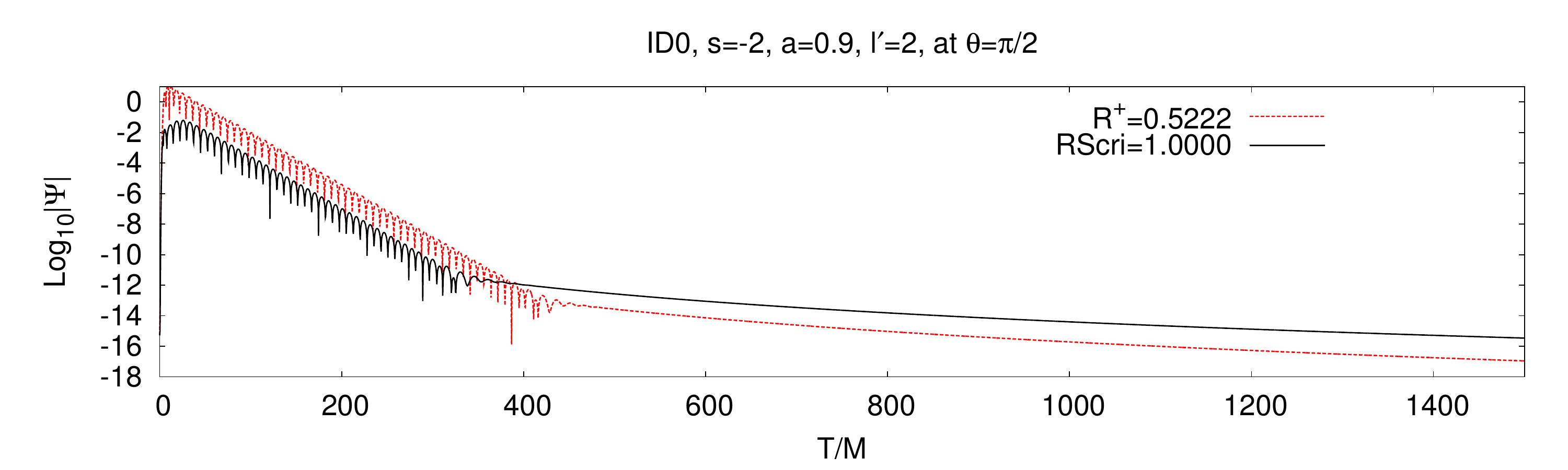}    
  \caption{Evolution of the perturbation field at the horizon and
    $\scri^+$ ($\theta=\pi/2$). The field is characterized by the
    quasi normal mode ringdown and a power law tail.
    The plot refers to a simulation of an axisymmetric gravitational
    perturbation ($s=-2$ and $m=0$) with ID0,
    $l'=2$ and $a=0.9$.}  
  \label{fig:waves}
\end{figure}

In this section the initial data employed, the code's convergence
properties, and the methodology used in simulation data analysis are
discussed. The black hole mass is set to $M=1$ from now on in this
paper, and the spin parameter is identified with the angular momentum,
$a=J/M^2$.    

Initial data are given specifying an angular profile, a radial profile, and
the time derivative of the field. The angular profile is usually
prescribed by a spin-weighted spherical harmonic multipole,
i.e.~$\psi(\theta)\propto {}^sY_{l' m'}(\theta)$ (pure multipole initial data).
Four initial data configurations are considered corresponding to
different choices of radial profiles and the time derivative of the
field. They are named ID0, ID1, ID2, and ID3, and defined by
\begin{align}
 \text{ID0} \ / \ \text{ID2} : & 
 \begin{cases}
   \psi(0,R) & = G(R) \ / \ 1 \\ 
   \p_T\psi(0,R) & = 0    \\
 \end{cases} 
\\
 \text{ID1}\ / \ \text{ID3} : & 
\begin{cases}
  \psi(0,R) & = 0 \\ 
  \p_T\psi(0,R) & = G(R) \ / \ 1 \ , \\
\end{cases} 
\end{align}
with 
\be
G(R)=e^{-\frac{w}{2} \; (R-R_0)^2} \ .
\ee
These initial data are commonly used in the literature in which they are
referred to as stationary (ID0, ID2) or non-stationary (ID1, ID3), and 
compact (ID0, ID1) or non-compact (ID2, ID3) support initial
data. Note that, strictly speaking, a Gaussian is not of compact support
but we verified that, with the parameters employed, the field value is
below the round-off level at the horizon and $\scri^+$.  
For example setting $w=3000$ and $R_0=0.8$, $G(R)\sim10^{-38}$ at the
horizon (for $a=0.9$ $R_+\approx0.5222$) and at $R=1$.  

We performed evolutions with the different initial data and with different
spin fields for black hole parameters $M=1$ and $a=\{0,0.5,0.9\}$. 
The qualitative outcome of all the simulations 
is shown in Fig.~\ref{fig:waves} for the exemplary case $s=-2$,
ID0 and $a=0.9$. The figure displays the radiation field
extracted at the horizon and at $\scri^+$ for $\theta=\pi/2$.
The solution has the well-known structure composed of an
initial transient, the QNM phase, and the
power law decay. 
We have checked frequencies and damping times of the QNMs in
some cases 

and found discrepancies below
or of the order of $1\%$ with the tables of Berti et al.~\cite{Berti:2009kk}. 
For example, in the case of a $s=-2$ perturbation
with $l'=2$ on Schwarzschild ($a=0$) the ringing frequency
and the damping time are $(\omega,\tau)=(0.3876,11.2361)$, to be
compared with $(\omega,\tau)=(0.3737,11.2410)$. For $l'=3$ we
find $(0.6133,10.7958)$ in good agreement with $(0.5994,10.7875)$. For
$l'=4$ we find $(0.8220,10.6248)$ to be compared with $(0.8091,
10.6202)$, and similarly for other cases.

\subsection{Convergence}
\label{subsec:conv}

Let us discuss the convergence tests performed to confirm the correct
implementation of the code. The results confirm the expected
exponential convergence of the PS code, as far as the solution does
not develop high gradients and/or reaches amplitudes comparable to
machine accuracy. At late times the radial profile of the
solution develops strong gradients, essentially due to the different
power indices $\mu$ at the horizon and $\scri^+$. The convergence rate is
then slower, and progressively more coefficients are necessary to
describe the solution. The particular shapes of these gradients depend also on 
$s$. Notably, in the cases $s>0$, the computations are more
efficiently performed with finite differencing in the radial direction.

\begin{figure}[t]
  \centering
  \includegraphics[width=0.49\textwidth]{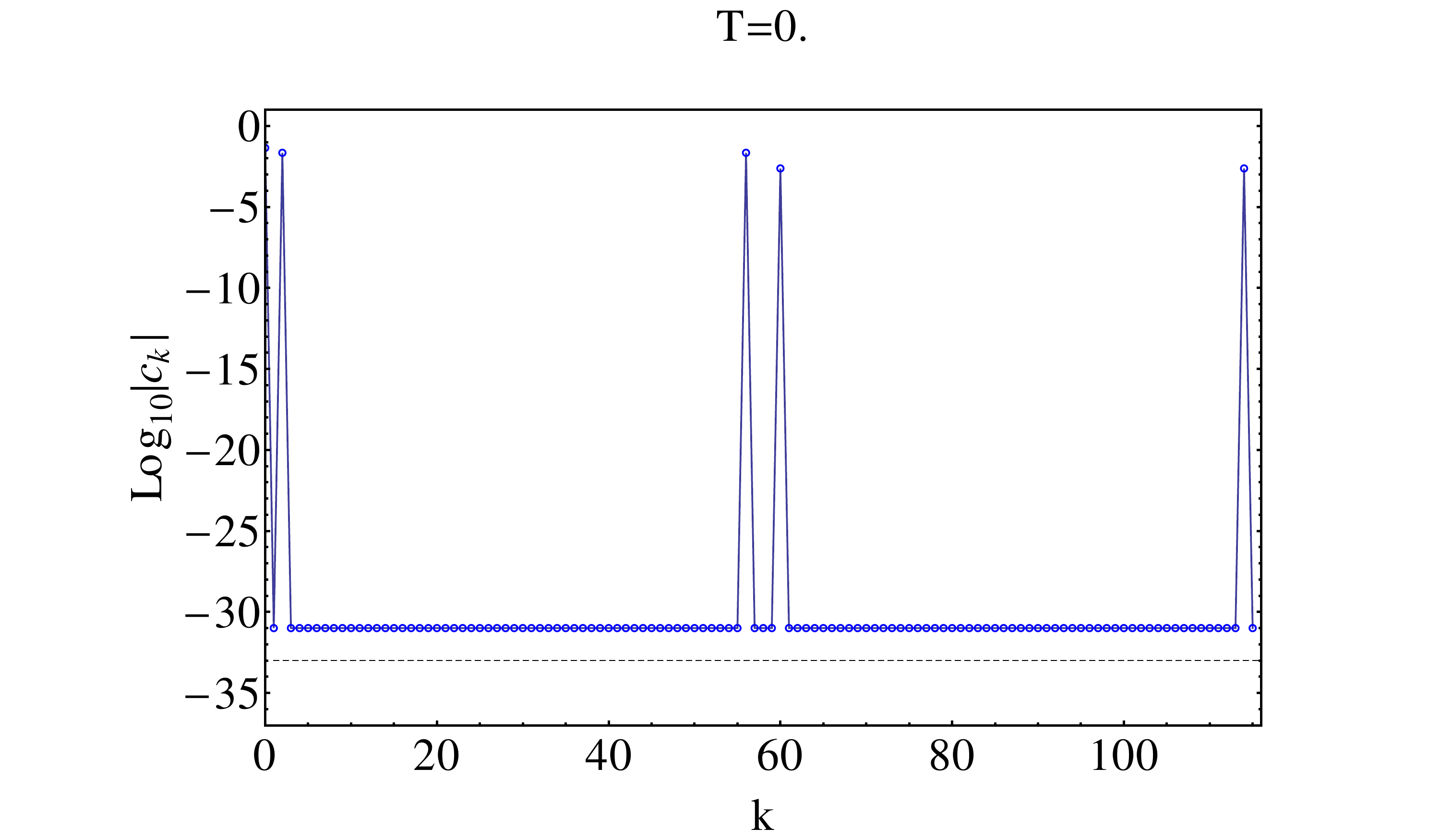}    
  \includegraphics[width=0.49\textwidth]{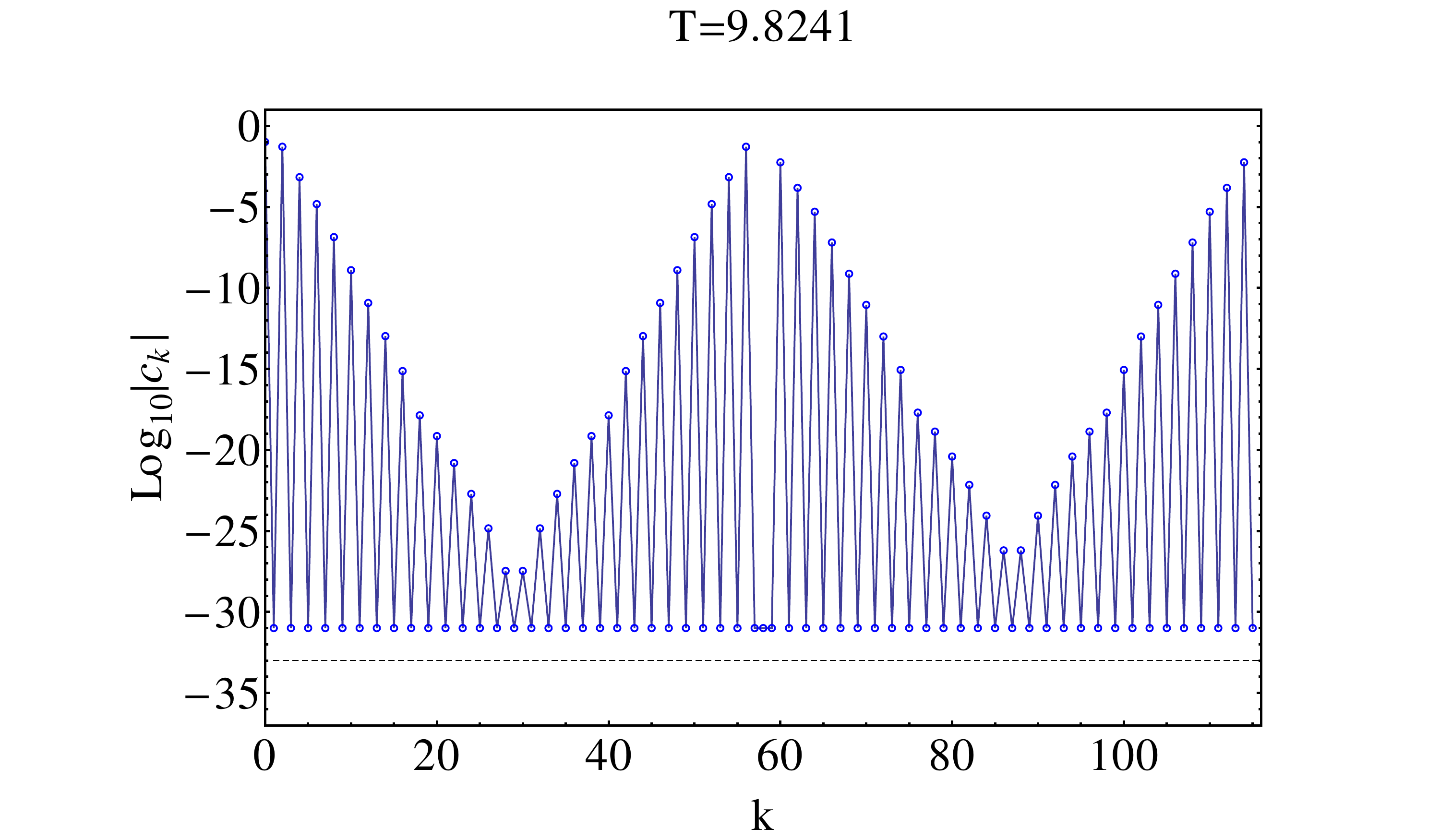}   \\
  \includegraphics[width=0.49\textwidth]{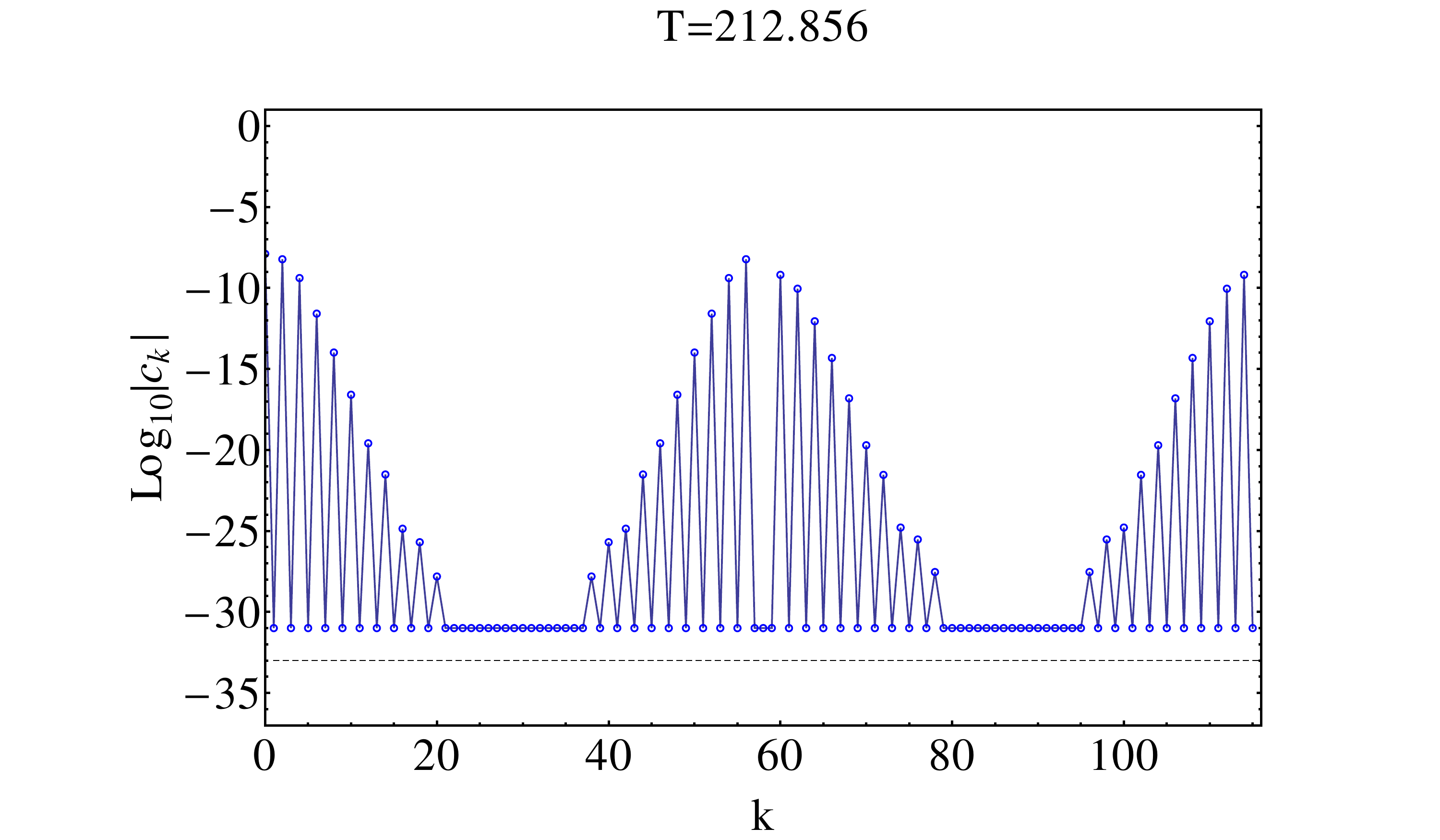}   
  \includegraphics[width=0.48\textwidth]{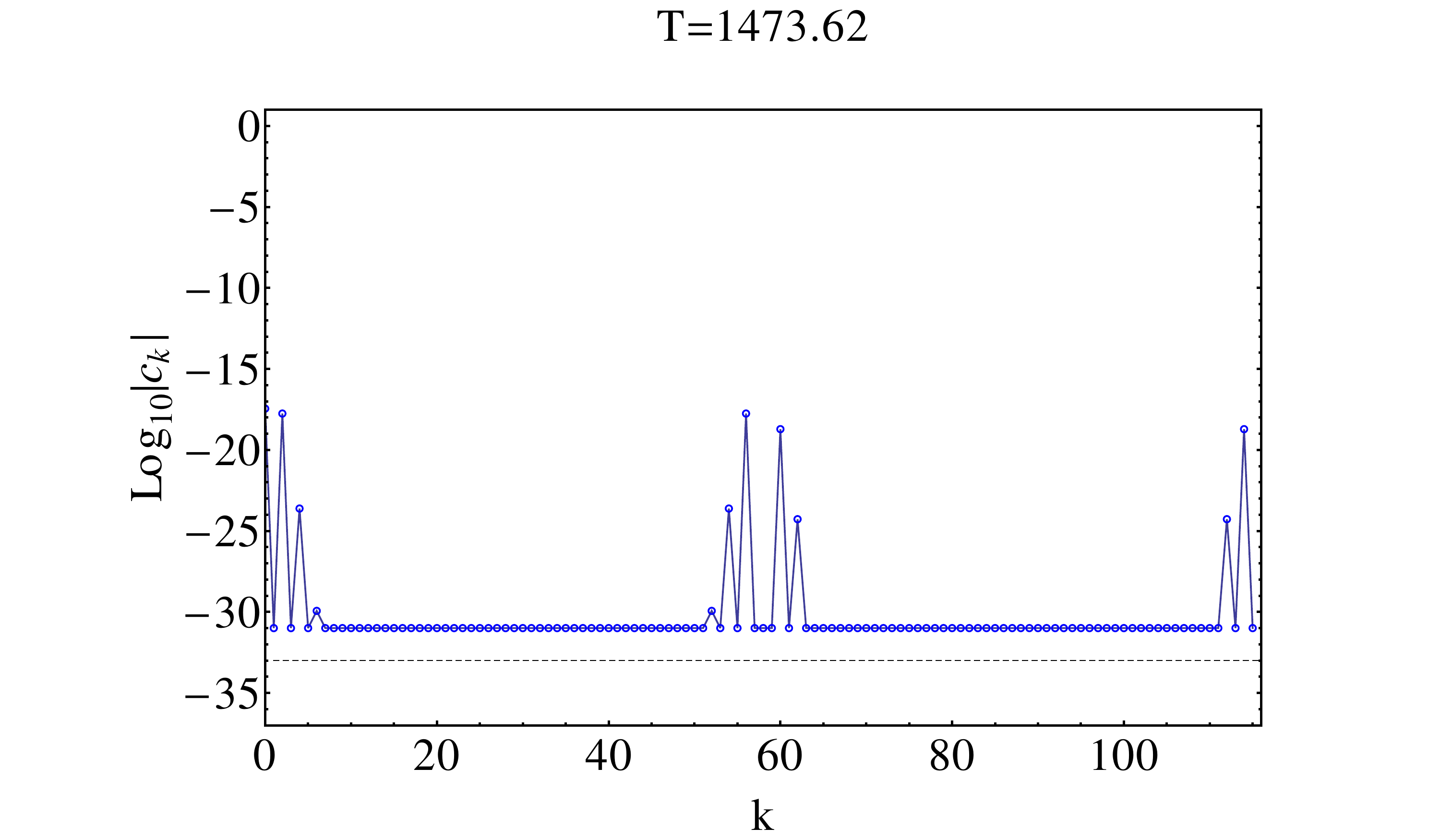}   
  \caption{Evolution of the spectral Fourier coefficients of $\Re\{\psi\}$. 
    Given $n_\theta=29$ angular grid points, the extended (double covering) function 
    is represented by $2n_\theta=58$ complex Fourier coefficients. The real parts are the 
    first $58$, the imaginary parts the second $58$. High-frequencies correspond
    to coefficients in the middle of each part. 
    The left top panel shows that an initially pure multipole can be represented with
    a few points, but, as a consequence of mode mixing, higher mode 
    coefficients are needed to describe the solution (top  right panel). 
    At late times only a few, low-frequency, coefficients are needed
    because only the lowest multipoles, with slower decay rates,
    have amplitudes above round-off level (bottom panels).} 
  \label{fig:Fou_Coeffs}
\end{figure}

\begin{figure}[t]
  \centering
  \includegraphics[width=0.49\textwidth]{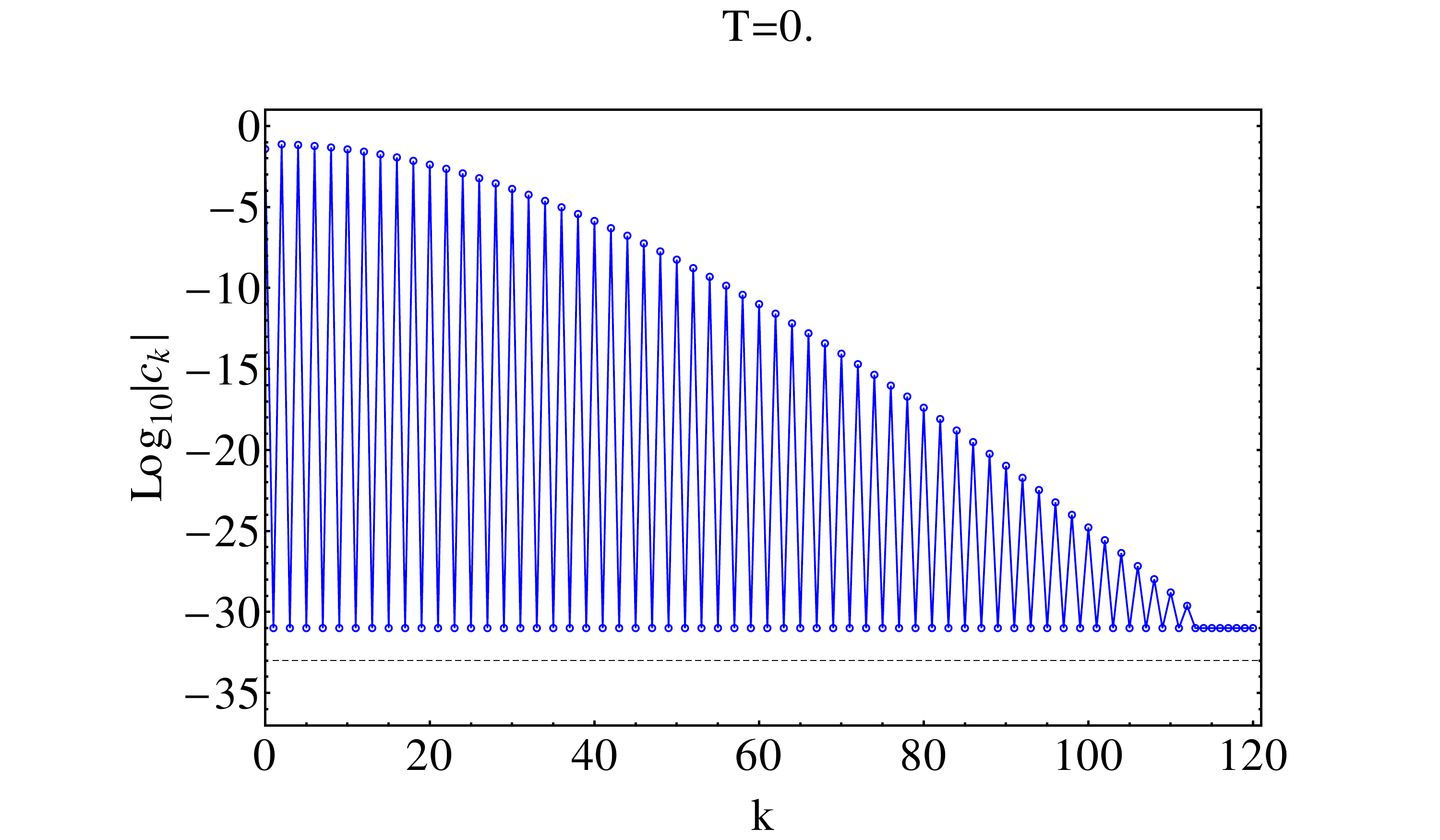}    
  \includegraphics[width=0.49\textwidth]{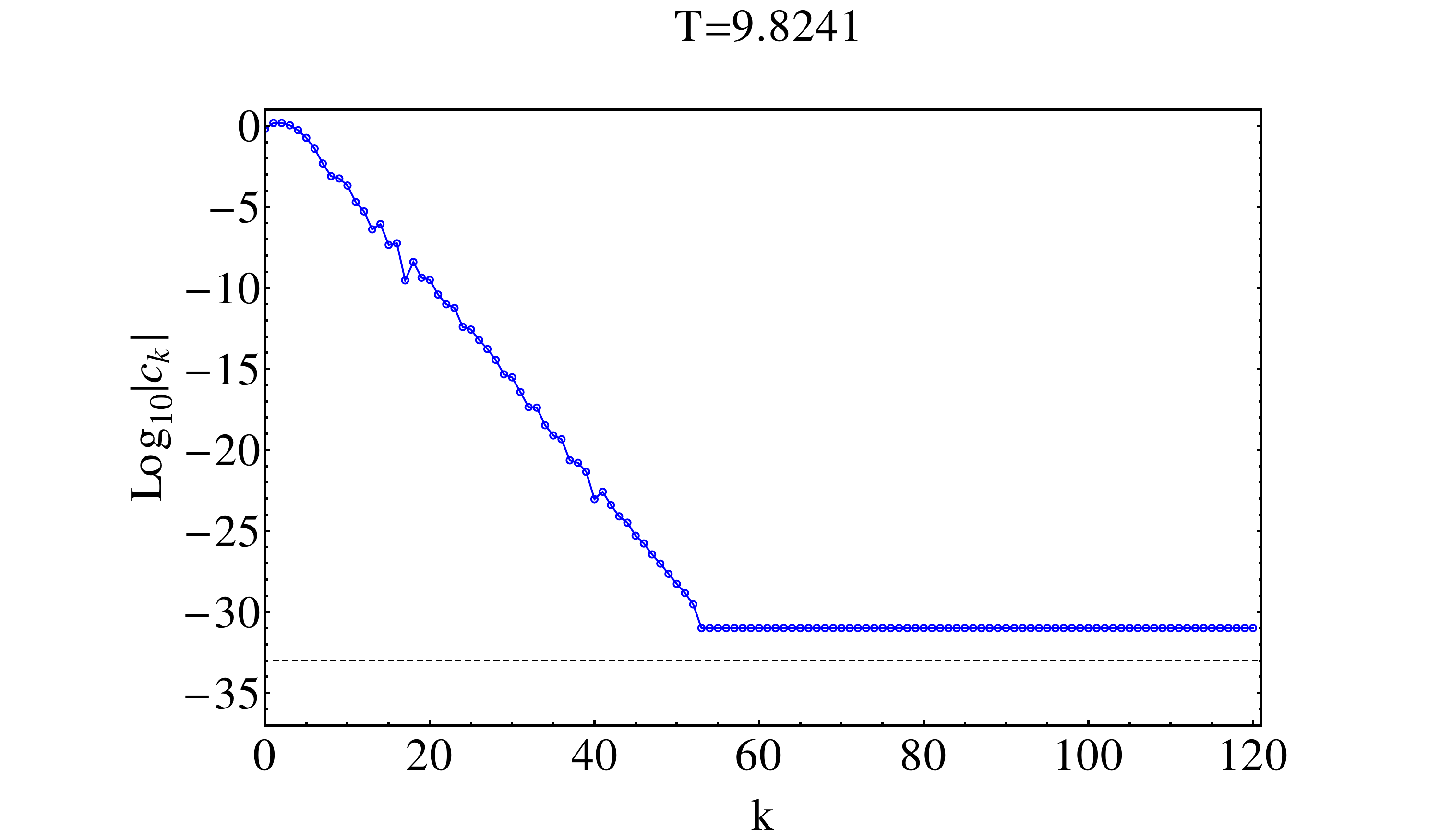} \\
  \includegraphics[width=0.49\textwidth]{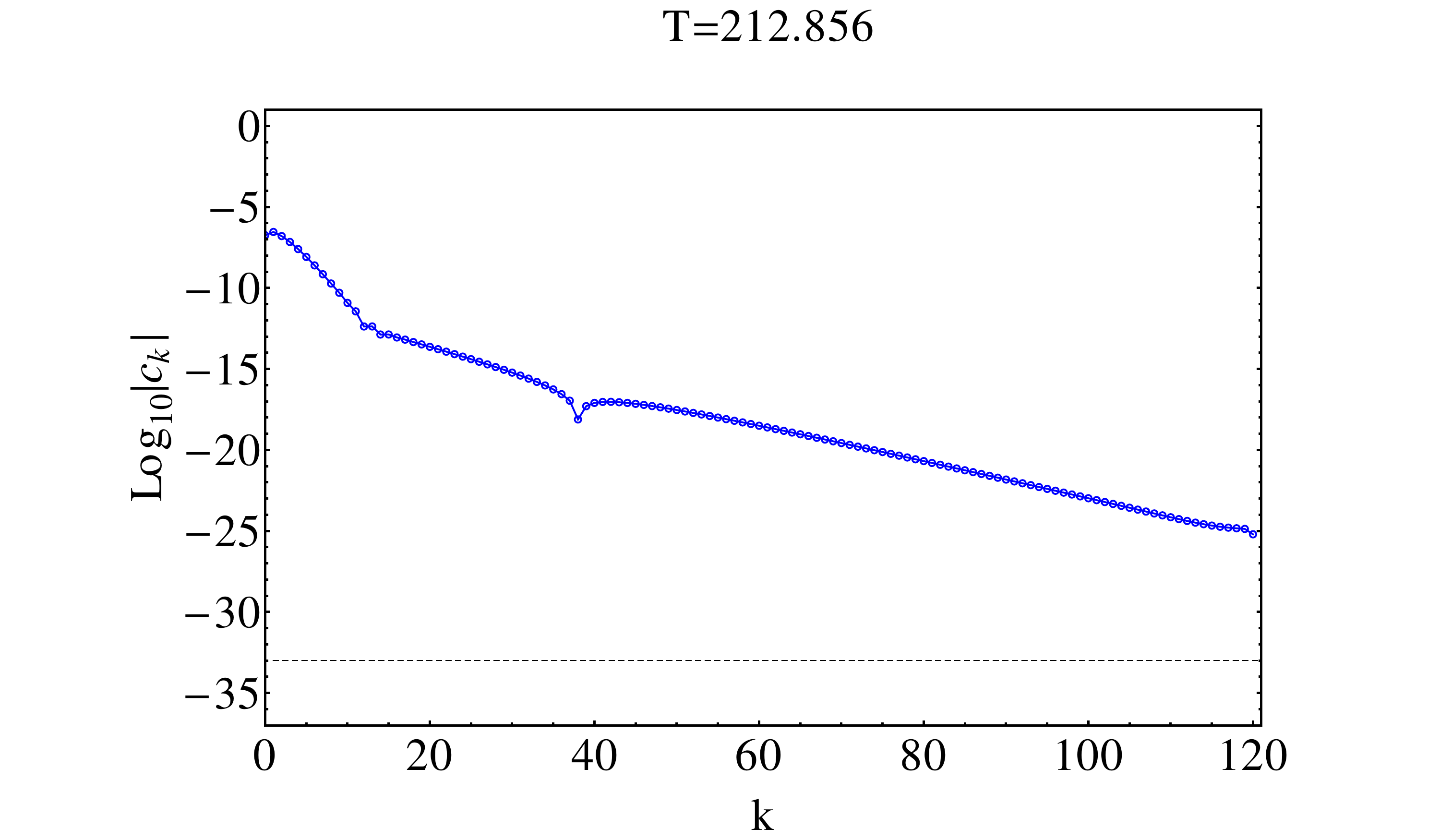} 
  \includegraphics[width=0.49\textwidth]{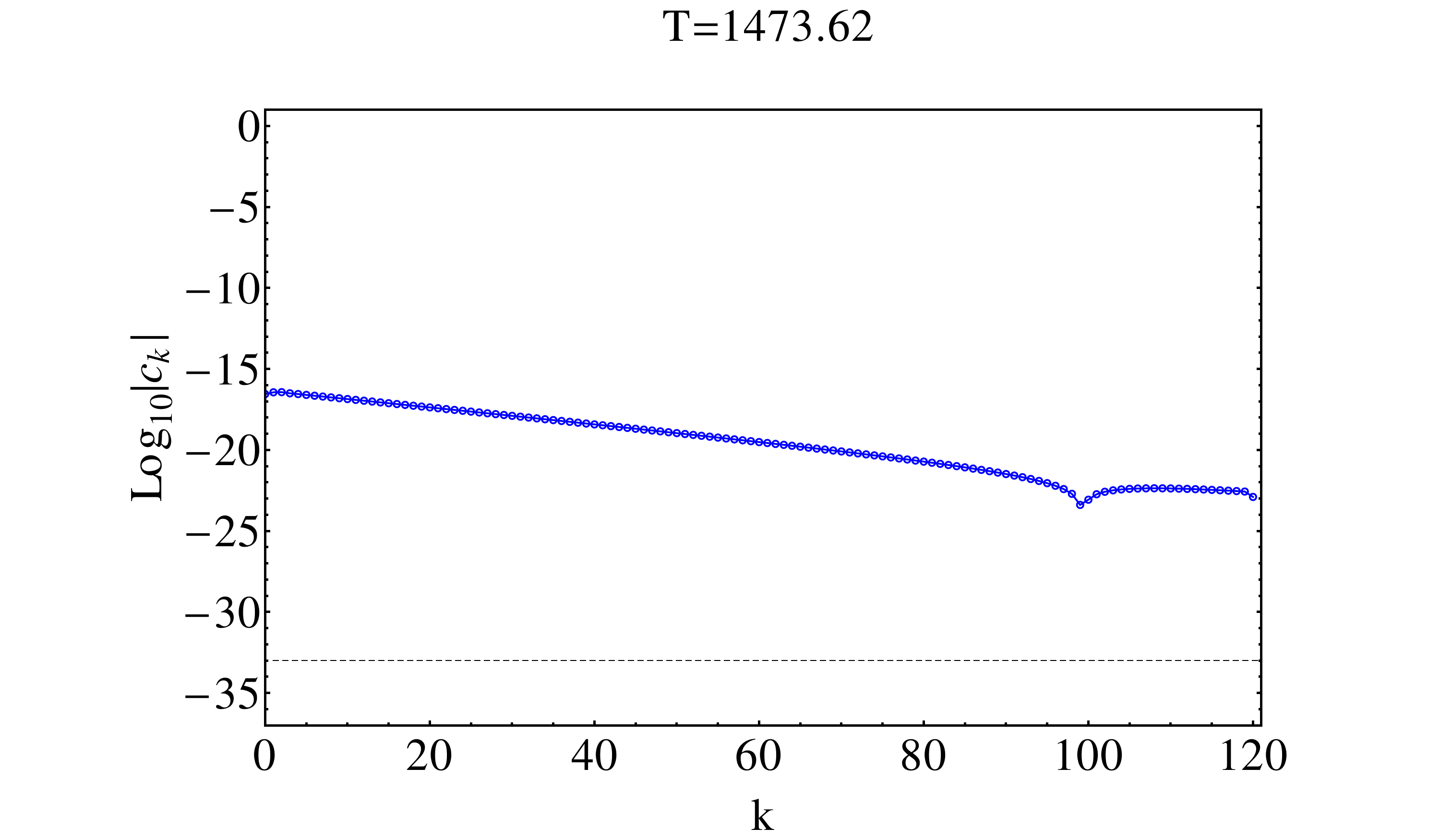} 
  \caption{Evolution of the spectral Chebyshev coefficients of $\Re\{\psi\}$ 
    using $n_r=121$ points.
    The top left panel illustrates that a narrow initial Gaussian needs already $n_r=110$
    points to be represented at round-off level. The top right panel
    shows that spectral convergence is achieved and at early times the
    solution is represented to round-off level.
    Due to the intrinsic form of the solution at late times, illustrated
    in Fig.~\ref{fig:step_function}, progressively more coefficients
    are needed to reach round-off in the tail phase (bottom left
    panel). As long as $|c_0|/|c_{n_r}|\geq 10^6$ (bottom right panel) the
    solution is not corrupted by high frequency noise.}  
\label{fig:Ch_Coeffs}
\end{figure}

\begin{figure}[t]
  \centering
  \includegraphics[width=\textwidth]{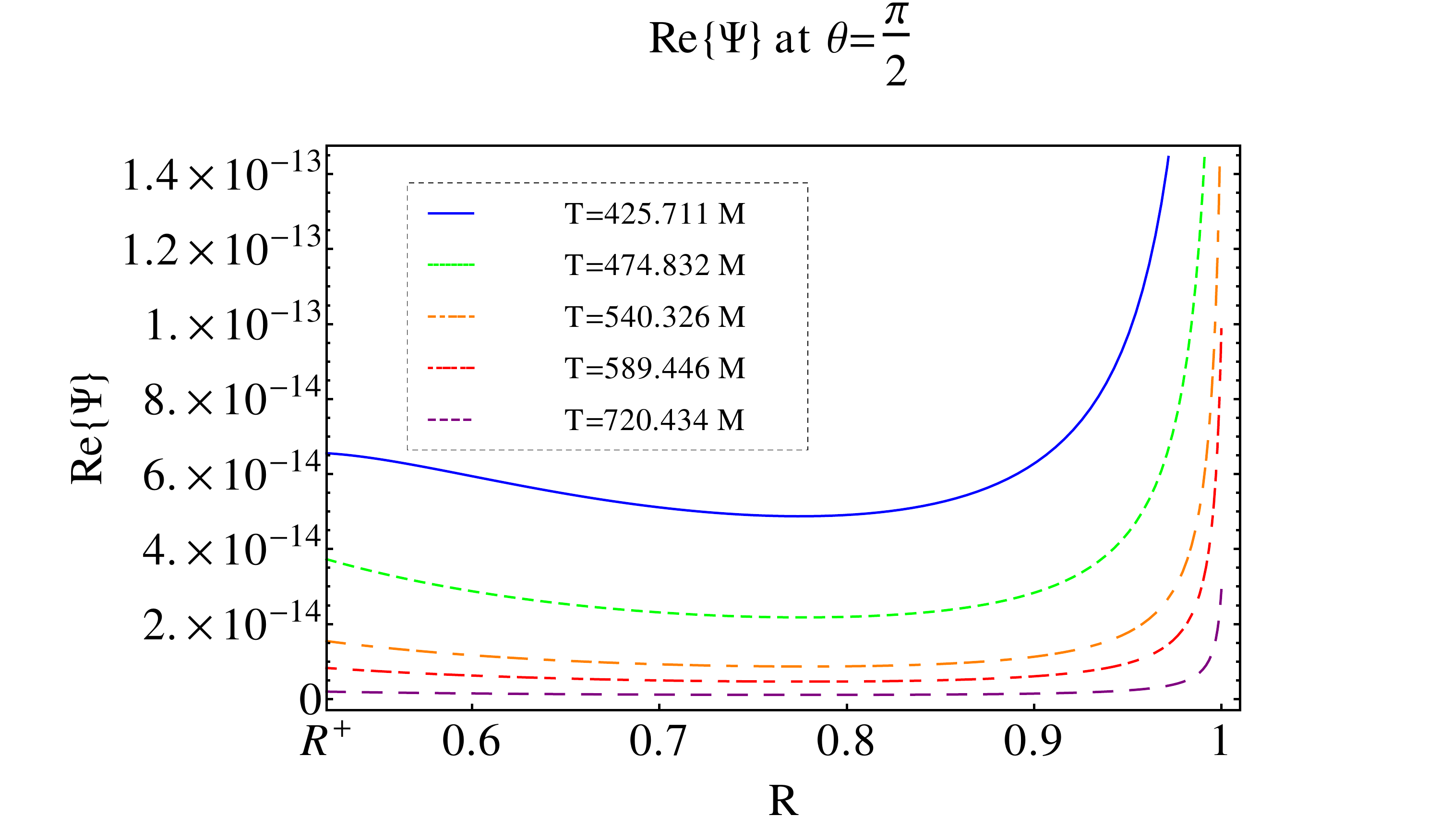}    
  \caption{
    Late time radial profile of the solution.
    Due to the slower decay rate at $\scri^+$ the solution
    approaches a step-function like form at late times. This corrupts
    the spectral convergence of the Chebychev expansion.
    The figure refers to $s=-2$, $a=0.9$, $l'=2$ and ID0 data.
    The problem is worse for the cases $s>0$ because
    the differences between the decay rates at $\scri^+$ and finite
    radii are larger.}   
  \label{fig:step_function} 
\end{figure}

Spectral convergence is monitored by looking at the 
expansion coefficients, $c_k$, of the spectral approximation. The
highest accuracy achievable in the  
description of a {\it smooth} function is determined by round-off
errors. Spectral convergence implies that the magnitudes of the
coefficients $c_k$ decay exponentially in $k$. 
We discuss here convergence of the PS code for a representative case, 
namely a $s=-2$, $a=0.9$ simulation with ID0, $l'=2$, and quadruple
precision. We experimentally find that large CFL factors can be used
without encountering stability problems, 
consistent with the small coordinate speeds of
 Eq.'s~(\ref{eq:values_speeds_a0})-(\ref{eq:values_speeds_a9}).
For example here we employ
$C_{\rm CFL}=100$, to be compared with the inverse of the (absolute
value of the) maximum speed $1/|c_{R,max}|\approx1/0.0827\approx12.1$. 
This behaviour is likely to be related to the use of the Chebychev
grid (clustering at the extrema), the Runge-Kutta scheme, and the fact
that the coordinate light speeds in the bulk of the domain are very
small, see Fig~\ref{fig:speeds}.
Note that this is favorable for long-term evolutions and late decay studies. 
Figures~\ref{fig:Fou_Coeffs} and~\ref{fig:Ch_Coeffs} show
the Fourier and Chebyshev expansion coefficients, respectively, of the
variable $\Re(\psi)$. The horizontal dashed line at
$10^{-33}$ indicates the nominal round-off level for the quadruple
precision employed.   

The plot in Fig.~\ref{fig:Fou_Coeffs} refers to a simulation with
$n_\theta=29$ angular points which amounts to $2n_\theta=58$ complex
coefficients for the extended function (double covering.) The coefficients' real parts
are the first $58$, the imaginary parts the other $58$. The coefficients
corresponding to highest frequencies are placed in the middle of the
real and imaginary sequences.
The round-off level is reached in these regions, i.e.~$k\sim36$ and $k\sim94$.
At time $T=0$ all but five coefficients are zero. This is because the
spin-weighted spherical harmonic 
${}^{-2}Y_{20}\propto \sin(\theta)^2 = (1 - \cos(2 \theta))/2$ and its
derivative can be exactly described using the five functions $\{1,\cos(\pm2
\theta),\sin(\pm2\theta) \}$. During the evolution (e.g.~$T\sim9.8$, top right panel 
of Fig.~\ref{fig:Fou_Coeffs}) all other even-indexed frequency modes are excited, as a consequence 
of mode mixing. High frequency modes die off faster because they have longer 
QNM ringing phases and their subsequent power law decays are faster. That is why 
progressively fewer coefficients are necessary to describe the
solution at late times (see two bottom panels). In the late stages of the simulations only 
the lowest modes survive and only very few coefficients are required.
Note that due to axisymmetry ($m=0$) only even-indexed coefficients are
different from zero in $\Re(\psi)$. 

The evolution of the Chebyshev expansion coefficients is shown in
Fig.~\ref{fig:Ch_Coeffs}. The initial Gaussian with $w=3000$ requires $n_r\sim100$
in order to be represented at round-off level. Every second coefficient
is vanishing because the Gaussian is set to the center of the domain
making it an entirely even function. 
At early times we observe spectral convergence, but a slower rate of 
convergence takes progressively over, and more coefficients (large
$k$) are required to describe the function. The use of $n_r \sim100$ does
not allow to reach round-off quadruple precision at late times.
This behaviour is due to the intrinsic form of the solution. 
Because the decay rate at $\scri^+$ is slower than at the horizon, the
solution develops large gradients for $R\to1$, approaching a
step-function-like form at late times. This is clearly described in
Fig.~\ref{fig:step_function}.
Experimentally we observe that until $|c_0|/|c_{n_r}|\geq 10^6$ the
simulations are stable, while for $|c_0|/|c_{n_r}|<10^5$ we observe
some high-frequency noise which, eventually, corrupts the simulation
and precludes the assessment of the decay rates. 
In the cases $s<0$, we found that
the use of $n_r \sim121\div141$ points and quadruple precision allows 
to simulate up to $T\sim3000\div4000$ and measure the decay rates 
with acceptable accuracy (see Sec.~\ref{subsec:LPI}). It may be possible to introduce a modification
of the RT coordinates that counters the ``piling up'' of late tails near the outer
boundary to preserve accuracy and convergence for longer times.

For $s>0$ resolving the radial profile of the late time solution is 
more challenging. The differences in the decay rates at
$\scri^+$ and finite radii are much larger than for $s<0$ cases, and a
large number of spectral coefficients is required. 
As an example one can consider the decay rates of the overall field of
$s=\pm2$ perturbations with $l'=2$ and ID1. 
For $s=-2$ the rates at finite radii and $\scri^+$ are $\mu=7$ and
$\mu=6$ which differ by only one power. 
The corresponding rates for $s=+2$ are $\mu=8$ at the horizon and
$\mu=2$ at $\scri^+$, which amounts to a difference of six powers and
produces large spatial gradients.  
In these situations about $n_r\sim400$ points are
needed in order to have stability at late times and measure cleanly the tail decays. 
Because of the quadruple precision, one of such simulations can take up to a
month on a modern workstation (the code is serial). 
However, stable simulations at a more reasonable computational cost can
be performed using FD in radial direction. Typical runs are performed
with $n_r=801$ (or $n_r=1601$), and take about ten hours to
reach $T\sim1500M$, which, in turn, is sufficient to compute the decay rates. 
Obviously, also in this case, convergence is not completely
under control and it is not possible to give a proper error estimate. 
However, we argue that these results are, at least, {\it qualitatively}
correct because they are compatible with those obtained with the PS
code. We compared the FD results, for several $s\leq0$ cases and few $s>0$
cases, with high resolution PS runs and found very good
agreement. 
As discussed in detail in the next section, the decay rates for $s>0$ computed 
with the PS derivative and high resolution typically agree
within $0.8\%$ with those computed with the FD derivative. Finally, because all the 
$s>0$ results agree with the analytical expectation, we can verify
their correctness a posteriori.

\subsection{Local-power-index calculation}
\label{subsec:LPI}

\begin{figure}[t]
  \centering
  \includegraphics[width=\textwidth]{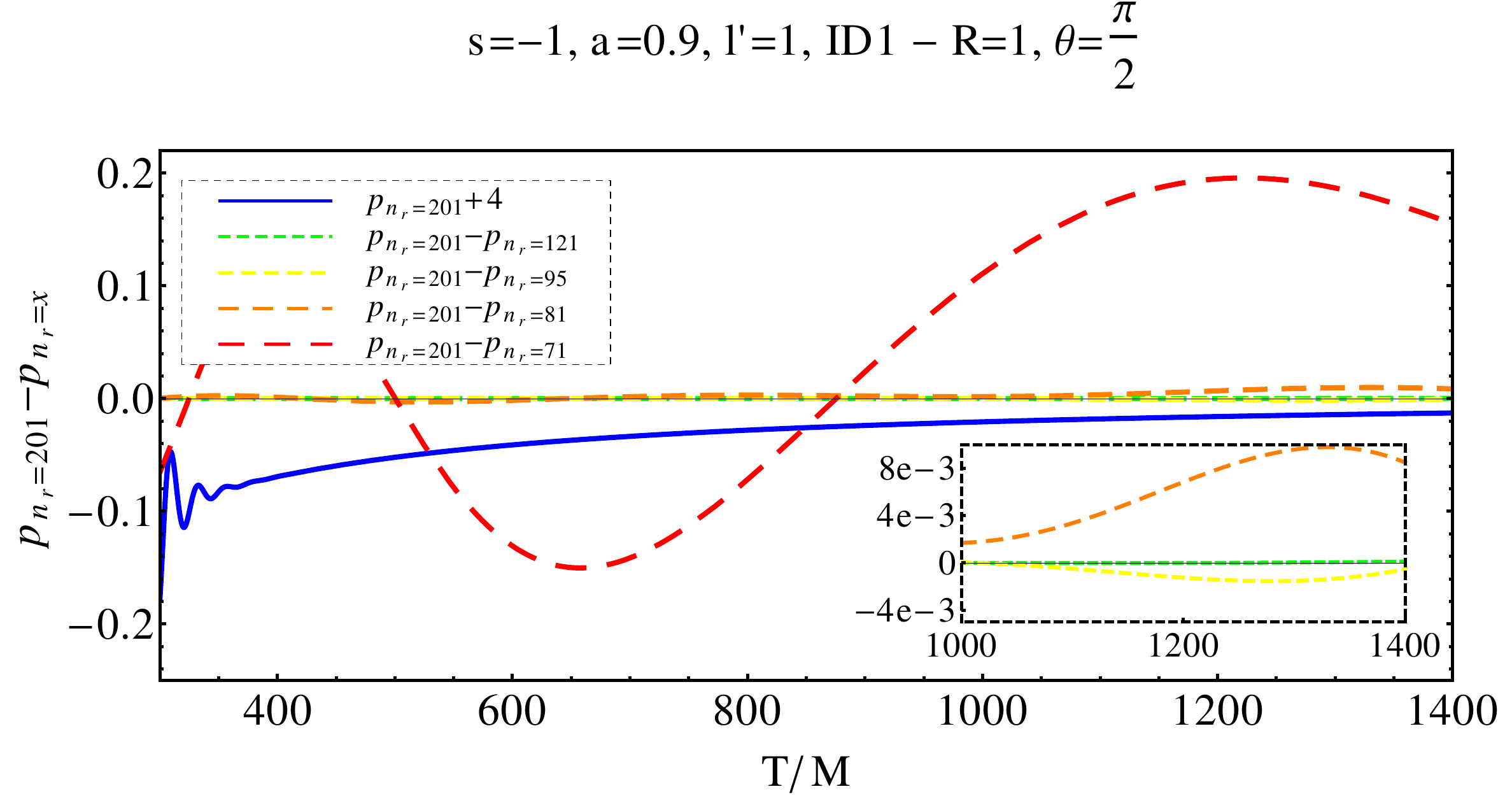}  
  \caption{Assessment of LPI accuracy with PS radial differentiation. The solid blue line shows 	
    the difference between the LPI of the $l=1$ projection at $\scri^+$ computed from a 
    $n_r=201$ simulation and the analytical asymptotic value ($p=-4$). The dashed lines show the 
    differences between the LPIs computed from simulations with different $n_r$. The plot 
    shows that the deviation from the expected value is less than $0.5\%$ at late times 
    (solid blue line). The self-differences with the
    target resolution $n_r=201$ are below $0.05\%$ for $n_r=121$, of
    order $0.5\%$ for $n_r=81$, and of order $5\%$ for $n_r=71$.
    The data refer to simulations with $n_\theta=29$ of a $s=-1$
    perturbation with $l'=1$ and ID1. Extraction is done at $R=1$ and $\theta=\pi/2$. The 
    runs were done with long-double precision.}  
  \label{fig:LPI_Accuracy}
\end{figure}

\begin{figure}[t]
  \centering
  \includegraphics[width=0.49\textwidth]{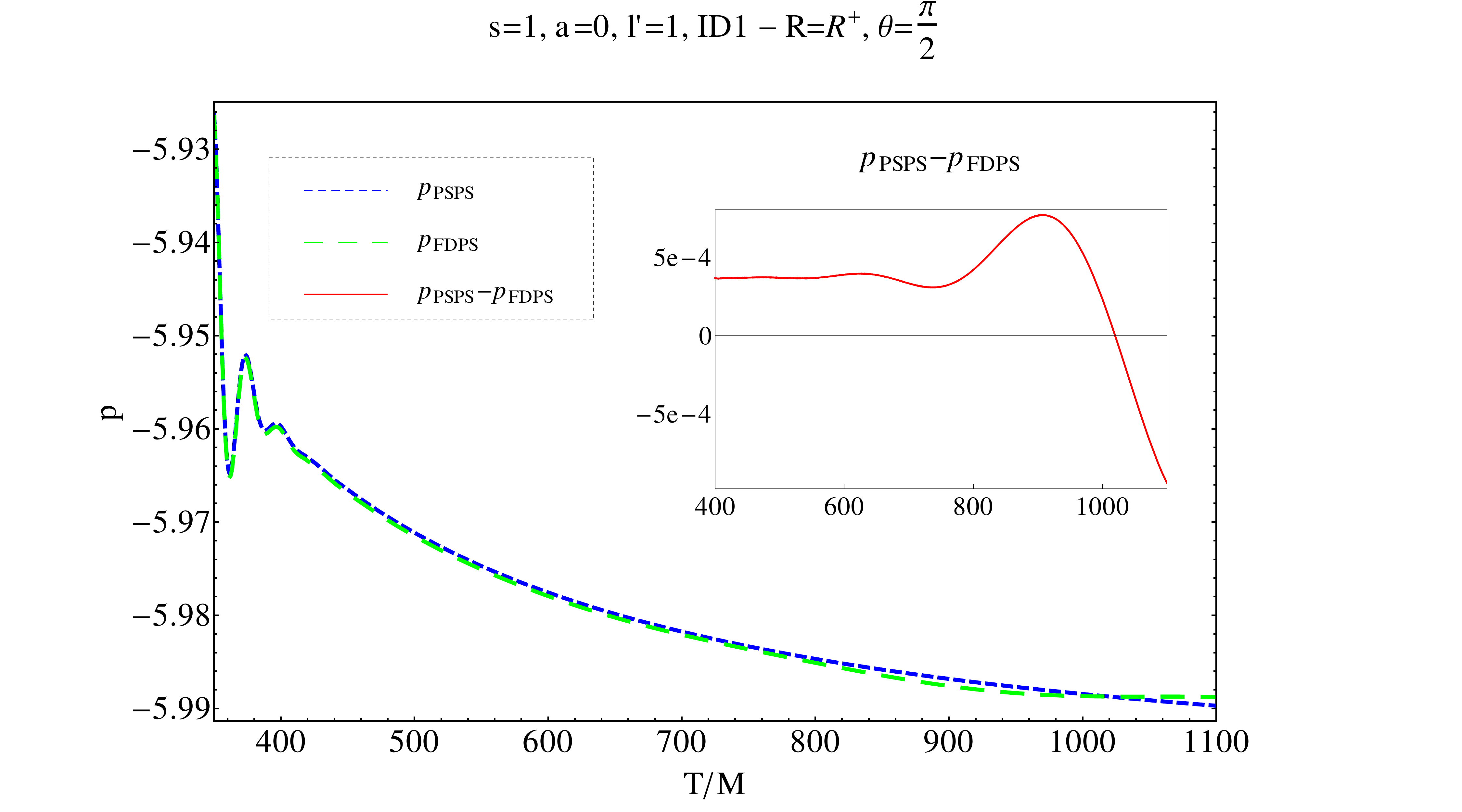} 
  \includegraphics[width=0.49\textwidth]{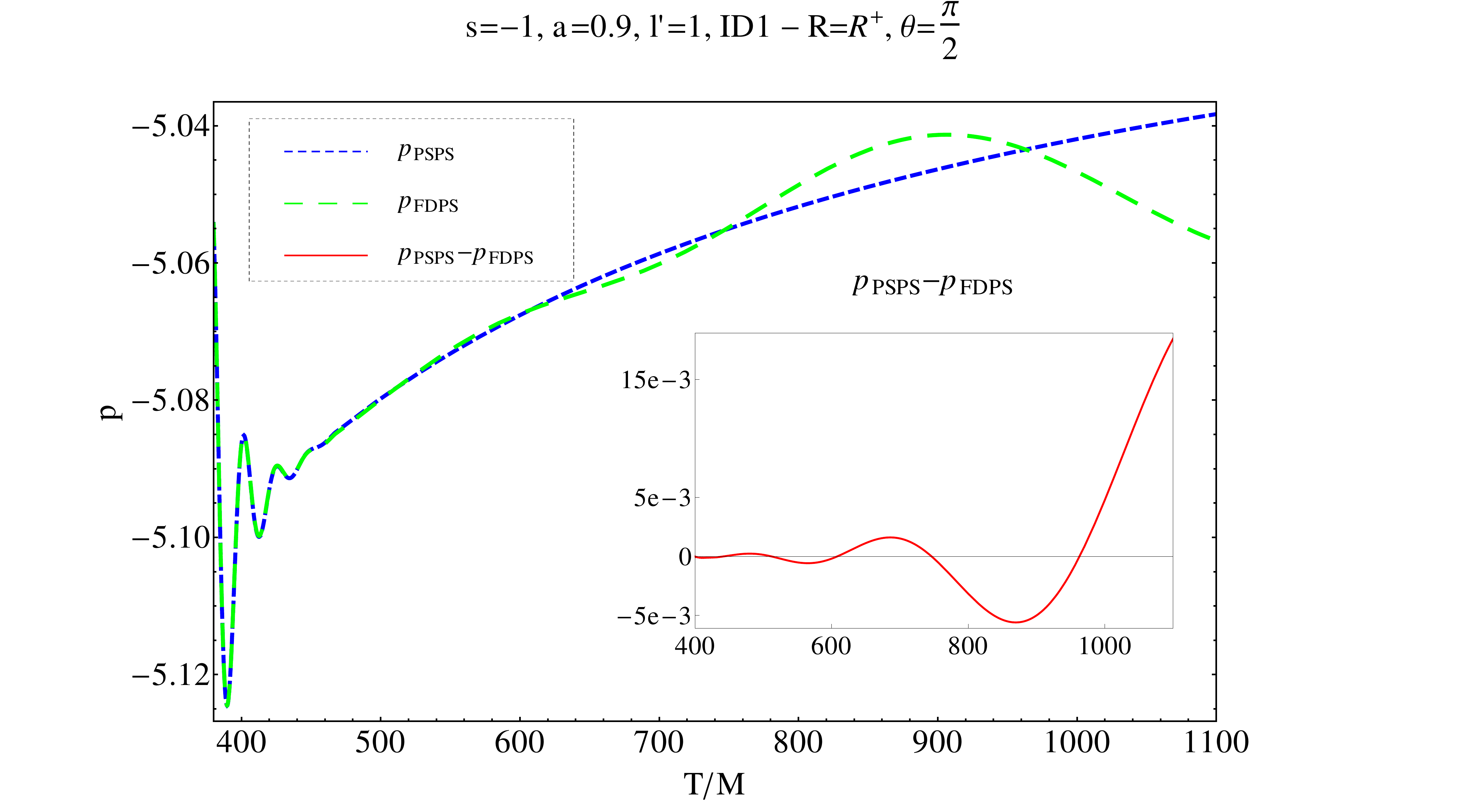}  
  \caption{Assessment of LPI accuracy with FD radial differentiation.

    The left panel compares the LPIs of the overall field of a  $s=+1$ 
    perturbation with $l'=1$ and ID1 as obtained from simulations with
    PS radial differentiation with $n_r=401$ and with FD radial 
    differentiation with $n_r=801$. Both LPIs are clearly approaching
    the analytic prediction ($p=-6$) (dashed blue and green
    lines). The inset shows their difference $\Delta p=p_{\rm PS}-p_{\rm FD}\sim10^{-3}$ at $T=1000M$. 
    The right panel compares the LPIs of the overall field of a  $s=-1$ 
    perturbation with $l'=1$ and ID1 as obtained from simulations with
    PS radial differentiation with $n_r=141$ and with FD radial 
    differentiation with $n_r=801$. Both LPIs are clearly approaching
    the analytic prediction ($p=-5$) (dashed blue and green
    lines). The inset shows their difference $\Delta p=p_{\rm PS}-p_{\rm FD}\sim10^{-2}$ at $T=1000M$. 
    Both data refer to simulations with $n_\theta=29$.
    Extraction is done at $R=R_+$ and $\theta=\pi/2$.}
  \label{fig:LPI_Accuracy_PSFD}
\end{figure}

The power law decay of the field is monitored by the local power index
(LPI)~\cite{Burko:1997tb}, calculated as 
\be
p = \frac{\p \log|\psi|}{\p\log T} 
= T \frac{\Re(\psi)\Re(\partial_T \psi) +
  \Im(\psi)\Im(\partial_T\psi) }{\Re(\psi)^2+\Im(\psi)^2}  \ , 
\ee
where the reduction variable $\partial_T\psi$ is employed directly
without taking a numerical derivative. Asymptotically in time, the LPI
approaches (minus) the decay rate $p\to-\mu$.
The LPIs of the projected modes, $p_l$, are
calculated analoguosly considering the projections of the
fields. For a field of spin $s$ composed of a single azimuthal
$m$-mode, $\psi=\psi_m\,e^{i m \varphi}$, 
the projections are computed as
\be
\psi_{lm}(T,R) = \langle {}^{s}Y_{lm}|\psi\rangle= 
\int_0^{2\pi} \, d\varphi\, \int_0^{\pi} \, d\theta \, \sin(\theta) \, 
\psi_m(T,R,\theta) e^{i m \varphi} \, {}^sY_{lm}^*(\theta,\varphi) \ .
\ee
Note that the integration over $\varphi$ is analytical and reduces to an
overall factor $2\pi$. 
The integration over $\theta$ needs to be very accurate to resolve the
excited higher modes $l>l'$ which have small amplitudes. Using the 
Riemann sum or the trapezoidal rule on $(0,\pi)$ is too inaccurate
for $n_\theta=29$ angular points. We performed a spectral
integration in the following way. Given a periodic function on $[0,2\pi]$ and
its spectral expansion $f(\theta)\sim\sum_k c_k e^{i k \theta}$, the
primitive is approximated by $F(\theta)=\int d \theta 
f(\theta)\sim\sum_k c_k e^{i k \theta} / (ik) $ 
and the integral can be computed as $F(\pi)-F(0)$. The primitive
function must be computed at the points $\theta=0,\pi$ which are not
included in the numerical grid, so we use interpolation. 


The accuracy of this technique was checked by testing the
orthonormal relations
$\langle{}^sY_{lm}|{}^sY_{jm}\rangle=\delta_{lj}$, which
yielded the expected \textit{double} precision (employed in the
post-processing) for the projections up to $l=5$ and $n_\theta=29$. 

Finally, we comment on the accuracy of the computed LPIs.
Figure~\ref{fig:LPI_Accuracy} illustrates how the accuracy of the LPIs
is assessed. The plot refers to the $l=1$ projected mode of a $s=-1$
perturbation with $l'=1$ and ID1. PS derivatives are employed in
radial direction, and $n_\theta=29$. The solid blue line shows the
difference between the LPI at $\scri^+$ computed from a $n_r=201$
simulation and the asymptotic value $p=-4$ analytically
predicted~\cite{Hod:2000fh}. The deviation from the expected value is
less than $0.5\%$ at late times. This is a general result in all
$s\leq0$ simulations. The dashed lines show the differences between
the LPIs computed from the $n_r=201$ (target) simulation and others
with different $n_r$. The difference with $n_r=121$ is below $0.05\%$,
with  $n_r=81$ of order $0.5\%$, and with $n_r=71$ of order $5\%$.
Analogously to this case, we find that $n_r\sim121\div141$ points
give a reasonably accurate LPI up to $T\sim1500M$ for most of the
simulations. 

As discussed in Sec.~\ref{subsec:conv} for $s>0$ perturbations
derivatives are computed using FD in the radial direction for
efficiency. Specifically, sixth order stencils are employed with no
need for artificial dissipation. The CFL factor is $C_{\rm CFL}=20$.
To assess the accuracy of the LPIs, the outcomes of FD 
and high-resolution PS simulations are compared for a few test cases.
A typical result is shown in Fig.~\ref{fig:LPI_Accuracy_PSFD}, which
refers to a $s=\pm1$ perturbation with $l'=1$ and ID1. 
For $s=+1$ (left panel) the PS data are computed with $n_r=401$
simulations, while the FD data with $n_r=801$. In both cases the LPIs
agree asymptotically with the predicted analytical values ($p=-6$) (dashed blue and green
lines) within $0.16\%$ at $T\sim 1100M$. They differ from each other by
$\sim10^{-3}$ ($\sim0.02\%$), as shown by the solid red line in
the inset.  
Also for $s=-1$ (and in general for $s\leq0$) the FD
differentiation gives accurate results, while not used because for
$s\leq0$ the PS code can be used more efficiently. The right panel of
Fig.~\ref{fig:LPI_Accuracy_PSFD} shows that PS data with $n_r=141$
differ from FD data with $n_r=801$ by $\sim0.4\%$ at $T\sim 1100M$
(see inset). In this case the FD simulation looses accuracy  
earlier and the green line deviates from the blue line. Still, the LPI is only 
$1.2\%$ off the predicted asymptotic analytical value ($p=-5$) at
$T=1100M$.

\section{Power law tails for axisymmetric perturbations}
\label{sec:decaym0}

In this section we report the measured decay rates for
$s=0,\pm1,\pm2$ axisymmetric perturbations. 
In all cases we have considered pure
spin weighted spherical harmonics multipole initial data with an
axisymmetric profile, i.e.~$\psi(\theta)\propto {}^sY_{l' m'}(\theta)$ with
$m'=0$. The black hole angular momentum is fixed to $a=0.5$ or
$a=0.9$. 
The parameters of the function $G(R)$ are $w=3000$ and $R_0=0.8$. 
For $s\leq0$ simulations we used PS differentiation in
radial direction, while for $s>0$ FD for the reason
discussed in Sec.~\ref{sec:exp}. We used grids of $141\times29$ and
$801\times29$ points for the spectral and finite difference
simulations, respectively. The CFL factor in the PS (FD) simulations is
$C_{\rm CFL}=100$ ($C_{\rm CFL}=20$ .) 
Quadruple (long-double) precision was
employed in the $s\leq0$ ($s>0$) cases.

The decay rates of the projected modes, $\mu_l$, are given in
Tabs.~\ref{tab:s0_LPI},~\ref{tab:s1_LPI}
and ~\unskip\ref{tab:s2_LPI}
for $s=0,\pm1,\pm2$ perturbations. The tables refer to the case
$a=0.9$. 
In the discussion of the decay rates we use the terms up/down-modes as
introduced by~\cite{Burko:2010zj}. Down-modes refer to the below
diagonal part of the tables, i.e.~$l<l'$, up-modes refer to the upper
diagonal and the diagonal part of the tables, i.e.~$l\geq l'$.  
The notation in these tables
follows basically the one used by~\cite{Racz:2011qu} but shall be
clarified here. In all cases the asymptotic in time decay rates at 
the horizon and finite radii are different from those at future null
infinity. Thus two $\mu_l$ values are reported separated by a vertical line
like $x|z$, where $x$ refers to the horizon and finite radii and $z$
to $\scri^+$. The LPIs at finite radii $R+<R<1$ vary
monotonically between $-x$ and $-z$ at early times and reach the value
$-x$ only asymptotically at late times. In practice the $x$ value in
the tables is taken close to the horizon. In the cases $s>0$ and $m=0$ 
there are three distinct asymptotic LPIs corresponding to different
decay rates at the horizon, finite radii (``finite radii'' in these
cases means also ``excluding the horizon''), 
and $\scri^+$. The tables have thus three entries $x|y|z$. The 
decay rate at finite radii $y$ is taken at a radius
$R_+<\tilde{R}<1$ where the asymptotic value is reached first. 
Typically $\tilde{R}\sim 1.05\div1.2\, R_+$, 
so it is very close to the horizon, but the exact position is different
for each data set. The larger the distance $|R-\tilde{R}|$, 
the greater the differences between the LPIs $|p(R)-p(\tilde{R})|$ 
at early times, but, asymptotically in 
time, $p(R)\rightarrow p(\tilde{R})$ for any finite radius
$R_+<R<1$. For $R>\tilde{R}$ 
the LPIs vary monotonically between the values $-y$ and $-z$ (as long as
an up-mode does not show splitting). For $R<\tilde{R}$ the transition
from $y$ to $z$ depends on the value of $s$ (details are given in the
following.)  

In some cases there are uncertainties in the assessment of the LPI
due to inaccuracies. These can either originate from the loss of
convergence in the radial direction due to the unfavorable form of the
solution as discussed in Sec.~\ref{subsec:conv} or to unresolved
higher modes~\unskip\footnote{ %
  Because the decay rate $\mu_{l^*}$ of a given projected mode $l^*$
  depends on several other modes $l>l^*$ excited by the initial data
  $l'$, the accuracy of $\mu_{l^*}$ also depends on how well the
  other modes $l>l^*$ are actually resolved.}.
The symbol $\xx$ is employed to indicate that the loss of
accuracy at late times prevents an unambiguous determination of the
LPI. In cases of LPI splitting we state the decay 
rate at finite radii in bold phase, ${\bf y}$, to emphasize that in 
our relatively short-time simulations we can not confirm its 
validity for every observer. In particular, the asymptotic decay
rate in up-modes with splitting can not always be assessed for observers
$R\to1$.
(In absence of splitting the LPI measurement is position dependent at
early times, but in this case there is an unambiguous trend towards
the {\it same} asymptotic decay rate at any finite radii.) 
In case of exclusion due to symmetries we use the horizontal
line $-$ and if we did not simulate that case we use the slash $/$.

\subsection{Decay rates for $s=0$ perturbations}
\label{subsec:lpi_s0}

\begin{figure}[t]
  \centering
  \includegraphics[width=0.49\textwidth]{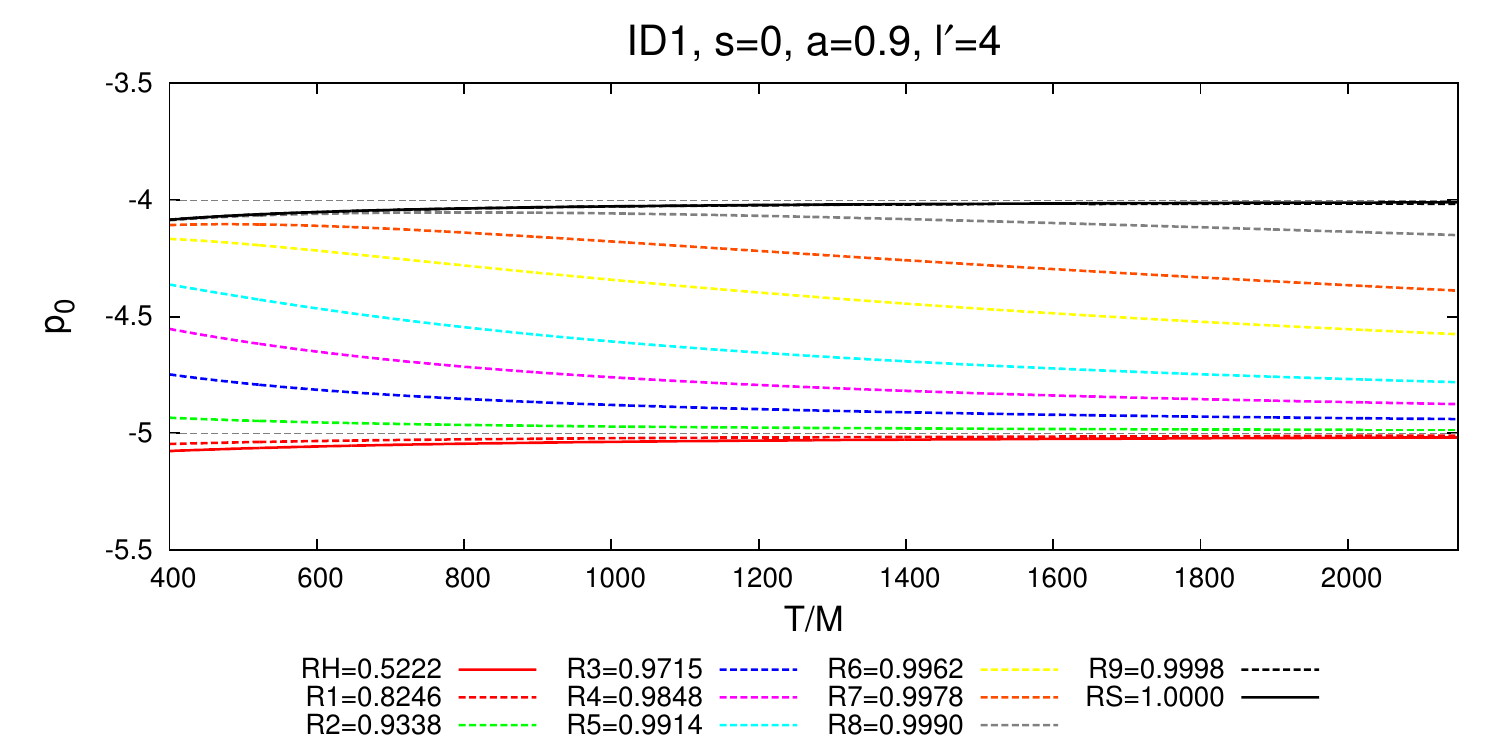}  
  \includegraphics[width=0.49\textwidth]{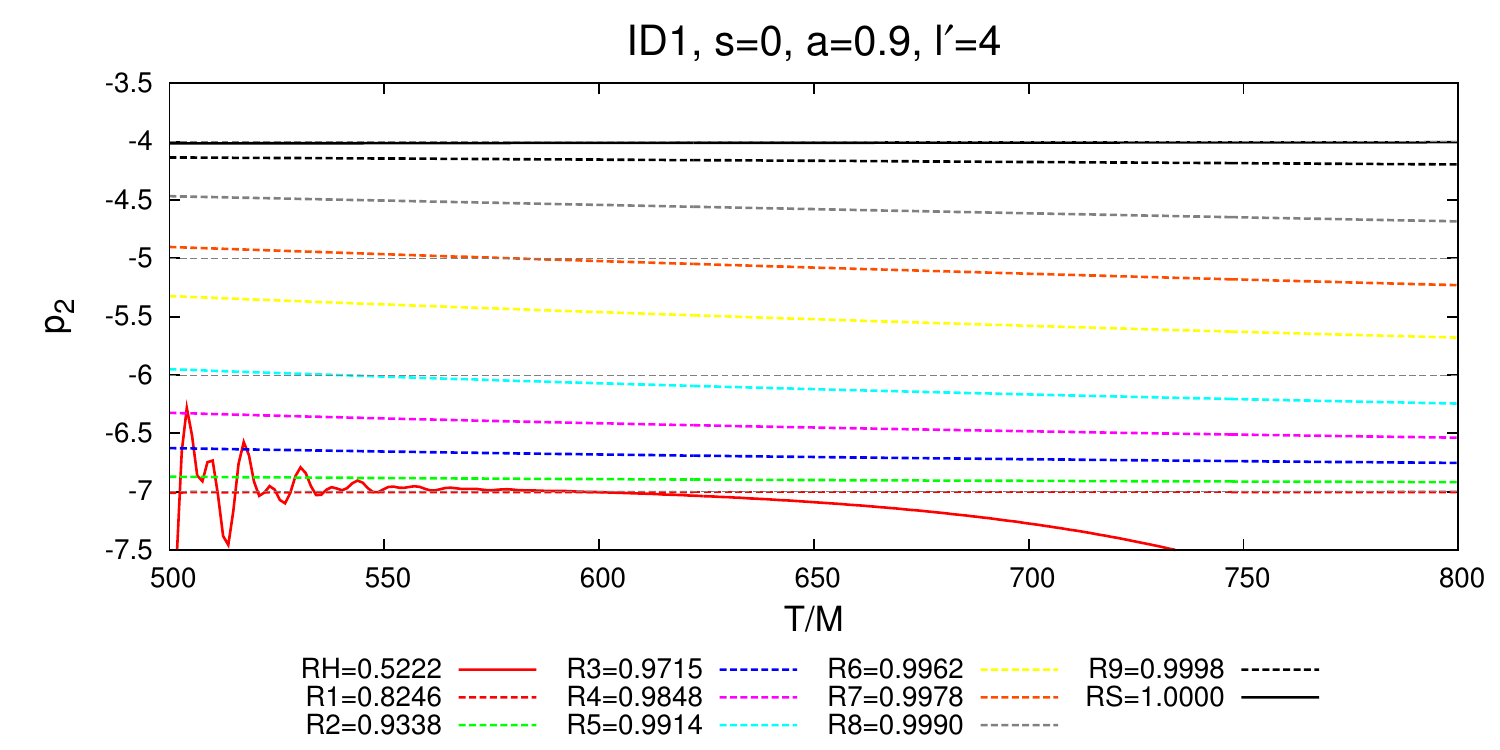}    
  \caption{LPIs of the projected modes $l=0$ (left) and
    $l=2$ (right) for a $s=0$ perturbation with ID1 and $l'=4$. 
    Different lines refer to different extraction radii.
    The left panel shows the decay rates approach the values $\mu_0=5|4$ at
    finite radii and $\scri^+$  respectively. Larger radii need longer
    time to reach their asymptotic decay rate $\mu_0=5$. The transition
    between the decay at $R_+$ and $\scri^+$ is monotonic.
    In the right panel there is a clear indication for the asymptotic
    values $\mu_2=7|4$ but the LPIs at close horizon radii depart from
    being a constant integer. Such cases are marked by
    brackets in the tables, here as $(7)|4$.}
  \label{fig:s=0_l'=4_l=02}
\end{figure}

\begin{figure}[t]
  \centering    
  \includegraphics[width=\textwidth]{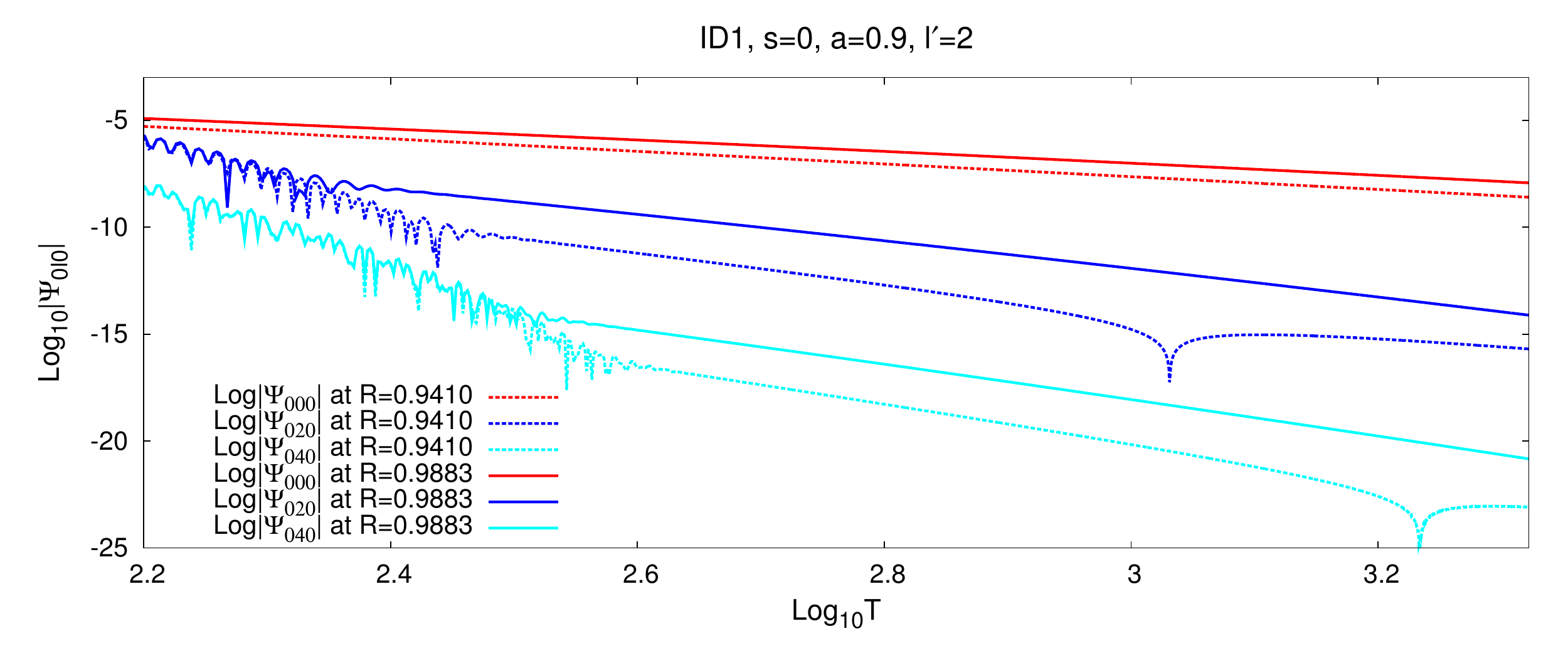}
  \caption{Evolution of the projected modes $l=0,2,4$ of a $s=0$
    perturbation  
    with ID1 and $l'=2$ at two extraction radii. The dashed lines 
    refer to the modes $l=0,2,4$ at $R=0.9410$ and the solid lines at
    $R=0.9883$. Note that the abscissa is logarithmic. 
    The plot illustrates the difference between far away observers (solid) 
    and observers closer to the horizon (dashed). The observer at $R=0.9410$ 
    measures a split of $p_2$ at $T\approx10^3$ and of $p_4$ at
    $T\approx10^{3.2}$, while the observer at $R=0.9883$ does not
    measure the splitting during the simulation 
    time. The corresponding LPIs are reported in
    Fig.~\ref{fig:ID1_a09_s0_li2_l2_UpModeSplit}.} 
  \label{fig:Log_UpModeSplit}
\end{figure}

\begin{figure}[t]
  \centering    
  \includegraphics[width=\textwidth]{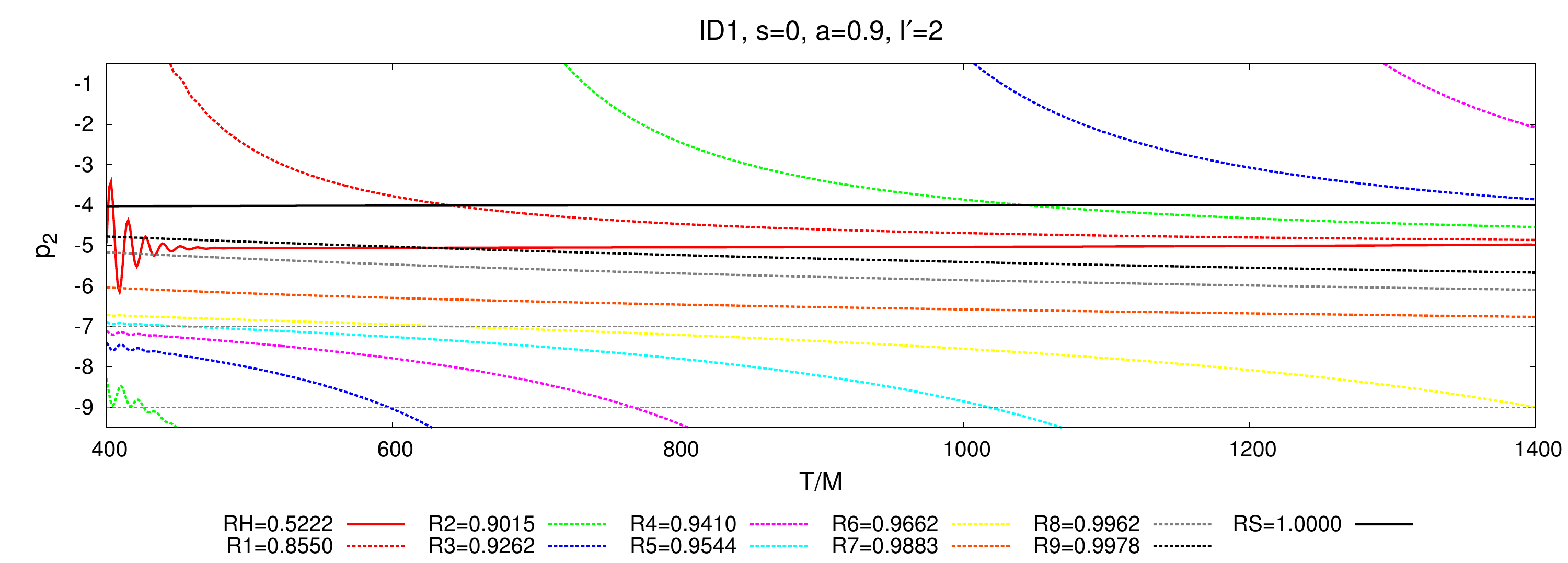} 
  \caption{Splitting of the LPIs in the up-mode $l=2$ for $s=0$ with ID1 and $l'=2$
    (corresponding to figure \ref{fig:Log_UpModeSplit}). The different lines refer to
    $10$ different extraction radii from $R=R_+$ to $R=1$.
    The decay rates at the horizon and $\scri^+$ are $\mu_2=5|4$. At radii
    in between the LPI splits, e.g.~at $R=0.9410$  
    (purple dashed line) the LPI is not a constant integer 
    during the evolution time. Only
    asymptotically in time the LPI at any finite
    radius will approach the value $p_2=-5$.  For the observer at
    $R\sim0.9883$ (dashed orange line) it is possible to measure an intermediate decay
    rate $p_2=-7$. Such cases are marked in bold face in the
    tables, indicating that at intermediate times different decay rates
    than the stated ones
    can be measured by observers far away from the horizon.}  
  \label{fig:ID1_a09_s0_li2_l2_UpModeSplit}
\end{figure}

\begin{table}[t]
  \centering
    \caption{Decay rates $\mu_l$ for $s=0$ with ID0 and ID1 at finite radii$|$future null infinity. 
      Brackets point to
      uncertainties in the LPI assessment due to possible inaccuracies or
      not verifiable splitting, $\times$ to ambiguous or immeasurable values, 
      $-$ to modes excluded by symmetry and
      $/$ to not simulated cases. Bold values denote splitting in
      time, i.e at intermediate times $p_l\neq-\mu_l$ for
      $R\lesssim\mathcal{J}^+$.}
    \label{tab:s0_LPI} 
      \begin{tabular}[t]{|c||c|c|c|c|c|c|}
        \multicolumn{7}{c}{ID0} \\
        \hline
        $l'$  &   $l=0$   &   $l=1$   &   $l=2$   &   $l=3$   &   $l=4$   &   $l=5$   \\     
        \hline\hline
        0     &   3$|$2   &    $-$    &   5$|$4   &    $-$    &   7$|$6   &    $-$    \\
        1     &    $-$    &   5$|$3   &    $-$    &   7$|$5   &    $-$    & (9)$|$(7) \\
        2     &   4$|$3   &    $-$    &{\bf6}$|$4 &    $-$    &{\bf8}$|$6 &    $-$    \\
        3     &    $-$    &   6$|$4   &    $-$    & $\xx|5$   &    $-$    & $\xx|$7   \\
        4     &   6$|$5   &    $-$    & (8)$|$5   &    $-$    & $\xx|$6   &    $-$    \\
        5     &     /     &     /     &     /     &     /     &     /     &     /     \\ 
        \hline
        \multicolumn{7}{c}{ID1} \\
        \hline
        $l'$  &   $l=0$   &   $l=1$   &   $l=2$   &   $l=3$   &   $l=4$   &   $l=5$   \\     
        \hline\hline
        0     &   3$|$2   &    $-$    &   5$|$4   &    $-$    &   7$|$6  &    $-$    \\
     1     &    $-$    &   5$|$3   &    $-$    &   7$|$5   &    $-$    & (9)$|$7   \\
     2     &   3$|$2   &    $-$    &{\bf5}$|$4 &    $-$    &{\bf7}$|$6 &    $-$    \\
     3     &    $-$    &   5$|$3   &    $-$    & $\xx|5$   &    $-$    & $\xx|$7   \\
     4     &   5$|$4   &    $-$    & (7)$|$4   &    $-$    & $\xx|$6   &    $-$    \\
     5     &    $-$    &   7$|$5   &    $-$    & (9)$|$5   &    $-$    & $\xx|$7   \\ \hline
  \end{tabular} 
\end{table}

In this section we report the decay rates for the projected modes of $s=0$
perturbations. The decay rate of the unprojected field is the same as that of 
the lowest projected mode.
For $s=0$ RT~\cite{Racz:2011qu} have presented a thorough
analysis of decay rates for all four initial data types ID0, ID1, ID2
and ID3 (and also other types). Thus for $s=0$ we restrict ourselves
to measure decay rates for ID0 and ID1 to confirm the correctness of
our code.  

As shown in Tab.~\ref{tab:s0_LPI} our decay rates are in perfect
agreement with the work of~\cite{Racz:2011qu} and
of~\cite{Jasiulek:2011ce,Zenginoglu:2012us,Burko:2010zj} for ID1. 
Figure~\ref{fig:s=0_l'=4_l=02} illustrates the LPI calculation for the
$l=0$ and the $l=2$ projected modes of an $l'=4$ simulation. 
From this example one sees that the decay rates of the $l=0$ projection
are clearly identifiable as $\mu_0=5|4$ (left panel). Instead, the
$l=2$ LPI lines bend down at radii close to the horizon and assume non-integer
values. This is likely due to inaccuracies in higher mode projections
so that in these cases we state the corresponding decay rates in the table
with brackets, e.g.\ $\mu_2=(7)|4$.    

We can unambiguously verify the LPI splitting only for the $l'=2$ case.
Higher values of $l'$ in the initial data seem to produce splitting,
but we can not assess unambiguously the values. In agreement with
previous results we do not observe splitting for the two lowest initial
modes $l'=0,1$.  
The LPI splitting is described by Fig.~\ref{fig:Log_UpModeSplit} and
Fig.~\ref{fig:ID1_a09_s0_li2_l2_UpModeSplit}, which refer 
to $l'=2$ with ID1. The splitting is clearly identifiable by looking
at $\log-\log$ plots. In Fig.~\ref{fig:Log_UpModeSplit} the splitting
for close horizon observers is visible as cusps in the tail phases of
the $l=2$ and $l=4$ modes (dashed blue and cyan lines). Observers far
away from the horizon (solid lines) will eventually 
develop the cusp but at later simulation times.
In Fig.~\ref{fig:ID1_a09_s0_li2_l2_UpModeSplit} the corresponding LPIs
are shown for different extraction radii. The asymptotic decay rates at the horizon
and $\scri^+$ are, respectively, $\mu_2=5$ and $\mu_2=4$, as
indicated by the solid red and solid black lines. The LPIs at finite
radii close to $R_+$ (dashed lines in the upper part of the plot)
approach the value $p_2=-5$ the faster the closer they are to $R_+$. On
the contrary, LPIs at finite radii $R>0.98$ have values
$p_2\simeq-7$ (see the black, grey and orange dashed lines) and are
expected to approach the value at $R_+$ only asymptotically. For these
radii the simulated time is not sufficient to observe the splitting
towards their asymptotic value. The cusp in the tail of the $l=2$ mode
at $R=0.9410$ in Fig.~\ref{fig:Log_UpModeSplit} corresponds to the
dashed purple line in Fig.~\ref{fig:ID1_a09_s0_li2_l2_UpModeSplit}
which bends down at early times and ``falls'' from the top of the plot
at later times. The splitting behaviour basically corresponds to a
sign change in the field, which could be probably explained with some
analytical argument. Note also that changing the location of the
initial Gaussian pulse has a strong effect on the splitting behavior
as discussed in~\cite{Zenginoglu:2012us} (see the latter reference for
more insights).

\subsection{Decay rates for $|s|=1$ perturbations}
\label{subsec:lpi_s1}

\begin{figure}[t]
  \centering
    \includegraphics[width=0.49\textwidth]{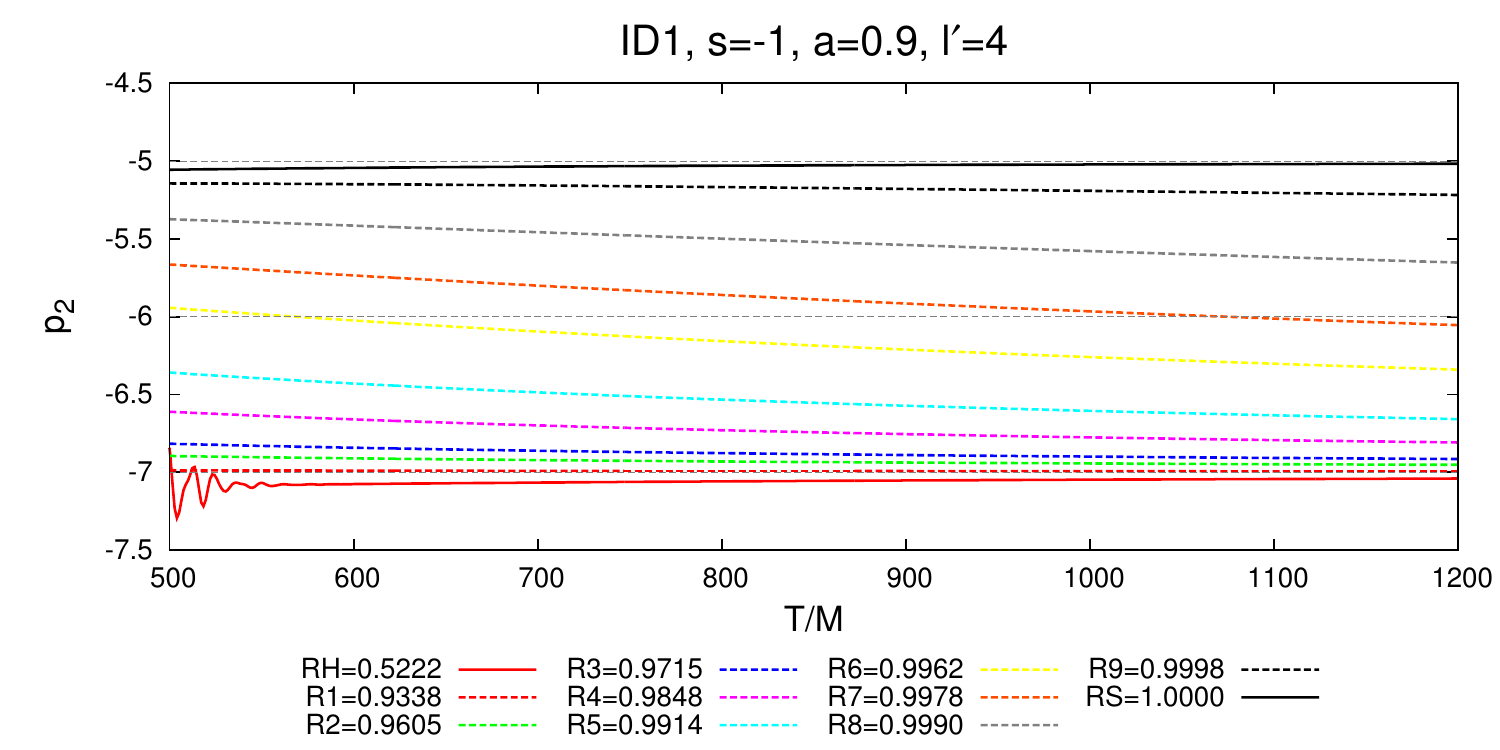}
    \includegraphics[width=0.49\textwidth]{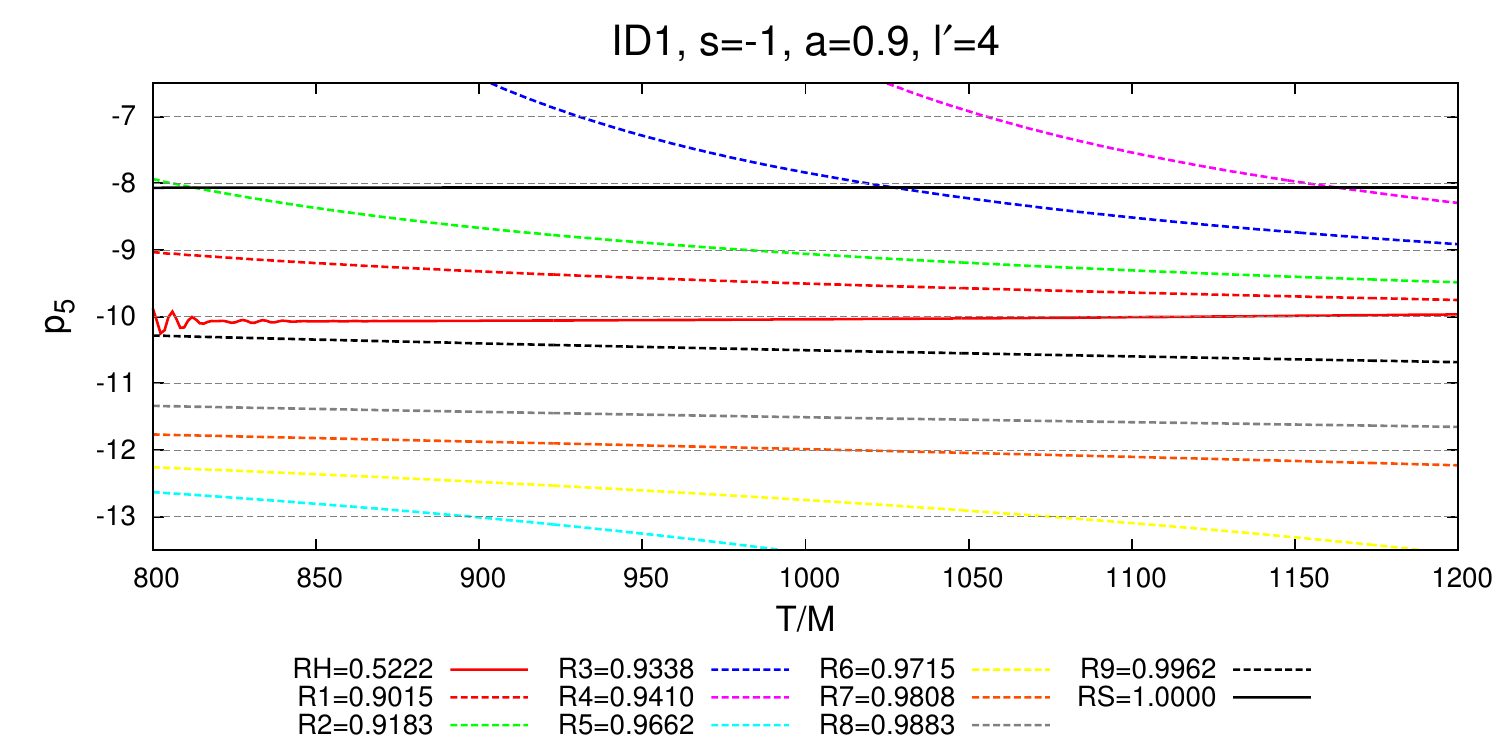}
    \caption{LPIs of the projected modes $l=2$ (left panel) and $l=5$
      (right panel) of an initial $s=-1$ perturbation with $l'=4$ and
      ID1. Different lines refer to different extraction radii. 
      The left panel shows the decay rates $\mu_2=7|5$ at finite
      radii and $\scri^+$ respectively. The transition is monotonic.
      The right panel shows that LPIs for the $l=5$ projection split
      in time. 
      At finite radii including the horizon and $\scri^+$ the decay
      rates are $\mu_5={\bf10}|8$, respectively.
      Only asymptotically in time far out observers like
      at $R=0.9962$ (dashed black line) or at $R=0.9883$ (dashed grey
      line) will observe the LPI lines bending down and approaching
      $p_5=-10$ from above. Such splitting in time is indicated by
      bold face in the tables.}   
    \label{fig:s=-1_l'=4_l=25_DecayRates}
\end{figure}

\begin{figure}[t]
  \centering
    \includegraphics[width=0.49\textwidth]{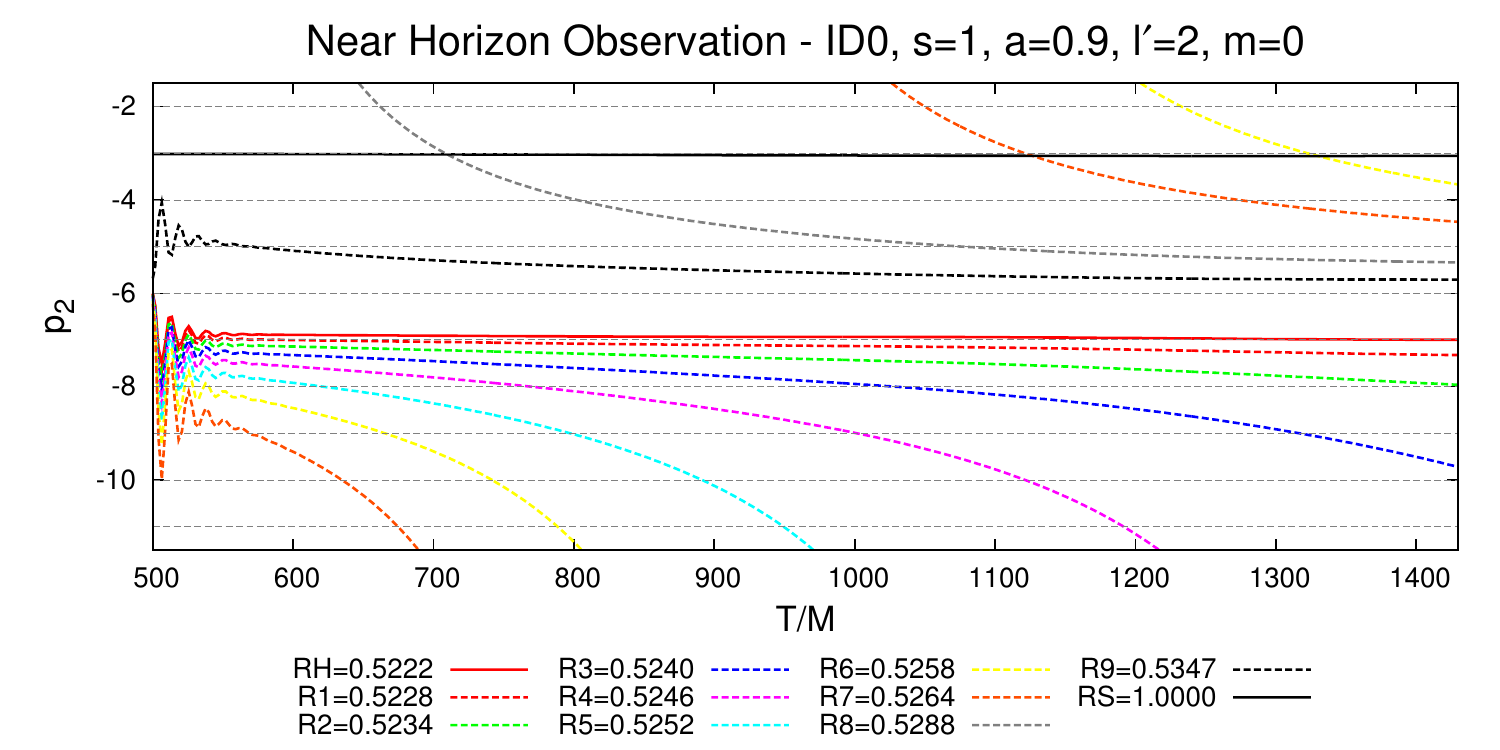}    
    \includegraphics[width=0.49\textwidth]{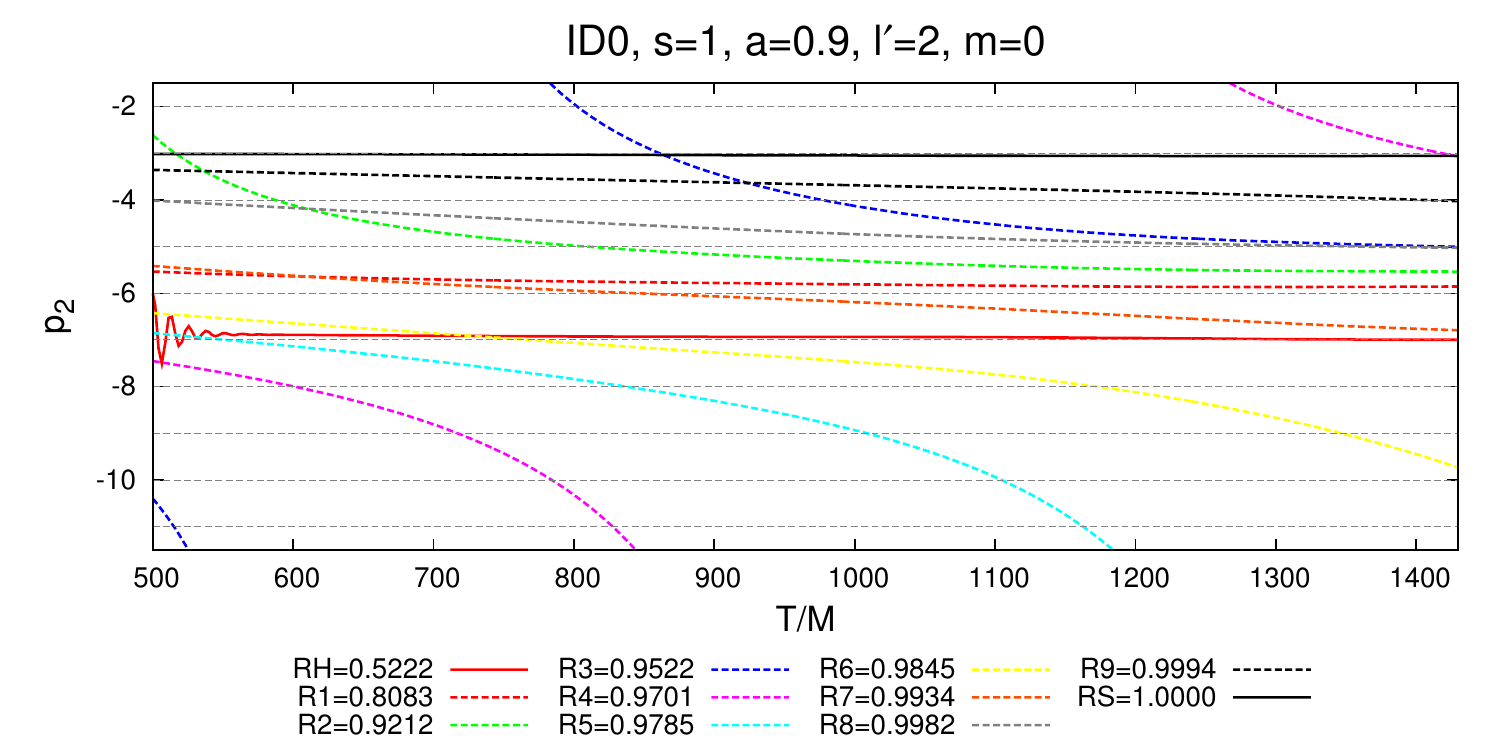}   
    \caption{LPIs of the projected mode $l=2$ of a $s=+1$ perturbation
      with $l'=2$ and ID0 measured by observers close to the horizon
      (left panel) and far away observers (right panel). From both
      panels one can see the splitting of LPIs into three distinct
      asymptotic values. The decay rates at $R_+$, finite radii excluding $R_+$, 
      and $\scri^+$ are $\mu_2=7|{\bf6}|3$. This splitting in space
      is specific to $s>0$, $m=0$ perturbations. 
      The left panel illustrates the behaviour of the LPIs
      when going from $R_+$ to $R=0.5347$. Observers
      close to $R_+$ measure initially decay rates close to $p_2=-7$
      but asymptotically they will see the LPI bend down and approach the finite radii 
      rate $p_2=-6$ from above (in this case observed first at $\tilde{R}\approx0.6010$). 
      Thus the splitting in space results also
      in a splitting in time for close horizon observers of $s=+1$,
      $m=0$ perturbations. 
      The right panel illustrates the splitting in time for far away
      observers already shown for $s=0$ and $s=-1$ perturbations in
      Fig.~\ref{fig:ID1_a09_s0_li2_l2_UpModeSplit} and the right panel
      of Fig.~\ref{fig:s=-1_l'=4_l=25_DecayRates}.}   
    \label{fig:s=1_l'=2_l=2_DecayRates}
\end{figure}

In this section we present our results for $s=\pm1$ perturbations. 
Table~\ref{tab:s1_LPI} summarizes the decay
rates of the projected modes.   

Similarly to the $s=0$ case, most of the LPIs can be calculated very
accurately. Looking for example at the left panel of
Fig.~\ref{fig:s=-1_l'=4_l=25_DecayRates}, which refers to the $l=2$
projection of a $s=-1$ perturbation with ID1 and $l'=4$, one can see
that the decay rates $\mu_2=7|5$ can be unambiguously inferred from
the plot. Table~\ref{tab:s1_LPI} reveals certain general patterns. For example the
$\mu_l$ at finite radii are always increasing~\unskip\footnote{ %
  Note that increasing values of $\mu_l$ in the tables means
  actually the LPIs $p_l$ get more negative, i.e.~the decay is faster.} 
by one if one steps from one column to the next, i.e.~to the higher projected
mode. Thus the overall field at finite radii is always dominated by
the lowest excitable mode, the $l=1$ mode for $s=|1|$. This is not
true for the decay at $\scri^+$, where higher projected modes can have
the same decay rates as the $l=1$ mode and can thus dominate the asymptotic
signal. Neglecting for a moment the two lowest rows, $l'=1,2$, 
$\mu_l$ also increases by one if one steps one row down.

Another interesting result is about the effect of different initial
data. Switching from compact to non-compact support ID results in a
slower decay. To see this one can compare the tables for
ID0 with the tables for ID2 and the tables for ID1 with the tables for
ID3. Changing from stationary to non-stationary ID leaves the
results for the lowest allowed initial mode $l_0$ (the first row in
the tables) unchanged. However, the second and the third rows seem to be
interchanged. From the third row on the difference for finite radii
decay rates seems, again, to be just a decrement by one when switching
from non-stationary to stationary ID. 

There are no other numerical investigations of projected modes' decay
rates for $s\neq0$, $a=0.9$ including the horizon and $\scri^+$, but
we can compare our results to the theoretical predictions
of~\cite{Hod:2000fh}. In particular, the Green's function calculation
provides a result immediately comparable with our ID1 case (see
also~\ref{app:green}). The decay rates of the overall field 
as stated in Tab.~1 of~\cite{Hod:2000fh} can
be compared with the first columns of our tables for ID1.  Also,
though not stated explicitly, the decay rates of projected modes for
ID1 can be read off the effective Green's functions in Eq.~(15) (finite
radii) and Eqs.~(35-36) ($\scri^+$) of~\cite{Hod:2000fh}. In summary,
as shown in the tables, our results for ID1 are in agreement with the
predictions by~\cite{Hod:2000fh} in all tested configurations, except
that an apparent difference can be observed for the second lowest
allowed initial mode $l'=l_0+1$. Comparing with Hod's results, the
values in this row are larger by one than the predicted values.
However, this discrepancy can be traced back to a missing case
distinction in the final parametrization of the result
in~\cite{Hod:2000fh}, see~\ref{app:green}.  With this small
correction, there is complete consistency between the analytical and
numerical results.

The splitting of LPIs recently reported for $s=0$
perturbations~\cite{Racz:2011qu,Jasiulek:2011ce,Zenginoglu:2012us} and
verified in Sec.~\ref{subsec:lpi_s0}, is also
present in $s\neq0$ cases. It is convenient to refer to this as a {\it
splitting in time} since intermediate decay rates $p_l\neq-\mu_l$ can
be measured at 
early times while for $T\to\infty$ a unique asymptotic decay rate $\mu_l$ is
obtained at any finite radii. An example of splitting in time is
reported in the right panel of
Fig.~\ref{fig:s=-1_l'=4_l=25_DecayRates}. The qualitative behaviour is
equivalent to the $s=0$ case previously described, e.g.\ the far away observer 
at $R=0.9962$ (dashed black line) measures at early times an
intermediate LPI $p_5\approx-11$ different from the asymptotic value
$\mu_5=-10$. Only observers at $R_+$ (solid red line) and close
to it (dashed red and green lines) measure the rate $p_5=-10$ during
the simulation time. Such splitting in time is only present in the
up-modes: for instance the left panel of
Fig.~\ref{fig:s=-1_l'=4_l=25_DecayRates} 
illustrates the LPIs of the down-mode $l=2$. No splitting is visible
and the LPIs vary monotonically between the values $p_2=-5$ (solid
black line) and $p_2=-7$ (solid red line) at $\scri^+$ and $R_+$,
respectively.  

The splitting in time differs from the splitting in the case of $s>0$
and $m=0$ predicted by~\cite{Hod:2000fh,Barack:1999ya}. In this case the LPI
has three distinct asymptotic values at the horizon, at finite radii,
and at $\scri^+$, which could be referred to as a {\it splitting in space}.
The splitting in space is observed in any projected mode
as well as in the overall field while the splitting in time is only
observed in the up-modes.   

The right panel of Figure~\ref{fig:s=1_l'=2_l=2_DecayRates} shows both splittings of LPIs
in the projected $l=2$ mode of a $s=+1$ perturbation with $l'=2$ and ID0.
The decay rates at $R_+$ and $\scri^+$ are, respectively, $\mu_2=7$ and
$\mu_2=3$, as read off from the solid red and black lines.  
The splitting in space is clearly visibile at the radius
$R\approx0.8083$ which is approaching the value $\mu_2=6$ (dashed red line). The left
panel of Figure~\ref{fig:s=1_l'=2_l=2_DecayRates} corresponds to the
same simulation as the right panel but focuses on observers very close to
the horizon. The latter measure at intermediate times LPIs close to the
horizon decay rate $p_2\lesssim-7$ (dashed red and green
lines). Instead observers well-off the horizon measure already at
early times the LPI $p_2=-6$. Note that the transition between both
takes place in a very small spatial interval because already observers
at $R=0.5288$  measure approximately the asymptotic decay rate
$p_2\approx-\mu_2=-6$ (dashed grey line).

\begin{table}[t]
    \caption{Decay rates $\mu_l$ for $s=-1$ (left) at finite radii$|$null infinity and for $s=+1$
      (right) at the horizon$|$finite radii$|$null infinity. 
      Brackets point to
      uncertainties in the LPI assessment due to possible inaccuracies or
      not verifiable splitting, $\times$ to ambiguous or immeasurable values.
      Bold values denote splitting
      in time, i.e.~at intermediate times $p_l\neq-\mu_l$ for
      $R\lesssim1$.} 
    \label{tab:s1_LPI} 
  \begin{minipage}[b]{0.49\linewidth}
    \centering      
    \begin{tabular}[t]{|c||c|c|c|c|c|}
      \multicolumn{6}{c}{ID0} \\
      \hline
      $l'$ &   $l=1$   &      $l=2$ &      $l=3$ &    $l=4$     &       $l=5$   \\     
      \hline
      1    &   5$|$4   &      6$|$5 &      7$|$6 &      8$|$7   &       9$|$8   \\
      2    &   5$|$4   & {\bf6}$|$5 & {\bf7}$|$6 & {\bf8}$|$7   &  {\bf9}$|$8   \\
      3    &   6$|$5   &      7$|$5 & {\bf8}$|$6 & {\bf9}$|$7   & {\bf10}$|$8   \\
      4    &   7$|$6   &      8$|$6 &      9$|$6 &    $\xx|$(7) &     $\xx|$(8) \\
      5    &   8$|$7   &      9$|$7 &     10$|$7 &   (11)$|$7   &     $\xx|$8   \\ 
      \hline
      \multicolumn{6}{c}{ID1} \\
      \hline
      $l'$ &   $l=1$   &      $l=2$ &      $l=3$ &      $l=4$   &       $l=5$   \\     
      \hline
      1    &   5$|$4   &      6$|$5 &      7$|$6 &      8$|$7   &       9$|$8   \\
      2    &   6$|$5   &      7$|$5 &      8$|$6 &      9$|$7   &      10$|$8   \\
      3    &   5$|$4   &      6$|$5 & {\bf7}$|$6 & {\bf8}$|$7   &  {\bf9}$|$8   \\
      4    &   6$|$5   &      7$|$5 &      8$|$6 & {\bf9}$|$7   & {\bf10}$|$8   \\
      5    &   7$|$6   &      8$|$6 &      9$|$6 &     10$|$7   &     $\xx|$8   \\
      \hline
      \multicolumn{6}{c}{ID2} \\
      \hline
      $l'$ &   $l=1$   &      $l=2$ &      $l=3$ &      $l=4$   &       $l=5$   \\     
      \hline
      1    &   4$|$3   &      5$|$4 &      6$|$5 &      7$|$6   &       8$|$7   \\
      2    &   4$|$3   & {\bf5}$|$4 & {\bf6}$|$5 & {\bf7}$|$6   &  {\bf8}$|$7   \\
      3    &   5$|$4   &      6$|$4 & {\bf7}$|$5 & {\bf8}$|$6   &  {\bf9}$|$7   \\
      4    &   6$|$5   &      7$|$5 &      8$|$5 &    $\xx|$6   &     $\xx|$7   \\
      5    &   7$|$6   &      8$|$6 &      9$|$6 &    $\xx|$6   &     $\xx|$7   \\ \hline
      \hline
      \multicolumn{6}{c}{ID3} \\
      \hline
      $l'$ &   $l=1$   &      $l=2$ &      $l=3$ &       $l=4$   &       $l=5$   \\     
      \hline
      1    &   4$|$3   &      5$|$4 &      6$|$5 &       7$|$6   &       8$|$7   \\
      2    &   5$|$4   &      6$|$4 &      7$|$5 &       8$|$6   &       9$|$7   \\
      3    &   4$|$3   &      5$|$4 & {\bf6}$|$5 &  {\bf7}$|$6   &  {\bf8}$|$7   \\
      4    &   5$|$4   &      6$|$4 &      7$|$5 &({\bf8})$|$6   &({\bf9})$|$7   \\
      5    &   6$|$5   &      7$|$5 &    (8)$|$5 &     (9)$|\xx$ &     $\xx|\xx$ \\ \hline
    \end{tabular} 
  \end{minipage}
  \hspace{0.5cm}
  \begin{minipage}[b]{0.49\linewidth}
    \centering      
    \begin{tabular}[t]{|c||c|c|c|c|c|}
      \multicolumn{6}{c}{ID0} \\
      \hline
      $l'$ &   $l=1$   &   $l=2$        &      $l=3$      &    $l=4$           &    $l=5$    \\     
      \hline\hline        
      1    & 6$|$5$|$2 & 7$|$6$|$3      &  8$|$7$|$4      &  9$|$8$|$5         & $\xx|$9$|$6 \\
      2    & 6$|$5$|$2 & 7$|${\bf6}$|$3 &  8$|${\bf7}$|$4 &  9$|${\bf8}$|$5    & $\xx|\xx|$6  \\
      3    & 7$|$6$|$3 & 8$|$7$|$3      &  9$|${\bf8}$|$4 & 10$|${\bf9}$|$5    & $\xx|\xx|$6  \\
      4    & 8$|$7$|$4 & 9$|$8$|$4      & 10$|$9$|$4      & 11$|$({\bf10})$|$5 & $\xx|\xx|$6  \\
      5    & 9$|$8$|$5 &10$|$9$|$5      &11$|$10$|$5      & 12$|$11$|$5        & $\xx|\xx|$6  \\ 
      \hline
      \multicolumn{6}{c}{ID1} \\
      \hline
      $l'$ &   $l=1$   &   $l=2$        &      $l=3$      &    $l=4$           &    $l=5$    \\     
      \hline\hline        
      1    & 6$|$5$|$2 & 7$|$6$|$3 & 8$|$7$|$4      &  9$|$8$|$5      & $\xx|$9$|$(6)  \\
      2    & 7$|$6$|$3 & 8$|$7$|$3 & 9$|$8$|$4      & 10$|$9$|$5      & $\xx|$10$|$6   \\
      3    & 6$|$5$|$2 & 7$|$6$|$3 & 8$|${\bf7}$|$4 &  9$|${\bf8}$|$5 & $\xx|\xx|$6    \\
      4    & 7$|$6$|$3 & 8$|$7$|$3 & 9$|$8$|$4      & 10$|${\bf9}$|$5 & $\xx|$(10)$|$6 \\
      5    & 8$|$7$|$4 & 9$|$8$|$4 &10$|$9$|$4      & 11$|$10$|$5     &  $\xx|\xx|$6    \\ 
      \hline
      \multicolumn{6}{c}{ID2} \\
      \hline
      $l'$ &   $l=1$   &   $l=2$   &   $l=3$        &    $l=4$   &    $l=5$     \\     
      \hline        
      1    & 5$|$4$|$1 & 6$|$5$|$2 & 7$|$6$|$3      &  8$|$7$|$4 & $\xx|$8$|$5  \\
      2    & 5$|$4$|$1 & 6$|$5$|$2 & 7$|$6$|$3      &  8$|$7$|$4 & $\xx|$8$|$5  \\
      3    & 6$|$5$|$2 & 7$|$6$|$2 & 8$|${\bf7}$|$3 &  9$|\xx|$4 & $\xx|\xx|$5  \\
      4    & 7$|$6$|$3 & 8$|$7$|$3 & 9$|$8$|$3      & 10$|\xx|$4 & $\xx|\xx|$5  \\
      5    & 8$|$7$|$4 & 9$|$8$|$4 &10$|$9$|$4      & 11$|\xx|$4 &
      $\xx|\xx|$5  \\ 
      \hline
      \multicolumn{6}{c}{ID3} \\   
      \hline
      $l'$ &   $l=1$   &   $l=2$   &      $l=3$     &   $l=4$        &    $l=5$     \\     
      \hline\hline        
      1    & 5$|$4$|$1 & 6$|$5$|$2 & 7$|$6$|$3      & 8$|$7$|$4      & $\xx|$8$|$5  \\
      2    & 6$|$5$|$2 & 7$|$6$|$2 & 8$|$7$|$(3)    & 9$|$8$|$(4)    & $\xx|$9$|$(5)\\
      3    & 5$|$4$|$1 & 6$|$5$|$2 & 7$|${\bf6}$|$3 & 8$|${\bf7}$|$4 & $\xx|\xx|$5  \\
      4    & 6$|$5$|$2 & 7$|$6$|$2 & 8$|$7$|$3      & 9$|${\bf8}$|$4 & $\xx|\xx|$5  \\
      5    & 7$|$6$|$3 & 8$|$7$|$3 & 9$|$8$|$3      &10$|$9$|\xx$    &
      $\xx|\xx|$5      \\ 
      \hline
    \end{tabular} 
  \end{minipage}
\end{table}

\subsection{Decay rates for $|s|=2$ perturbations}
\label{subsec:lpi_s2}

\begin{figure}[t]
  \centering
  \includegraphics[width=\textwidth]{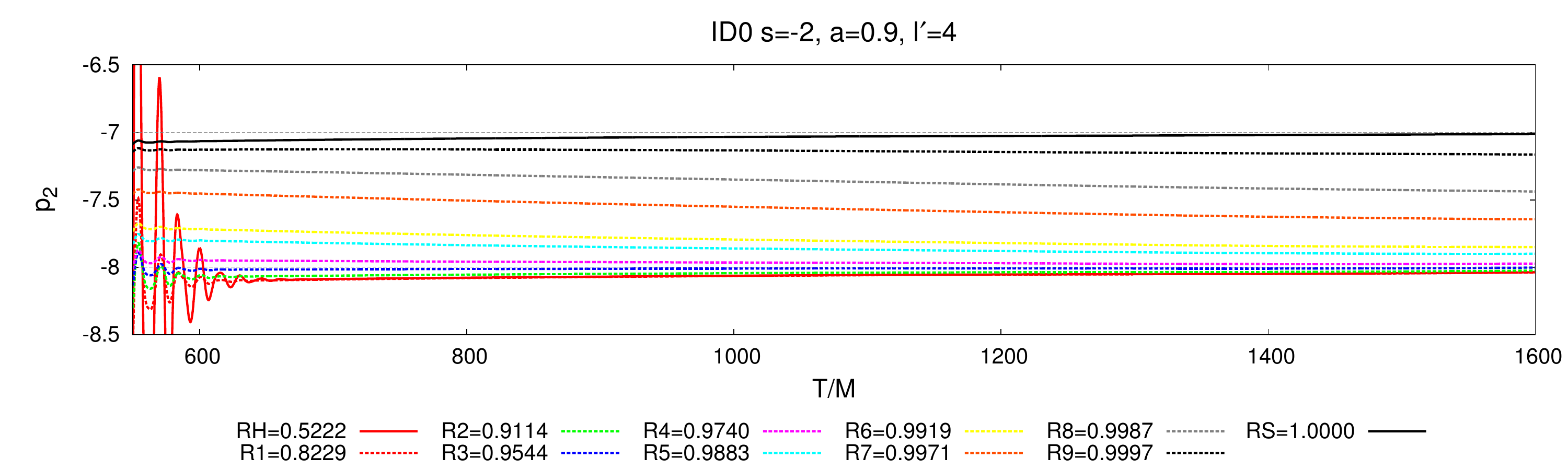}    
  \caption{LPIs of the projected mode $l=2$ of a $s=-2$ perturbation
    with $l'=4$ and ID0. Different lines refer to different extraction
    radii. 
    The asymptotic decay rates extracted from this plot are
    $\mu_2=8|7$ corresponding to the decay rates at finite radii
    including the horizon and future null infinity. The transition
    of the LPIs from $\scri^+$ to $R_+$ is smooth as for 
    $s=0,-1,+2$ perturbations.}  
  \label{fig:s=-2_l'=4_l=2_DecayRates}
\end{figure}

\begin{figure}[t]
  \centering   
   \includegraphics[width=0.49\textwidth]{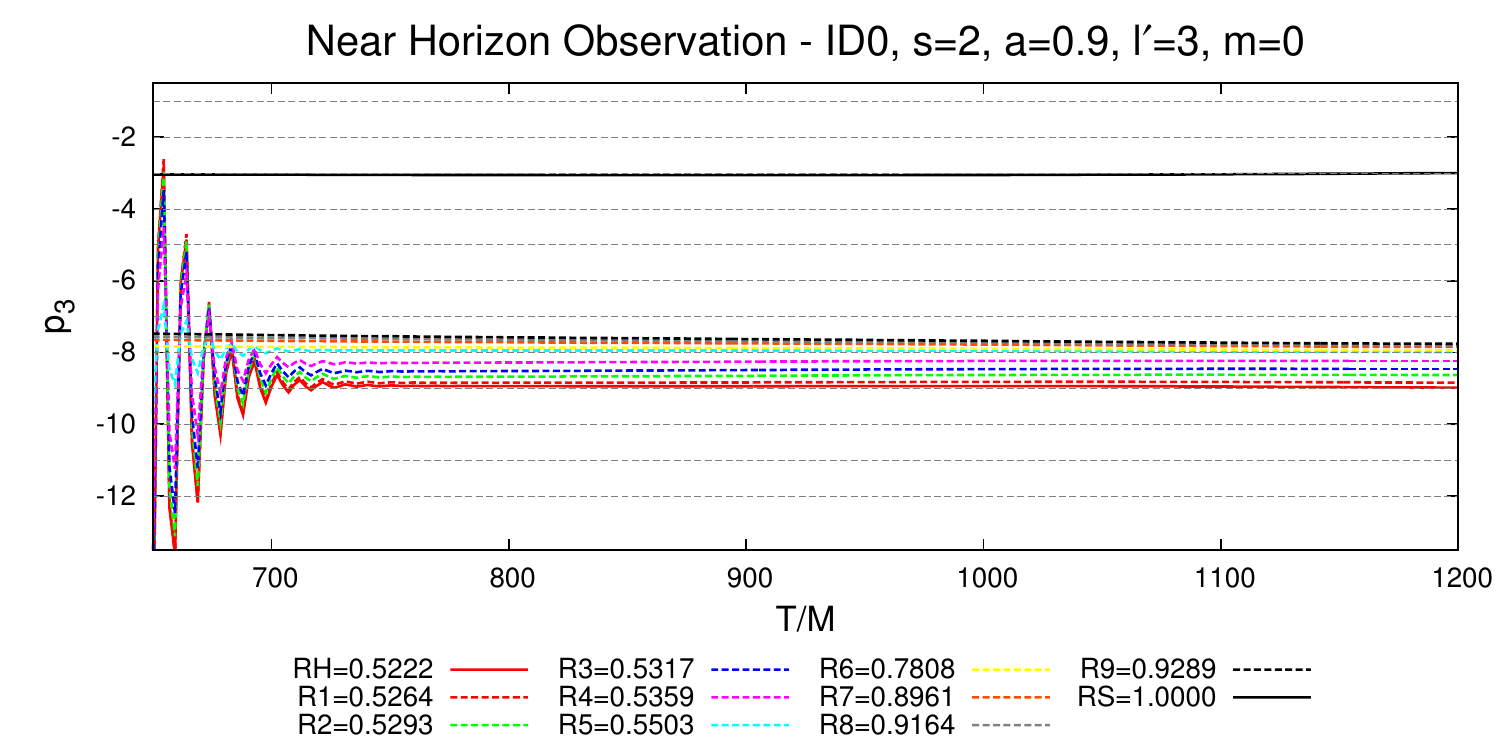}
   \includegraphics[width=0.49\textwidth]{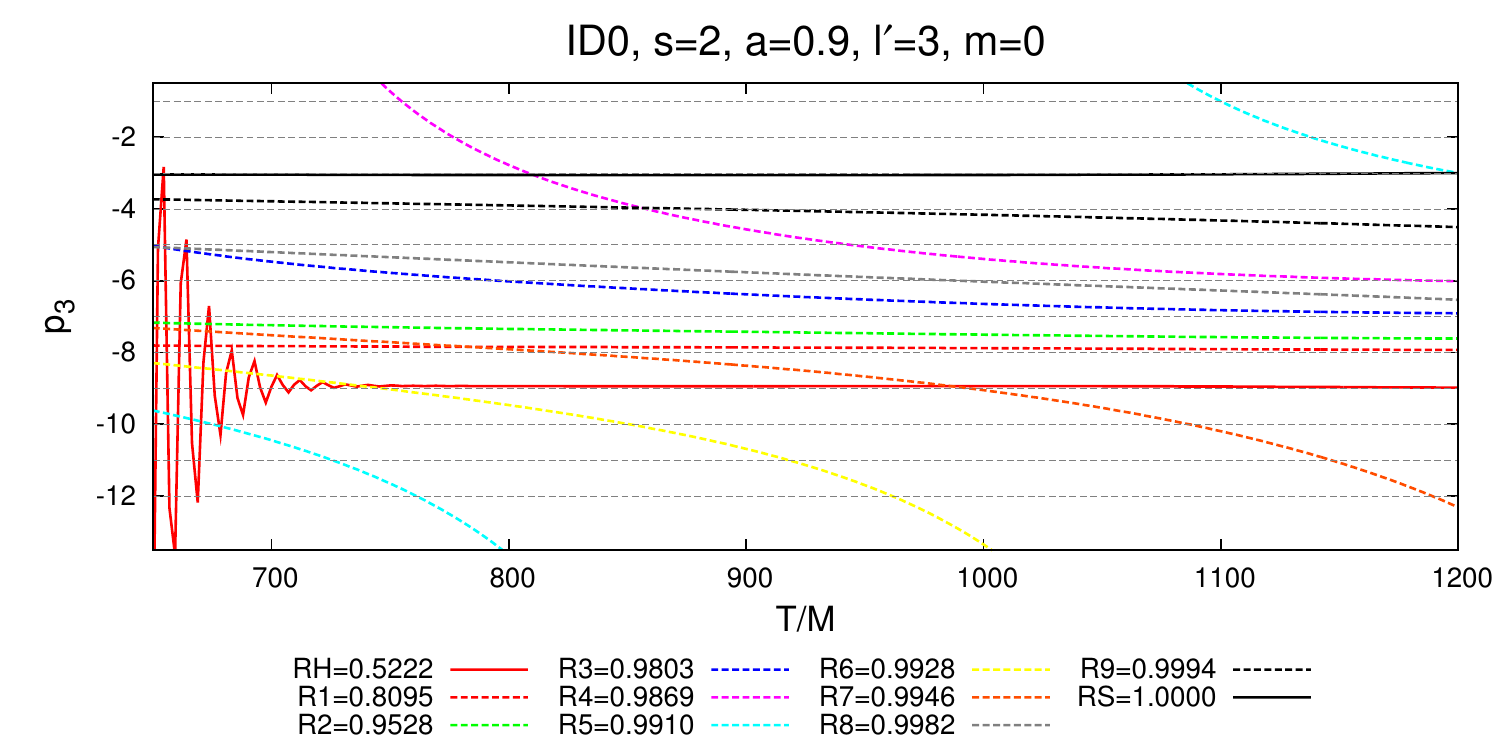}      
   \caption{LPIs of the projected mode $l=3$ of a $s=+2$ perturbation
      with $l'=3$ and ID0 measured by close to the horizon observers
      (left panel) and far away observers (right panel). From both
      panels one can see the splitting of LPIs into three distinct
      asymptotic decay rates at $R_+$, finite radii excluding $R_+$, 
      and $\scri^+$, here $\mu_3=9|{\bf8}|3$. This splitting in space
      is special to $s>0$, $m=0$ perturbations. 
      The left panel illustrates the behaviour of the LPIs
      when going from $R_+$ to $\tilde{R}\approx0.5503$. 
      Differently from $s=+1$, $m=0$ perturbations 
      (Fig.~\ref{fig:s=1_l'=2_l=2_DecayRates}) no splitting in time is
      observed.
      The right panel illustrates the splitting in time for far away
      observers already shown for $s=0$ and $|s|=1$ perturbations in
      Fig.~\ref{fig:ID1_a09_s0_li2_l2_UpModeSplit} and the right panels
      of Figs.~\ref{fig:s=-1_l'=4_l=25_DecayRates},~\ref{fig:s=1_l'=2_l=2_DecayRates}.} 
   \label{fig:s=pm2-splittings}
\end{figure}

In this section we present our results for $s=\pm2$ perturbations. 

Qualitatively, $s=-2$ simulations are in complete agreement with $s=0$
and $s=-1$ simulations. As an example,
Fig.~\ref{fig:s=-2_l'=4_l=2_DecayRates} shows the LPIs of the  
$l=2$ projection for $l'=4$ and ID0 at different extraction 
radii. From the plot we can extract the decay rates $\mu_2=8|7$, 
confirming that also for $s=-2$ there are two distinct asymptotic
decay rates at any finite radii including the horizon and at future
null infinity. The radius dependence is monotonic as already described
for the down-modes of $|s|=0,1$ perturbations. 

The LPI calculations for $|s|=2$ are summarized in
Tab.~\ref{tab:s2_LPI}. The decay rates follow exactly
the same patterns as described in Sec.~\ref{subsec:lpi_s1} for
$|s|=1$. At finite radii $\mu_l$ increases by one when switching to 
the next higher projection $l$ or the next higher initial mode $l'$   
(neglecting again the first two rows $l'=l_0,\, l_0+1$). The overall
field is asymptotically dominated by the lowest mode, i.e.~for $m=0$
the $l=2$ projection. Instead, at $\scri^+$ the $l=2$ mode is not 
always the dominant one, because higher modes $l>2$ can have the
same decay rate. 
Changing from compact to non-compact initial data results in the
decay rates decrease by one.
Changing from  stationary
to non-stationary initial data leads to the same modifications of the
decay rates as discussed for the case $|s|=1$. 
Comparing with the predictions of~\cite{Hod:2000fh} for ID1 again all
our results agree. See again~\ref{app:green} for the case $l'=l_0+1$.

The splitting in time of the LPIs of up-modes is also
observed for $|s|=2$. As an example the right panel of 
Fig.~\ref{fig:s=pm2-splittings} illustrates the splitting in time for
the $l=3$ projection of a $s=+2$ perturbation with $l'=3$
and ID0. The left panel of
Fig.~\ref{fig:s=pm2-splittings} refers to the same simulation as the
right panel but concentrates on the LPIs at radii very close to the
horizon.
The LPIs at $R_+$ and $\scri^+$ are, respectively, $p_2=-9$ and
$p_2=-3$, and they are marked by the solid red and black lines.  
The splitting in space is clearly visible as the LPI is $p_2=-8$ at the radius
$R\approx0.8095$ (dashed red line in right panel). 
It is interesting to investigate the close horizon behaviour in the
left panel. 
Differently from the $s=+1$, $m=0$ case (compare with
Fig.~\ref{fig:s=1_l'=2_l=2_DecayRates}, left panel), splitting in time
is {\it not} present for LPIs measured by close horizon observers
$R_+\lesssim R$. 
This is not a special feature of the particular case shown here. We
observe this monotonic transition behaviour in all the projections, as
well as the overall field, for all $s=+2$, $m=0$ perturbations.

\begin{table}[h]
    \caption{Decay rates $\mu_l$ for $s=-2$ (left) at finite radii$|$null infinity and for $s=+2$ (right) at the horizon$|$finite radii$|$null infinity.
      Brackets point to
      uncertainties in the LPI assessment due to possible inaccuracies or
      not verifiable splitting, $\times$ to ambiguous or immeasurable values.
      Bold values denote splitting in time, i.e.~at intermediate times $p_l\neq-\mu_l$ for
      $R\lesssim1$.} 
    \label{tab:s2_LPI} 
    \begin{minipage}[b]{0.49\linewidth}
    \centering      
    \begin{tabular}[t]{|c||c|c|c|c|}
      \multicolumn{5}{c}{ID0} \\
      \hline
      $l'$ &   $l=2$   &   $l=3$   &   $l=4$   &   $l=5$    \\     
      \hline\hline        
      2    &   7$|$6   &   8$|$7   &      9$|$8 &     10$|$9    \\
      3    &   7$|$6   &{\bf8}$|$7 & {\bf9}$|$8 &{\bf10}$|$9 \\
      4    &   8$|$7   &   9$|$7   &{\bf10}$|$8 &{\bf11}$|$9 \\      
      5    &   9$|$8   &  10$|$8   &     11$|$8 &    $\xx|$(9) \\  
      \hline
      \multicolumn{5}{c}{ID1} \\
      \hline
      $l'$ &   $l=2$   &   $l=3$   &   $l=4$       &    $l=5$      \\    
      \hline\hline        
      2    &   7$|$6   &   8$|$7   &   9$|$8       &   10$|$9      \\
      3    &   8$|$7   &   9$|$7   &  10$|$8   	   &   11$|$9      \\
      4    &   7$|$6   &   8$|$7   & {\bf9}$|$(8)  & {\bf10}$|$(9) \\      
      5    &   8$|$7   &   9$|$7   &  10$|$8   	   & {\bf11}$|$9     \\ 
      \hline
      \multicolumn{5}{c}{ID2} \\
      \hline
      $l'$ &   $l=2$   &   $l=3$   &   $l=4$   &    $l=5$   \\    
      \hline\hline        
      2    &   6$|$5   &   7$|$6   &   8$|$7   &    9$|$8   \\
      3    &   6$|$5   &{\bf7}$|$6 &{\bf8}$|$7 &{\bf9}$|$8  \\
      4    &   7$|$6   &   8$|$6   &{\bf9}$|$7 &{\bf10}$|$8 \\      
      5    &   8$|$7   &   9$|$7   &  10$|$7   &    $\xx|$8  \\ 
      \hline
      \multicolumn{5}{c}{ID3} \\
      \hline
      $l'$ &   $l=2$   &   $l=3$   &   $l=4$   &    $l=5$    \\    
      \hline\hline        
      2    &   6$|$5   &   7$|$6   &   8$|$7   &    9$|$8     \\
      3    &   7$|$6   &   8$|$6   &   9$|$7   &   10$|$8     \\ 
      4    &   6$|$5   &   7$|$6   &{\bf8}$|$7 &{\bf9}$|$8    \\      
      5    &   7$|$6   &   8$|$6   &   9$|\xx$   &  $\xx|\xx$ \\ \hline
    \end{tabular}   
  \end{minipage}
  \hspace{0.5cm}  
  \begin{minipage}[b]{0.49\linewidth}
    \centering      
    \begin{tabular}[t]{|c||c|c|c|c|}
      \multicolumn{5}{c}{ID0} \\
      \hline
      $l'$ &    $l=2$   &    $l=3$        &   $l=4$          &   $l=5$            \\     
      \hline\hline        
      2    &  8$|$7$|$2 &  9$|$8$|$3      & 10$|$9$|$4       &  $\xx|$10$|$5      \\
      3    &  8$|$7$|$2 &  9$|${\bf8}$|$3 & 10$|${\bf9}$|$4  &  $\xx|${\bf10}$|$5 \\
      4    &  9$|$8$|$3 & 10$|$9$|$3      & 11$|${\bf10}$|$4 &  $\xx|${\bf11}$|$5 \\ 
      5    & 10$|$9$|$4 & 11$|$10$|$4     & 12$|$11$|$4      &     $\xx|\xx|$5       \\ 
      \hline
      \multicolumn{5}{c}{ID1} \\
      \hline
      $l'$ &    $l=2$   &    $l=3$        &   $l=4$          &   $l=5$            \\     
      \hline\hline        
      2    &  8$|$7$|$2 &  9$|$8$|$3      & 10$|$9$|$4       &  $\xx|$10$|$5      \\
      3    &  9$|$8$|$3 & 10$|$9$|$3      & 11$|$10$|$4      &  $\xx|$11$|$5 \\
      4    &  8$|$7$|$2 &  9$|$8$|$(3)    & 10$|${\bf9}$|$4  &  $\xx|${\bf10}$|$5 \\ 
      5    &  9$|$8$|$3 & 10$|$9$|$3      & 11$|$10$|\xx$    &
      $\xx|\xx|$5       \\ 
      \hline
      \multicolumn{5}{c}{ID2} \\
      \hline
      $l'$ &    $l=2$   &    $l=3$        &   $l=4$          &   $l=5$            \\     
      \hline\hline        
      2    &  7$|$6$|$1 &  8$|$7$|$2      &  9$|$8$|$3       &  $\xx|$9$|$4      \\
      3    &  7$|$6$|$1 &  8$|$7$|$2      &  9$|$8$|$3       &  $\xx|$9$|$4 \\
      4    &  8$|$7$|$2 &  9$|$(8)$|$2    & 10$|\xx|$3       &  $\xx|\xx|$4 \\ 
      5    &  9$|$8$|$3 & 10$|$9$|$3      & 11$|$(10)$|$(3)  &$\xx|\xx|$4 \\ 
      \hline
      \multicolumn{5}{c}{ID3} \\
      \hline
      $l'$ &    $l=2$   &    $l=3$        &   $l=4$          &   $l=5$            \\     
      \hline\hline        
      2    &  7$|$6$|$1 &  8$|$7$|$2      &  9$|$8$|$3       &  $\xx|$9$|$4      \\
      3    &  8$|$7$|$2 &  9$|$8$|$(2)    & 10$|$9$|$(3)       &  $\xx|$10$|\xx$ \\
      4    &  7$|$6$|$1 &  8$|$7$|$(2)    &  9$|${\bf8}$|$3  &  $\xx|\xx|$4 \\ 
      5    &  8$|$7$|$2 &  9$|$8$|$2      & 10$|$(9)$|$(3)   &$\xx|\xx|$4  \\ 
     \hline
   \end{tabular} 
 \end{minipage}
\end{table}

\section{Late time decay of non-axisymmetric perturbations}
\label{sec:decaym2}

In this section we report numerical experiments for the late time
decay of non-axisymmetric perturbations. We consider $s=0$ and $s=-2$
perturbations and $a\in[0.9,1]$. We use again pure multipole ID1, 
$\psi(\theta)\propto{}^sY_{l'm'}$, with $m'=2$ ($m'=1$) for $s=-2$ ($s=0$) and 
$l'=1,2,3,4,5$. For $s>0$ and $a=0.9$ we discuss the cases $s=+1$, $l'=2$ and $s=+2$, 
$l'=3$, but no further systematic investigations are performed.

\begin{table}[t]
    \caption{Decay rates $\mu_l$ for $s=0$ and $m=1$ with ID1 at finite radii$|$null infinity.
      Brackets point to
      uncertainties in the LPI assessment due to possible inaccuracies or
      not verifiable splitting, $\times$ to ambiguous or immeasurable values, 
      $-$ to modes excluded by symmetry. Bold values denote splitting in time, i.e at 
      intermediate times $p_l\neq-\mu_l$ for $R\lesssim1$. Square brackets point 
      out different values compared to our $m=0$ tables. }
    \label{tab:m_neq_0_rates_s0} 
    \begin{minipage}[b]{\linewidth}
    \centering         
    \begin{tabular}[t]{|c||c|c|c|c|c|}     
        \hline
        $l'$  &   $l=1$   &   $l=2$   &   $l=3$   &   $l=4$   &   $l=5$   \\     
        \hline\hline       
	1     &   5$|$3   &    $-$    &   7$|$5   &    $-$    &   9$|$7   \\
	2     &    $-$    & [7]$|$4   &    $-$    & [9]$|$6   &    $-$    \\
	3     &   5$|$3   &    $-$    & {\bf7}$|5$&    $-$    & {\bf9}$|$7   \\
	4     &    $-$    & (7)$|$4   &    $-$    & $\xx|$6   &    $-$    \\
	5     &   7$|$5   &    $-$    &   9$|$5   &    $-$    & $\xx|$7   \\ \hline
    \end{tabular} 
  \end{minipage}  
\end{table}

\begin{table}[t]
    \caption{Decay rates $\mu_l$ for $s=-2$ and $m=2$ with ID1 at finite radii$|$null infinity.
      Bold values denote splitting in time, i.e at intermediate 
      times $p_l\neq-\mu_l$ for $R\lesssim1$. Square brackets point out 
      different values compared to our $m=0$ tables. }
    \label{tab:m_neq_0_rates_sm2} 
  \begin{minipage}[b]{\linewidth}
    \centering      
    \begin{tabular}[t]{|c||c|c|c|c|}
      \hline
      $l'$ &   $l=2$   &   $l=3$   &   $l=4$       &    $l=5$       \\    
      \hline\hline        
      2    &   7$|$6   &   8$|$7   &   9$|$8       &   10$|$9       \\
      3    & [7]$|$[6] & [8]$|$7   & [9]$|$8       &  [10]$|$9      \\
      4    &   7$|$6   &   8$|$7   & {\bf9}$|$8    & {\bf10}$|$9    \\      
      5    &   8$|$7   &   9$|$7   &  10$|$8   	   & {\bf11}$|$9    \\ \hline           
   \end{tabular} 
 \end{minipage}
\end{table}

\subsection{Late time decay for $a=0.9$}
\label{subsec:a09}

Let us first discuss the case $a=0.9$. The numerical settings are the 
same as those for axisymmetric perturbations, but using
$R_0\approx0.76$.
The field dynamics at late times is
characterized by power law tails, as in the axisymmetric case. 

The decay rates $\mu_l$ for $s=0$ are summarized in
Tab.~\ref{tab:m_neq_0_rates_s0}. Similarly, the decay rates for $s=-2$
are summarized in Tab.~\ref{tab:m_neq_0_rates_sm2}.  
The values different from the
axisymmetric case are explicitly indicated by square brackets
(compare with~Tab.~\ref{tab:s0_LPI},~\ref{tab:s2_LPI}.) In the $s=0$ case the decay of
the overall $m=1$ field can be different from the axisymmetric one, just
because the lowest allowed mode for even $l'$ perturbations with $m=1$ 
is $l=2$ instead of $l=0$. In the $s=-2$ case instead the lowest
allowed mode is $l_0=2$ for $m=2$ as well as for $m=0$. Still the
overall $m=2$ field is found to decay differently from the axisymmetric
analogue for $l'=3$. As indicated by the tables, the decay rates that
differ from the axisymmetric case are only those relative to $l'=2$
($l'=3$) for $s=0$ ($s=-2$) (second rows in $m\neq0$ tables).  See~\ref{app:green} for
the mathematical origin of this.

The $s=0$, $m=1$ results agree with e.g.~\cite{Racz:2011qu} and they
are consistent with various values reported in the
literature~\cite{Krivan:1996da,Scheel:2003vs,Burko:2007ju,Burko:2010zj,Jasiulek:2011ce}, where comparable.
The $s=-2$ results agree completely with the formulas extracted from
the effective Green function computed in~\cite{Hod:2000fh}.
Note that for $m\neq0$ the parameterization in Eq.~(15) and Eqs.~(35-36)
of~\cite{Hod:2000fh} yields the correct results also for
$l'=l_0+1$ (compare Sec.~\ref{sec:decaym0}).    

Furthermore, for $s>0$ the two simulations with $s=+1$, $l'=2$ and
$s=+2$, $l'=3$, agree with Hod's formula. 
For instance, in case of a $s=2$ and $m=2$ perturbation we measure $\mu_2=7|2$
as opposed to $\mu_2=9|8|3$ for $m=0$, which confirms the prediction 
that the splitting of LPIs in space is not present for $m\neq0$.

For what concerns splitting, we confirm that the splitting of LPIs in the $l=2,4$
projections of a $s=0$ and $l'=2$ perturbation observed for $m=0$ disappears in
the non-axisymmetric case $m=1$~\cite{Racz:2011qu}. 
By contrast, we observe splitting of LPIs in time in
the $l=3,5$ projections of a $l'=3$ perturbation, similarly to the $m=0$
case.

\subsection{Late time decay for $a\to1$}
\label{subsec:a1}

\begin{figure}[t]
  \centering   
   \includegraphics[width=0.49\textwidth]{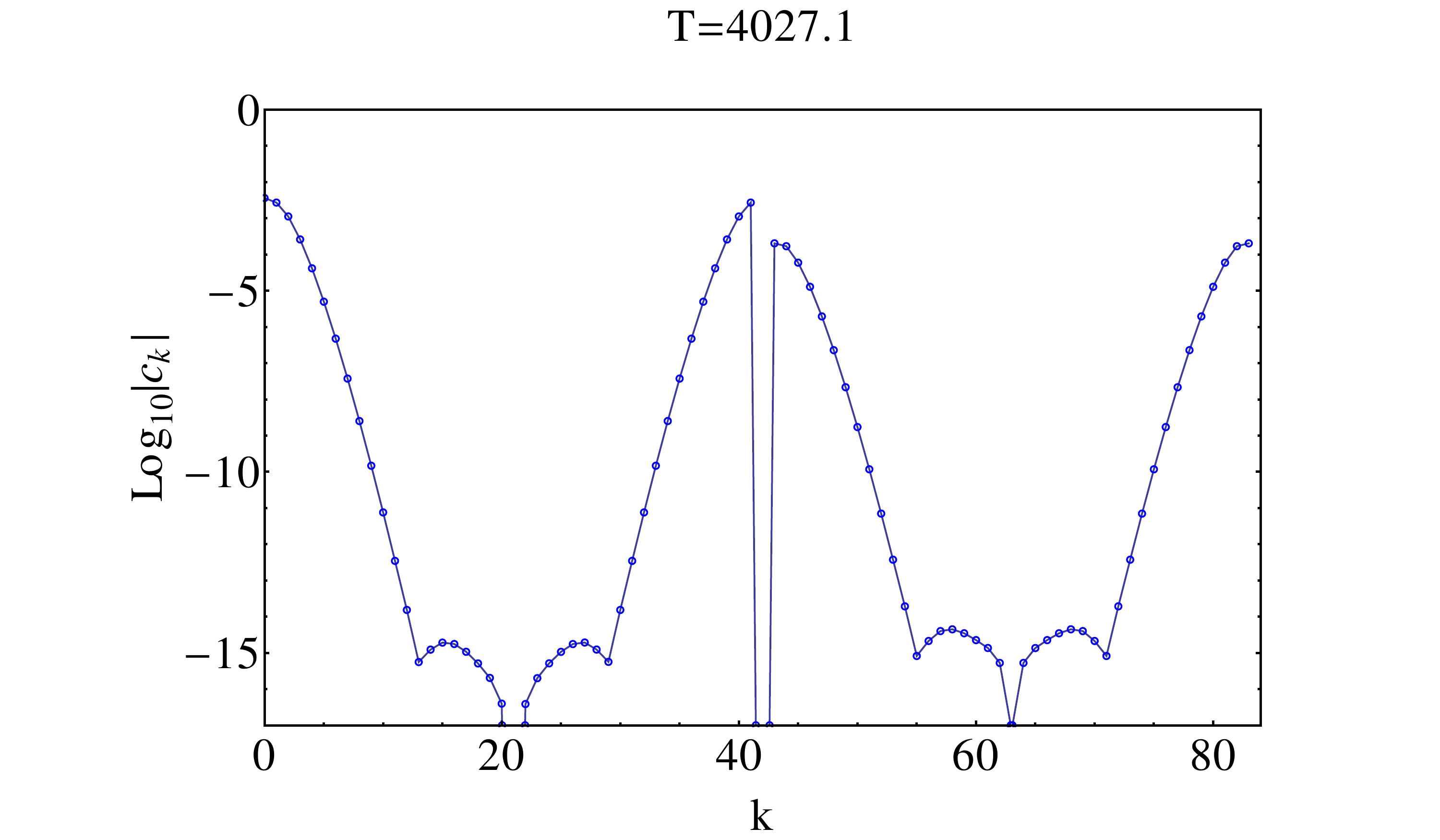}
   \includegraphics[width=0.49\textwidth]{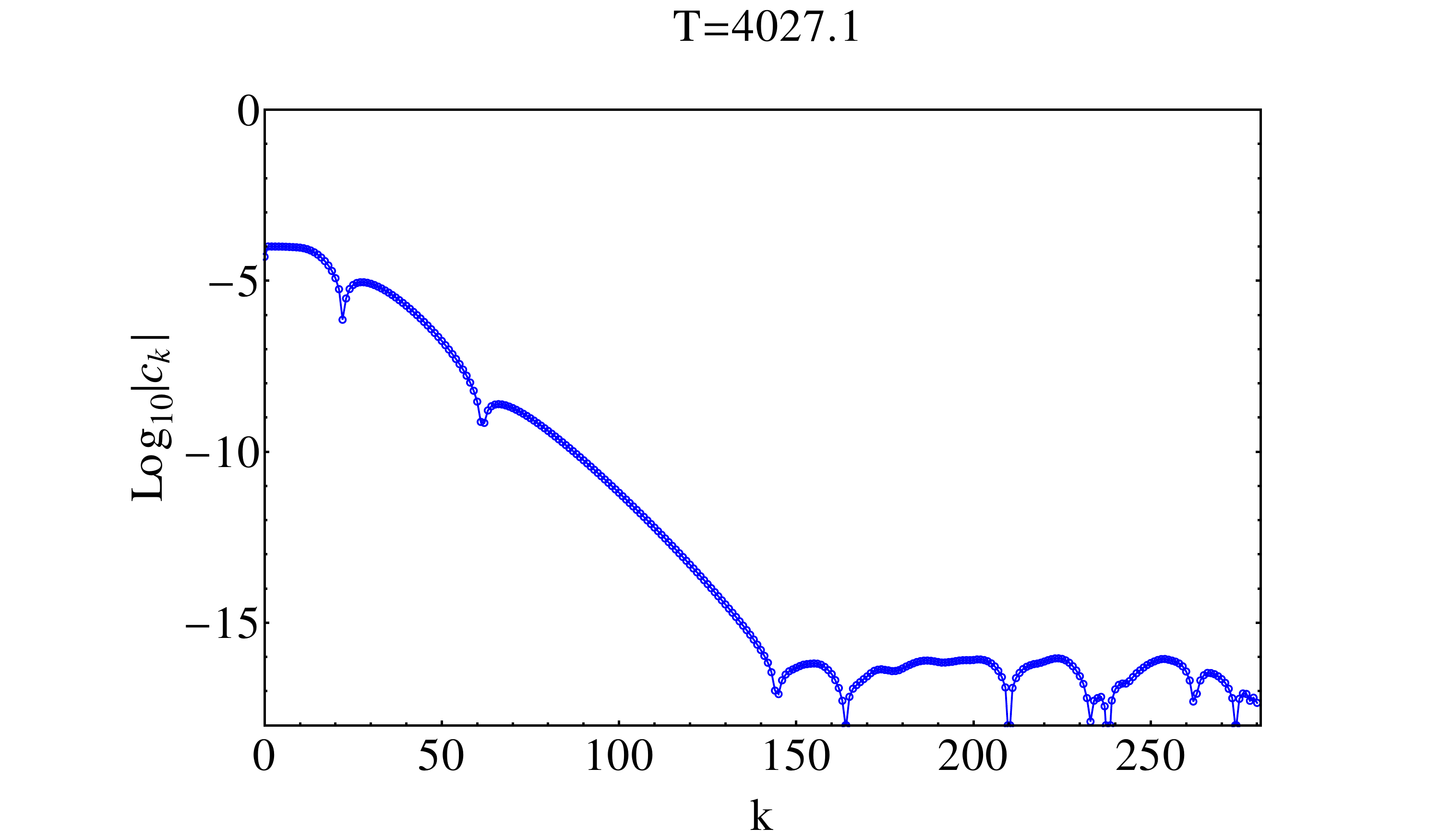}
   \caption{Spectral Fourier (left) and Chebyshev (right)
     coefficients of the field variable $\Re\{\psi\}$ for $s=-2$ and
     $a=0.9999$ at $T\approx4027M$.   
     The use of $n_{\theta}=21$ and $n_r=281$ points allows to
     achieve spectral convergence at late times in simulations of perturbations
     of almost extremal Kerr black holes. Long-double precision is
     employed here.}
 \label{fig:convergence_trappedmodes}
\end{figure}

\begin{figure}[t]
  \centering   
   \includegraphics[width=0.32\textwidth]{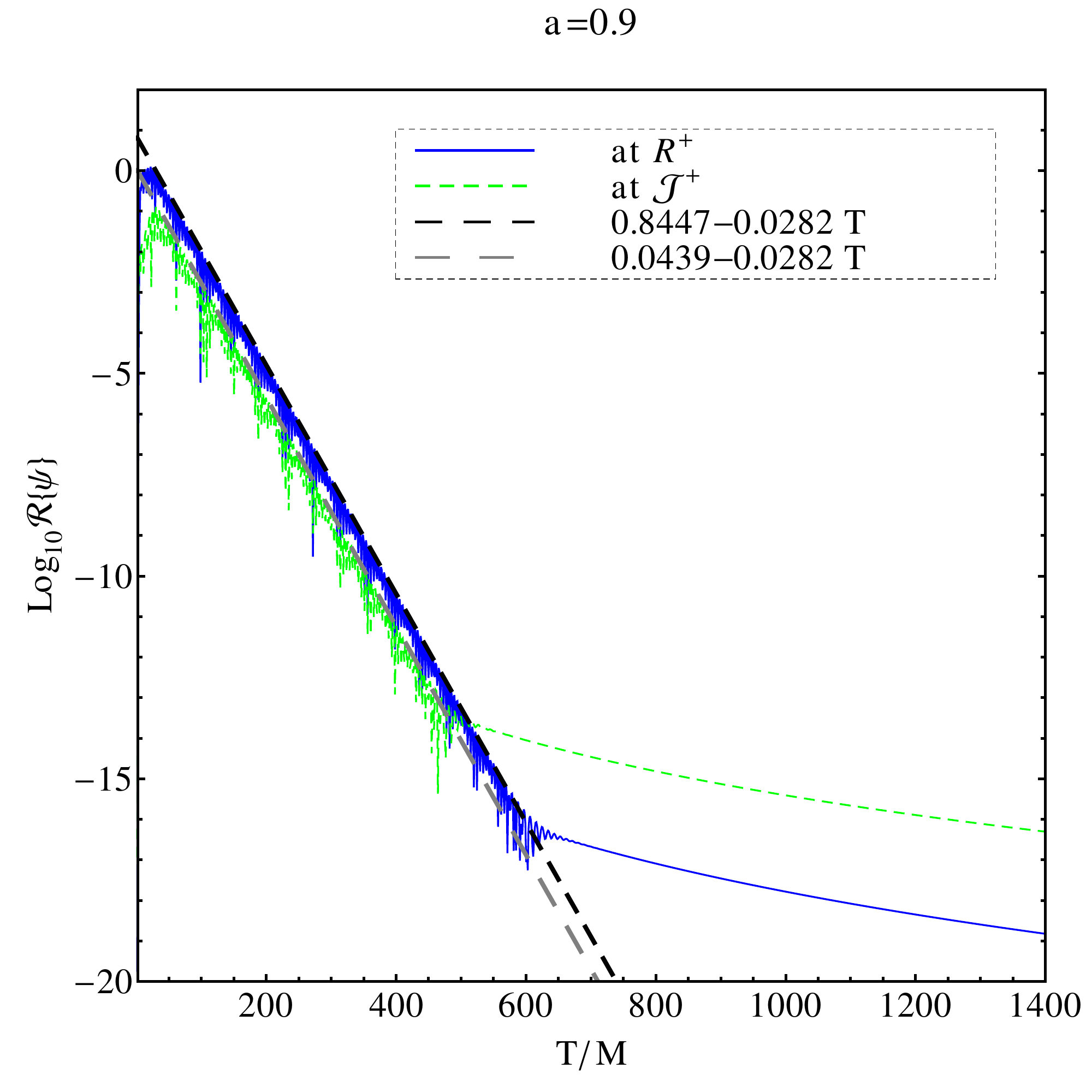}
   \includegraphics[width=0.32\textwidth]{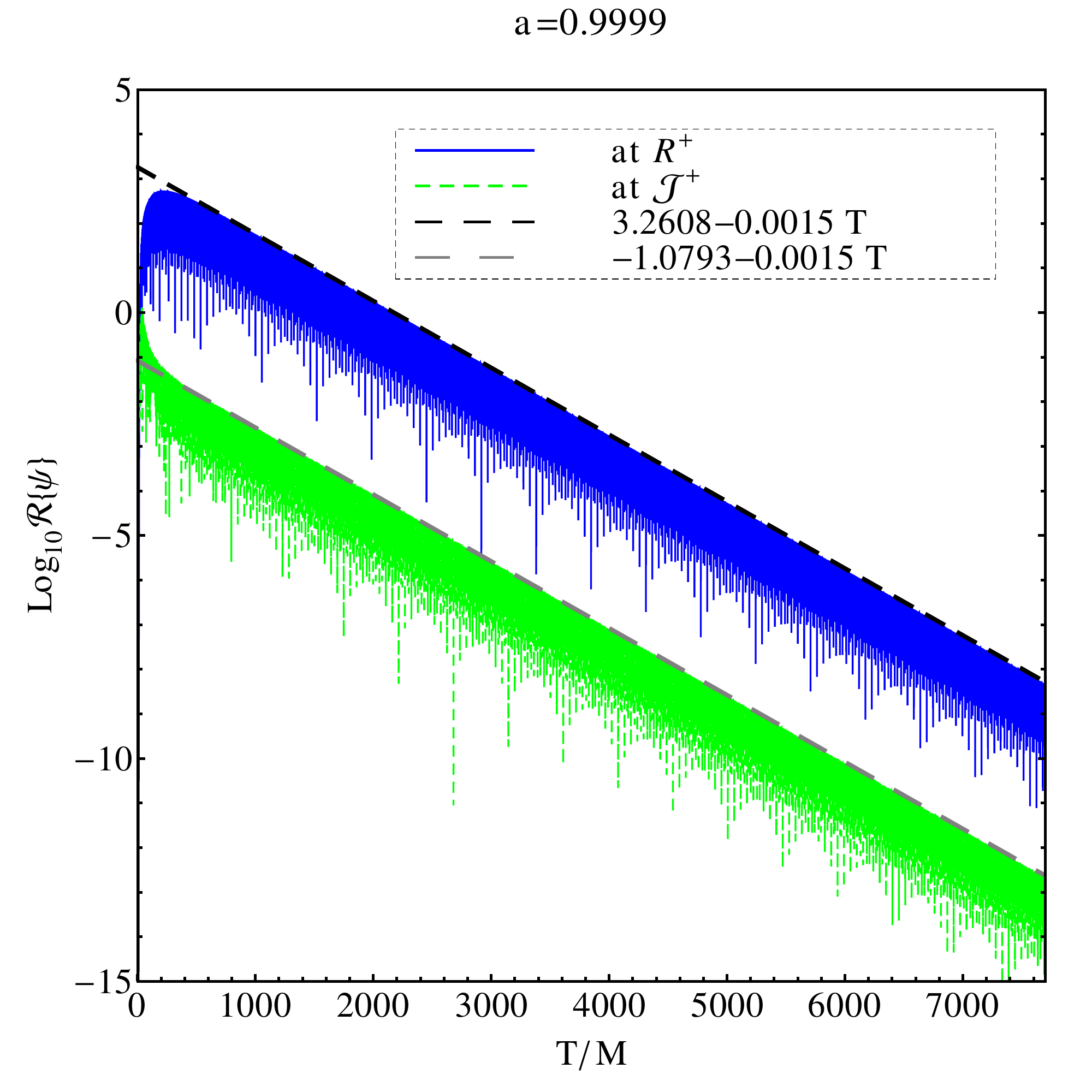}  
   \includegraphics[width=0.32\textwidth]{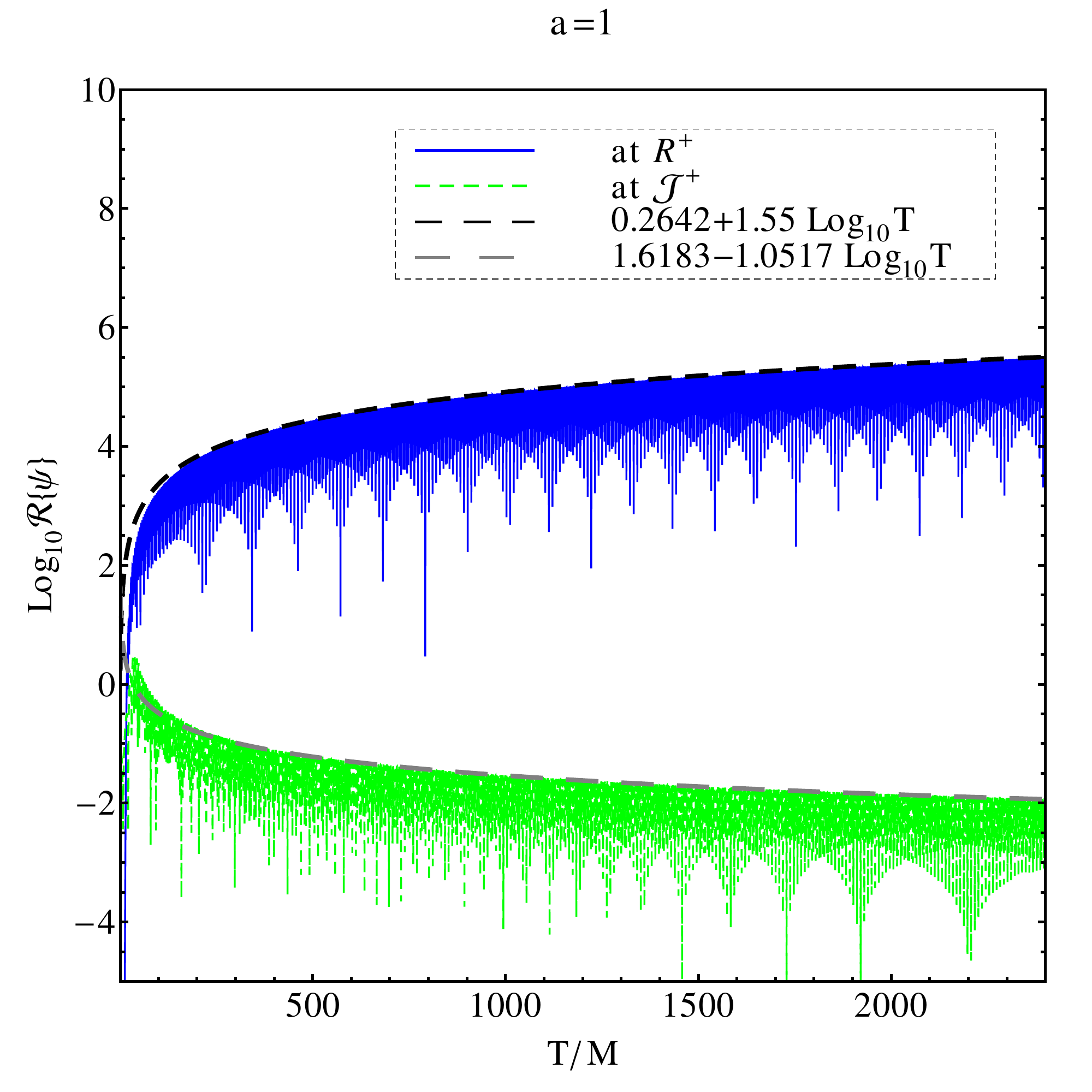}  
   \caption{Late time decay of $s=-2$ non-axisymmetric perturbations
     for highly spinning backgrounds. 
     The plots show the field variable $\Re\{\psi\}$ extracted at
     $R=R_+$ and $R=1$ and at $\theta=\pi/2$ for $a\in\{0.9,0.9999\}$ 
     and $\theta=1.1345$ for $a=1$.
     Initial data are ID1 and  $l'=m=2$.
     From left to right the field decays with a 
     power law tail ($a=0.9$), or with an oscillatory behaviour damped
     by either a slow exponential ($a=0.9999$) or a power law $1/T$
     ($a=1$).} 
  \label{fig:s=m2_trappedmodes}
\end{figure}

In this section we discuss the late time decay behaviour as the
angular momentum of the background  approaches the extremal value
$a=1$. We consider perturbations with $s=0$ and $s=-2$ and the
Gaussian in the initial data has parameters $R_0=0.9$ and $w=6000$.

Varying the angular momentum $a$ in the range
$\{0.9,\,0.99,\,0.999,\,0.9999,\,1\}$ in otherwise identical
simulations, we observe that the field develops an oscillatory
behaviour in space, mostly localized near the horizon and of
progressively larger amplitude for $a\to1$. The signal extracted 
at $R_+$ or $\scri^+$ is characterized by
a QNM phase which becomes longer (damping time increases); for $a=0.9999$ 
a power law tail is not observed up to $T\sim8000$ but may eventually 
arise at later times.
This behaviour is consistent with the findings
of~\cite{Andersson:1999wj,Glampedakis:2001js}, and can be
qualitatively understood in terms of modes trapped in the superradiance
resonant cavity.   

For $a\to1$ the amplitude of the oscillating field grows by a few orders of
magnitude near $R_+$ at the beginning of the simulation
(especially for $s=-2$). This leads to a step-like function which corrupts 
convergence if the number of radial grid points $n_r$ is too small
(see analogous discussion in Sec.~\ref{sec:num}).
Accurate simulations require PS radial derivatives and $n_r\sim281$
points for non-extreme cases. For the extreme case we set
$n_r=561$ for $s=0$ and $n_r=701$ for $s=-2$. 
On the other hand the large amplitudes and small damping times allow
for the use of long-double instead of quadruple (except for $a=0.9$ and
$s=-2$), which alleviates the computational costs.
Due to these facts we observe spectral convergence during the whole
simulation time, as shown in Fig.~\ref{fig:convergence_trappedmodes}
for $s=-2$ and $a=0.9999$. Both the Fourier coefficients
(left panel) and the Chebyshev coefficients (right panel) reach
round-off level in approximately straight lines at very late times
$T\sim4000$. Note that the Fourier coefficients contain both even- and
odd-indexed frequencies in contrast to the $m=0$ simulations shown in
Fig.~\ref{fig:Fou_Coeffs}.

The simulations' outcome is illustrated in 
Fig.~\ref{fig:s=m2_trappedmodes} for $s=-2$, $l'=m=2$, where we report the field
extracted at the horizon and $\scri^+$. On the simulated timescale,
the power law tail (left panel) is observed for values $a\leq0.999$. 
For larger values ($a=0.9999$, central panel) an oscillatory and
exponentially damped behaviour dominates the dynamics both at the horizon and
$\scri^+$. For $a=1$ the field remains oscillatory but does not decay
exponentially during the simulated time. At the horizon the field amplitude 
is still growing at $T=2500$, while at $\scri^+$ the decay
departs from an exponential law.
A similar qualitative behaviour is observed in the case $s=0$, $l'=m=1$.

It is instructive to compare the QNM (complex) frequencies
$\omega=(\omega_r,\omega_i)$ extracted
from the $s=-2$ simulations with the data by Berti et
al.~\cite{Berti:2009kk}  (listed there up to $a=0.9999$). 
The fundamental (longest lived) QNM has the frequency extracted 
by counting the periods of the signal for long times.
(Note that for $a\to1$ the frequencies of the 
overtones accumulate around this value.) 
Damping times (or $\omega_i$) are extracted by fitting the envelope
(maxima) of the field.
For $a=0.9$ we find $\omega\sim(0.6827,0.0649)$ both at 
$R_+$ and $\scri^+$ which agrees very well with the target values
$(0.6715,0.0649)$. For $a=0.99$ we find $\omega\sim(0.8747,0.0294)$ 
compared to $(0.8709,0.0294)$ and for $a=0.999$ we get 
$\omega\sim(0.9584,0.0105)$ compared to $(0.9558,0.0105)$.
For $a=0.9999$ we obtain $\omega\sim(0.9862,0.0035)$, again in agreement with
$(0.9857,0.0035)$. These results also agree with the analytical
formulas for QNM frequencies of near-extreme black holes
proposed in~\cite{Hod:2008zz,Hod:2009td}; the discrepancy is
$(\sim1\%,\sim5\%)$ for $a=0.99$ and $(\sim0.02\%,\sim0.5\%)$ for $a=0.9999$. 
The envelope of the field for $a=0.9999$ is well represented by the fit 
\be
\log_{10}|\Re{\psi}| =
  -1.079(3) - 0.0015064(1) T \ ,
\ee
where the error on the last significant digit of the coefficients is
reported in parentheses. 
Note that the fitting coefficient proportional to $T$ needs to be divided
by $\log_{10}(e)$ to obtain $\omega_i$.
In the extreme case we measure $\omega_r\sim0.9970$ at $\scri^+$ 
and $\omega_r\sim1.0015$ at $R_+$. It is interesting to notice 
that this is very close to the superadiance frequency 
$\omega_+=m\,a/(2\,M\,r_+)$ (similarly for $s=0$ we obtain
 $\omega_r\sim0.5011$ both at $R_+$ and $\scri^{+}$ in the extreme case). 
The envelope of the field is best fitted by the power law 
\be
\log_{10}|\Re{\psi}| =
  1.16(1) - 1.05(2) \log_{10} T \ , 
\ee
which confirms the expected $T^{-1}$ behaviour for
$s=-2$~\cite{Glampedakis:2001js}.

\section{Conclusion}
\label{sec:conc}

In this work we have presented a novel approach to time domain numerical
solutions of the Teukolsky equation (TE) for generic spin perturbations.
The approach is based on the use of the hyperboloidal slicing of the
Kerr spacetime, previously introduced by R{\'a}cz and T{\'o}th
(RT)~\cite{Racz:2011qu}, together with a proper rescaling of the 
field variable that leads to a regular equation for generic spin
perturbations (Eq.~\eqref{eq:rescale2}).
The RT coordinates are both horizon penetrating and compactify
null infinity ($\scri^+$) at a finite coordinate radius. A generalization is 
given by Eqs.~\eqref{eq:RT5}-\eqref{eq:RT6}, which, in particular,
allows for a generic compactification. The TE obtained from this
2-parameter family of coordinate transformations and the field
rescaling is regular. 

Accurate time-domain numerical solutions for generic spin perturbations 
can be performed employing standard discretization techniques. In
particular, long-term, stable 2+1 evolutions are obtained with the
method-of-line scheme (e.g.~Runge-Kutta time integrators) and either
high-order finite differencing or pseudo-spectral spatial derivatives.  
No boundary conditions are needed, and the computational domain
includes the horizon and $\scri^+$. 

As an application, we performed numerical experiments to investigate
the late time decays of generic spin perturbations. This problem
is a severe test for the robustness of the approach.

In a first series of experiments we investigated power law decays of
axisymmetric ($m=0$) perturbations of spins $s=0,\pm1,\pm2$. 
A variety of initial data have been used: stationary (ID0, ID2) or
non-stationary (ID1, ID3), and compact (ID0, ID1) or non-compact
(ID2, ID3) support. In all the cases we used pure multipole initial
data. 
The numerical results for $s=0$ agree with those reported in
the literature. In particular, for non-stationary and
compact-support initial data (ID1) the decay rates follow
Eq.~\eqref{eq:BK_mul}. 
For what concerns the decay rates of the projected modes of $s\neq0$ perturbations,
it is straightforward to write down {\it empirical} formulas summarizing our results
for $m=0$.
For instance we find for stationary and compact-support initial data (ID0)
\be
\label{eq:empif_h}
\mu_l =  \left\{ \begin{array}{l l}
  l' + l + 3 + \delta \ , & l' =  l_0 \\
  l' + l + 2 + \delta \ , & l' >  l_0 \ , \\ 
  \end{array} \right.    
\ee
at $R_+$. Above, $l_0=\max(|m|,|s|)=|s|$ and $\delta=0$ in
general, but $\delta=1$ in the special cases 
$m=0$ and $s>0$. The corresponding decay rates at $i^+$ are given by
the same expressions with $\delta=0$ in all cases. At $\scri^+$ the rates follow
\be
\label{eq:empif_s}
\mu_l =  \left\{ \begin{array}{l l}
  l - s + 2 \quad \ , & l\geq l'  \\
  l' - s + 1 \quad \ , & l<l'  \ . \\
\end{array} \right.    
\ee
Similarly for non-stationary, compact-support initial data (ID1) we obtain  
\be
\label{eq:empif_h}
\mu_l =  \left\{ \begin{array}{l l}
  l'  + l + 3 + \delta\ , &  l' = l_0,l_0+1 \\
  l'  + l + 1 + \delta\ , &  l' >    l_0+1  \ , \\ 
  \end{array} \right.    
\ee
at $R_+$. The corresponding decay rates at $i^+$ are given by
the same expressions with $\delta=0$ in all cases, while at $\scri^+$ we get  
\be
\mu_l =  \left\{ \begin{array}{l l}
    l-s+3 \ ,&    l+1 \geq l' , \; l'=l_0+1=l+1 \\
    l-s+2    \ ,&    l+1 \geq l' , \; l'\neq l_0 +1 = l+1 \\
    l' -s      \ ,&  l+1 < l'	
\end{array} \right.    
\ee
Decay rates for ID2 (ID3) can be easily deduced from those for ID0 (ID1) by
decreasing the $\mu_l$ by one 
\be
\mbox{ID0}\to\mbox{ID2},\; \mbox{ID1}\to\mbox{ID3} : \mu_l\to\mu_l-1 .
\ee
These empirical formulas describe all our numerical results for
$s\neq0$ and $m=0$.

The decay rates computed for $s\neq0$ can be compared 
with Hod's analytic predictions~\cite{Hod:1999rx,Hod:2000fh}. The
comparison is easiest to make for ID1. The results agree with the
analytical computation (see~\ref{app:green} for a review of the
calculation and a small correction for the case $l'=l_0+1$ with $m=0$).
For similar type of initial data, Krivan et al~\cite{Krivan:1997hc}
reported for $s=-2$ perturbations with the concerned $l'=3$ initial
mode $\mu=7.72$. This value was interpreted to asympotically reach
$\mu=7$ but could be also compatible with our finding of $\mu=8$.
The results of Burko and Khanna for $s=+2$~\cite{Burko:2002bt}  can
not be compared with ours due to the different coordinates used there. 

For the first time, to our knowledge, we have verified the LPI
splitting (in time) in upper modes of $s\neq0$ perturbations, as
well as the LPI splitting (in space) of $s>0$, $m=0$
perturbations. In the latter case, the LPIs reach their asymptotic
value first at a radius $\tilde{R}$ close to the horizon. The 
LPIs measured by $R_+<R<\tilde{R}$ show a qualitative difference for
$s=+1$ and $s=+2$ perturbations: in the former case a splitting in
time is observed very close to $R_+$. 

In a second series of experiments we investigated the late time decay
of non-axisymmetric ($m\neq0$) perturbations, mostly of spins
$s=0,-2$. Focusing on non-stationary and compact support initial data
(ID1), we considered the late time behavior of the field for
different values of the black hole spin. For $a=0.9$ we find power
law decays which agree with the literature results for $s=0$ and
with~\cite{Hod:1999rx,Hod:2000fh} for $s=-2$. 

If the black hole rotation approaches the extremal limit, the late
time decays follow a different law. Perturbations of a rapidly
rotating non-extremal background are at very late times still
dominated by a weakly damped QNM pattern.
In the extremal case, the field decay measured from the
simulations is consistent with a $T^{-1}$ power law.
These results agree with the findings
of~\cite{Andersson:1999wj,Glampedakis:2001js}, the case $s=-2$ is here
confirmed for the first time by a numerical time-domain experiment.

In conclusion, the method introduced here is found to be accurate and
robust for the solution of the TE in the time domain.
To the best of our knowledge these
results are the first numerical investigations of the late time decay at
future null infinity for $s\neq0$ perturbations on Kerr.
We also view these results as preparatory to the development of an 
accurate black-hole-binary test-mass laboratory \unskip
~\cite{Nagar:2006xv,Bernuzzi:2010ty,Bernuzzi:2010xj,Bernuzzi:2011aj,Barausse:2011kb}. 
Some of the results presented may be of interest for the self-force
literature.

\ack

We are indebted to A.~Zengino{\u g}lu for discussions and pointing out the
rescaling in Eq.~\eqref{eq:rescale2} for BL coordinates in an early
stage of this work. We thank L.~Barack and S.~Hod for helpful comments on the
manuscript.
Our work also benefitted from  
discussions with D.~Hilditch and A.~Weyhausen. 
This work was supported in part by  DFG grant SFB/Transregio~7
``Gravitational Wave Astronomy''. 
E.H.\ wishes to thank the 
council of Olf and W. Buchholz.
S.B.\ thanks J.D.~Bernuzzi for a careful reading of the manuscript.

\appendix

\section{Equation coefficients and RT metric}
\label{app:coefs}

The coefficients of Eq.~\eqref{eq:TE_RT} are
\begin{align*}
  C_0  & = \frac{-\frac{1}{2} a^2 \left(R^2-1\right)^2-M \left(R^2-1\right) R (s+1)+R^2 s \left(s \cot ^2(\theta )-1\right)}{R^3} \\
  C_{T} & = \frac{1}{R (R+1)
   \left(R^2+1\right)^3} \\
  & \left\{-a^2 (R-1) (R+1)^2 \left(2 M \left(3 R^6+5 R^4-7 R^2-1\right)-R^4-6 R^2-1\right) \right. \\
   & \left. +2 i a (R+1) \left(R^2+1\right)^3 s \cos (\theta ) \right. \\
   & \left. -2 \left[4 M^2 (R-1) R (R+1)^2 \left(R^4
   (s+2)+2 R^2 (s+3)+s\right) \right. \right.\\ 
   & \left.\left. \qquad  + M (R+1) \left(R^6 (7 s+3)-2 R^5 s+13 R^4 (s+1)-4 R^3 (s+2)+R^2 (5 s-7)-2 R s-s-1\right) \right. \right. \\
   & \left. \left. \qquad  +2 (R-1) R \left(R^4 s+R^3+2 R^2 (s+1)+R+s\right)\right]\right\} \\
  C_R & = -\frac{\left(R^2-1\right)}{2 R^2 \left(R^2+1\right)^3} \\
  & \left\{ a^2 \left(2 R^4+5 R^2+1\right) \left(R^2-1\right)^2  \right.\\
  & \left. +2 R \left[M \left(R^2-1\right) \left(R^4 (s+3)+2 R^2 (s+4)+s+1\right)+2 R \left(R^4 (s+1)+R^2 (2
   s+3)+s\right)\right]\right\} \\
  C_{\theta} & = -\frac{\cot (\theta )}{R} \\
  C_{\varphi} & = \frac{-a R^2+a-2 i R s \cot (\theta ) \csc (\theta )}{R^2} \\
  C_{TT} & =\frac{1}{2 R (R+1) \left(R^2+1\right)^2} \\
    & \left\{a^2 (-(R+1)) \left(32 M^2 R^6-\left(64 M^2+32 M+1\right) R^4+\left(32 M^2+32 M+6\right) R^2-1\right) \right.\\
    & \left. +a^2 (R+1) \left(R^2+1\right)^2 \cos (2 \theta ) \right. \\
    & \left. -8 R \left[16 M^3 (R-1) R^2
   (R+1)^2+8 M^2 R \left(R^3-R^2-3 R-1\right) \right. \right.\\
    & \left. \left. \qquad +M \left(R^3-11 R^2-5 R-1\right)-R (R+1)\right]\right\} \\  
  C_{RR} & = -\frac{\left(R^2-1\right)^2 \left(a^2 \left(R^2-1\right)^2+4 R \left(M \left(R^2-1\right)+R\right)\right)}{4 R \left(R^2+1\right)^2} \\
  C_{\theta\theta} & = -\frac{1}{R} \\
  C_{\varphi\varphi} & = -\frac{\csc ^2(\theta )}{R} \\
  C_{TR} & = -\frac{2}{\left(R^2+1\right)^2} \\
  & \left\{ a^2 \left(R^2-1\right)^2 \left(2 M \left(R^2-1\right)-1\right)+2 \left(4 M^2 R \left(R^2-1\right)^2+M \left(3 R^2+4 R+1\right) (R-1)^2-2
   R^2\right)\right\} \\
  C_{T\varphi} & = \frac{4 a \left(1-2 M \left(R^2-1\right)\right)}{R^2+1} \\
  C_{R\varphi} & = -\frac{a \left(R^2-1\right)^2}{R^3+R} \\
\end{align*}
The coefficients of Eq.~\eqref{eq:TE_RTm} are given by
\begin{align*}
  \tilde{C}_0  			& = C_0 + i \; m \; \Re\{C_{\varphi}\} - m \; \Im\{C_{\varphi}\} - m^2 \; C_{\varphi\varphi} \\
  \tilde{C}_{\theta} 		& = C_{\theta}  \\
  \tilde{C}_{\theta\theta} 	& = C_{\theta\theta}  \\
  \tilde{C}_R 			& = C_{R} + i \; m \; C_{R\varphi} \\
  \tilde{C}_{RR} 		& = C_{RR} \\
  \tilde{C}_{T}			& = C_{T} + i \; m \; C_{T\varphi} \\
  \tilde{C}_{TR}		& = C_{TR} \\
  \tilde{C}_{TT}		& = C_{TT}
\end{align*}
The Kerr metric in RT coordinates can be written 
\be
ds^2= (-\alpha^2+\beta_R\beta^R+\beta_{\varphi}\beta^{\varphi}) dT^2 + 2 (\beta_R \; dT \;dR + \beta_{\varphi} \;dT \; d{\varphi}) +  (\gamma_{RR} dR^2 + \gamma_{\theta\theta} d\theta^2 + \gamma_{\varphi\varphi} d\varphi^2 + 2 \gamma_{R\varphi} \; dR \; d\varphi) \; ,
\ee
where the $3+1$ metric functions are given by
\begin{align*}
 \beta_R & = \frac{4 R \left(8 M^2 (R-1) R+2 M (R+1) \left(\rho ^2 (R-1)^2-1\right)-\rho ^2 (R-1)\right)}{\rho ^2 (R-1)^3 (R+1)^2}\\
 \beta_{\theta} & = 0 \\
 \beta_{\varphi} &= -\frac{4 a M R \sin ^2(\theta )}{\rho ^2 \left(1-R^2\right)}\\
 \gamma_{RR} & = -\frac{4 \left(4 M R^2-(4 M+1) R-1\right) \left(16 M^2 R^2 (R-1)+4 M R (R+1) \left(\rho ^2 (R-1)^2-1\right)+\rho ^2
   (R-1)^3\right)}{\rho ^2 (R-1)^5 (R+1)^3} \\
 \gamma_{\theta\theta} & = \rho ^2 \\
 \gamma_{\varphi\varphi} & =  \frac{\Sigma ^2 \sin ^2(\theta )}{\rho ^2} \\
 \gamma_{R\varphi} &=-\frac{2 a \sin ^2(\theta ) \left(16 M^2 (R-1) R^2-4 M (R+1) R+\rho ^2 \left(R^3-R^2+R-1\right)\right)}{\rho ^2 (R-1)^3
   (R+1)^2} \\
 \alpha & = \left\{\left(\rho ^2 (R-1)^2 \left(R^2+1\right)^2 \left(a^2 \sin ^2(\theta ) \left(\rho ^2-\frac{4 M
   R}{R^2-1}\right)-\Sigma ^2\right)\right) \right.\\ 
   & \qquad \left. \left[a^2 \sin ^2(\theta ) \left(16 M^2 (R-1) R^2-4 M (R+1) R+\rho ^2
   \left(R^3-R^2+R-1\right)\right)^2  \right. \right.\\
    & \qquad \left. \left. +(R-1) (R+1) \Sigma ^2 \left(4 M R^2-(4 M+1) R-1\right) \right. \right. \\
    & \qquad \quad \left. \left. \left(16 M^2 R^2 (R-1)+4 M
   R (R+1) \left(\rho ^2 (R-1)^2-1\right)+\rho ^2 (R-1)^3\right)\right]^{-1}\right\}^{\frac{1}{2}} \qquad ,
\end{align*}
and the shorthands
\begin{align*}  
  r(R) 	 &= \frac{2 R}{1-R^2}  \\
  \Delta(R) &= r(R)^2 - 2 \, M \, r(R) + a^2 \\
  \Sigma(R,\theta) &= \sqrt{ \left(r(R)^2 + a^2  \right)^2 - a^2 \Delta(R) \sin(\theta)^2} \\
  \rho(R,\theta)   &= \sqrt{ r(R)^2 + a^2 \cos(\theta)^2 } 
\end{align*}    
are used.

\section{Green's function calculation}
\label{app:green}

In this appendix we review the analytical calculation
of~\cite{Hod:1999ci,Hod:1999rx,Hod:2000fh}, and specify it for the
case of ID1 pointing out explicitly a special case relative to $m=0$,
$s\neq0$ and pure multipole initial 
data $l'=l_0+1$ that was overlooked in the final parametric formulas
of~\cite{Hod:2000fh}. 
To avoid confusion we will follow the naming conventions
of~\cite{Hod:2000fh}, but keep $l'$ as index of the pure multipole
initial data. The calculation is in BL
coordinates, and employs the radial tortoise coordinate $y$ defined by 
$dy = (r^2+a^2)/\Delta dr$. The main variable $\Psi$ refers to the 2+1
decomposed fields rescaled by $\Delta^{-s/2} (r^2+a^2)^{-1/2} $
(compare with Eq.~(\ref{eq:rescale2}).)

The solution to the 2+1 TE wave equation for $t>0$ can be expressed in
terms of the retarded Green's function $G(y,\theta,y',\theta',t)$ and the
initial data as (see
Eq.~(6) in~\cite{Hod:2000fh}) 
\begin{align}
\label{eq:GreenSol}
\Psi(y,\theta,t) =  2 \pi \int \int_0^{\pi} &
\left\{B_1(y',\theta') \left[ G(y,\theta,y',\theta',t)
  \Psi_t(y',\theta',0) + G_t(y,\theta,y', 
\theta',t) \Psi(y',\theta',0)\right] \right.\\ \nonumber
 & \left. + B_2(y',\theta') G(y,\theta,y',\theta',t)
   \Psi(y',\theta',0) \right\} \ 
\sin\theta' \, d\theta' dy'  \ , 
\end{align}
where
$B_1(y,\theta)$ and $B_2(y,\theta)$ are background dependent complex 
functions (Eq.~(4-5) in~\cite{Hod:2000fh}), in particular 
$B_1(y,\theta) = 1-b_1(y)\,\sin^2\theta$.

For non-stationary, pure multipole ID1 $\Psi(y,\theta,0)=0$ with
$\Psi_t(y,\theta,0)\propto {}^sY_{l'm}(\theta)$ only the first of the
three terms in Eq.~\ref{eq:GreenSol} has to be evaluated, 
\begin{align}
\Psi(y,\theta,t) =  2 \pi \int \int_0^{\pi} 
B_1(y',\theta') G(y,\theta,y',\theta',t) \Psi_t(y',\theta',0) \;
\sin\theta' \, d\theta' dy' \ .
\end{align}
The Green's function $G(y,\theta,y',\theta',t)$ can be expressed in terms
of its Fourier transform and expanded in the spin weighted spheroidal
harmonics ${}^sS_{lm}(\theta,a\omega)$ basis, 
\begin{equation}
\label{eq:GreenFun}
 G(y,\theta,y',\theta',t)=(2\pi)^{-2} \int\limits_{-\infty+i \,c}^{\infty+i \, c} 
\sum\limits_{l=l_0}^{\infty} 
\tilde{G}_l(y,\theta,y',\theta',\omega) 
      {}^sS_{lm}(\theta,a\omega) 
      {}^sS_{lm}(\theta',a\omega)\, e^{-i \, \omega \, t} d\omega \ .
\end{equation}
The Green's functions $\tilde{G}_l(y,\theta,y',\theta',\omega)$ admit
an analytic solution in terms of two linearly independent 
solutions of the radial Teukolsky equation, see Eq.~(18) of~\cite{Hod:1999ci}.
Given $\tilde{G}_l(y,\theta,y',\theta',\omega)$, to obtain the Green's
function it remains to plug it back into Eq.~\ref{eq:GreenFun} and
evaluate the integral. As discussed in~\cite{Leaver:1986gd} the
integration contour is chosen in the lower half of the complex
$\omega$ plane. The integral consists of three distinct
contributions:  
(i)~an integral along the semicircle contributing to the
early time solution, 
(ii)~the residues (from the poles of
$\tilde{G}_l(y,\theta,y',\theta',\omega)$) in the lower half of the
plane, corresponding to the QNM contribution; 
(iii)~an integral along the branch cut
$\tilde{G}_l(y,\theta,y',\theta',\omega)$ on the negative imaginary
$\omega$ axis, contributing to the late time solution. In the
following we consider only the latter and write the Green's function as 
$G(y,\theta,y',\theta',t) \approx G^C(y,\theta,y',\theta',t)$.
 
The calculation of the tail contribution to the Green's function, and
thus of the solution in Eq.~\ref{eq:GreenSol}, proceeds by considering a
small $\omega$ approximation, which corresponds to selecting the large
$r$ contribution of the (effective) scattering potential. 
Considering $\tilde{G}_l(y,\theta,y',\theta',\omega)$ in the
small-$\omega$ limit the expression for $G^C$ for observers at 
time like infinity 
reads (Eq.(12) of~\cite{Hod:2000fh})   
\begin{align}
\label{eq:GC}
 G^C(y,\theta,y',\theta',t) & = 
 \sum\limits_{l=l_0}^{\infty} K_{ls} \, (y\,y')^{l+1}
 \int\limits_{0}^{-i \, \infty} 
            {}^sS_{lm}(\theta,a\omega) 
            {}^sS_{lm}(\theta',a\omega) \, \omega^{2l+2}
 \, e^{-i \, \omega \, t} \; d\omega \\ \nonumber   
\end{align}
where $K_{ls}$ are constants solely depending on $l$ and $s$.

Next, appropriate approximations for ${}^sS_{km}(\theta,a\omega)$
have to be introduced. In general, the spheroidal spherical harmonics
are defined by the eigenvalue problem
\be
 L \ {}^sS_{lm} = \left(L_0 + L_1\right) {}^sS_{lm} = - A_{slm}
 {}^sS_{lm} \ ,
\ee
where $L$ is a differential operator which splits into a
$\omega$-independent part with $\theta$-derivatives ($L_0$) and
$L_1=(a\omega)^2 \cos^2\theta - 2 (a\omega) s \cos\theta$, the
$\omega$-dependent part. If $(a\omega)$ is small, one can view $L_1$ as a
perturbation of the operator $L_0$ and compute a perturbative
expansion in $(a\omega)$~\cite{Schiff:1968}.
The unperturbed eigenfunctions are the spin weighted spherical
harmonics, which are the eigenfunctions in the non-rotating case,
i.e.~${}^sS_{lm}={}^s Y_{lm} +\mathcal{O}(a\omega)$. The expansion
then reads
\be
\label{eq:Slmexp}
 {}^sS_{lm} = {}^sY_{lm} + \sum\limits_{i\neq l} \,
 \frac{\langle sim|L_1|slm\rangle}{\Delta A_{li}^{(0)}} \,  {}^sY_{im}  + ... 
\ee
where $\langle sim|F(\theta)|slm\rangle=\int d\Omega \ 
{}^sY^*_{im}{}^sY_{lm}F(\theta)$, and $\Delta A_{li}^{(0)}$ is the
difference between the $l$-th and $i$-th unperturbed eigenvalues
(independent of $\theta$ and $\omega$). The computation of the
expansion coefficients requires the evaluation of the integrals
$\langle sim|\cos\theta|slm\rangle$ and $\langle
sim|\cos^2\theta|slm\rangle$, the former are non-vanishing for
$i=l,l\pm1$, the latter for $i=l,l\pm1,l\pm2$.
The expansion in Eq.~\eqref{eq:Slmexp} can be written as 
\be
\label{eq:SlmexpC}
{}^sS_{lm} = \sum_{i=l_0} C_{li} \ (a\omega)^{|l-i|} \ {}^sY_{im}\ , 
\ee
where the coefficients $C_{li}$ depend on $(a\omega)$, but are at leading order
independent of $(a\omega)$ for $(a\omega)\ll1$. 

Substituting  the unperturbed eigenfunctions in Eq.~\eqref{eq:GC},
i.e.~${}^sS_{lm}\rightarrow {}^sY_{lm}$, leads to
\be
 G^C(y,\theta,y',\theta',t) = \sum\limits_{l=l_0}^{\infty}  K_{ls} \,
 (y\,y')^{l+1} \, {}^sY_{lm}(\theta) {}^sY_{lm}^*(\theta')
 \,\int\limits_{0}^{-i \, \infty} \omega^{2l+2} \, e^{-i \, \omega \,
   t} \; d\omega  
 \propto t^{-(2l+3)} \ ,  
\ee
which reproduces the Price law with $\mu=2l+3$.
In order to describe mode coupling and obtain the actual tails on
Kerr, one needs to consider the expansion in Eq.~\eqref{eq:Slmexp}.
Renaming some indexes, dropping the $s$ and $m$ indexes, and
rearranging some terms, Eq.~\eqref{eq:GC} reads 
\be
 G^{C} = 
 \sum_{l=l_0}^{\infty} 
 \sum_{i=l_0}^{\infty}
 \sum\limits_{k=l_0}^{\infty} 
 K_{ks} \, (y\,y')^{k+1} \,
 \int\limits_{0}^{-i \, \infty} 
d\omega \, e^{-i \, \omega \, t} \, \,\omega^{2k+2} \, 
C_{kl}  \,  (a\omega)^{|k-l|} \, Y_l(\theta) \,
C_{ki}  \,  (a\omega)^{|k-i|} \, Y^*_i(\theta') \ ,
\ee
from which one can read off the $l$-mode contribution $G^{C}_l$ defined
by $G^{C}=\sum_l G^{C}_l$. 

We define now an {\it effective} Green's
function for the $l$-mode that (i)~takes into account the non-vanishing
terms in the sum over $i$ to the (outer) integral $\int
d\theta'\sin\theta' B_1(\theta') {}^sY_{l'm}(\theta')$,
and (ii)~restricts to the lowest powers of
$(a\omega)$ that contribute to the sum over $k$.
Step (i) requires the evaluation of the integrals like $\langle 
sim|\sin^2\theta'|sl'm\rangle$. In particular, for $m\neq0$ one has
that the non-vanishing terms of $\langle sim|\sin^2\theta'|sl'm\rangle$
are those given by $i=l',l'\pm1,l'\pm2$.
For $m=0$ one has that the non-vanishing terms of
$\langle si0|\sin^2\theta'|sl'0\rangle$ are those given by
$i=l',l'\pm2$. The calculation at step (i) gives 
\begin{align}
  G^{C \, {\rm eff}}_l&= 
  \sum\limits_{k=l_0}^{\infty} 
  K_{ks} \, (y\,y')^{k+1}
 \int\limits_{0}^{-i \, \infty}
 d\omega \, e^{-i \, \omega \, t} \, \,\omega^{2k+2} \, 
 C_{kl}  \,  (a\omega)^{|k-l|} \, Y_l(\theta)\\
 &\left\{\right. 
 \left[ 1 - b_1(y') \right] 
 C_{kl'} (a\omega)^{|k-l'|} Y_{l'}^*(\theta') \nonumber\\
 & - b_1(y') \left[\right. 
   C_{kl'-2} (a\omega)^{|k-(l'-2)|} Y_{l'-2}^*(\theta') 
   + C_{kl'-1} (a\omega)^{|k-(l'-1)|} Y_{l'-1}^*(\theta') \nonumber\\
   & + C_{kl'+1} (a\omega)^{|k-(l'+1)|} Y_{l'+1}^*(\theta')
   + C_{kl'+2} (a\omega)^{|k-(l'+2)|} Y_{l'+2}^*(\theta')\left.\right]\nonumber
\left.\right\}
\end{align}

Note that of the terms in the curly brackets the
$C_{kl'\pm1}$ do not exist for $m=0$ and the $C_{kl'-2}$ and
$C_{kl'-1}$ do not exist, respectively, for $l'=l_0,l_0+1$ and $l'=l_0$. 
Step (ii) requires power counting for given combinations of $(l,l')$. 
The freedom of the powers in the curly brackets requires to analyze 
distinctly the three cases $l'=l_0$, $l'=l_0+1$ and $l'\geq l_0+2$. 
In each of these checking the first terms of the $k$-sum reveals 
that the lowest power is always obtained by the $k=l_0$ term (however 
$k>l_0$ terms may amount to the same power). 
Thus the lowest power of
$(a\omega)$ is 
\be
n_l = \begin{cases}
2l_0+2+|l_0-l'| + |l_0-l| & \ , \ l'=l_0  \ , \\
2l_0+2+|l_0-(l'-1)| + |l_0-l| & \ , \ l'=l_0+1 \ , \\
2l_0+2+|l_0-(l'-2)| + |l_0-l| & \ , \ l'>l_0+1 \ .
\end{cases}
\ee
The lowest power of $(a\omega)$ for $m=0$ is instead given by
\be
n_l = \begin{cases}
2l_0+2+|l_0-l'| + |l_0-l| & \ , \ l'=l_0  \ , \\
2l_0+2+|l_0-l'| + |l_0-l| & \ , \ l'=l_0+1 \ , \\
2l_0+2+|l_0-(l'-2)| + |l_0-l| & \ , \ l'>l_0+1 \ ,
\end{cases}
\ee
as a consequence of the vanishing integrals $\langle
si0|\sin^2\theta'|sl'0\rangle$ for $i=l'\pm1$.
It is important to note that for $l'\geq l_0+2$, other terms in the
sum than the $k=l_0$ can occur with the power $n_l$.
This is  because, while the $\omega^{2k+2}$ contribution will always be
greater than $\omega^{2l_0+2}$ for $k>l_0$, the $\omega$ powers that
occur with the $C$ coefficients can be smaller for $k>l_0$ than for
$k=l_0$.  

Finally, the power of the tails for each $l$-mode can be calculated
from the integrals $\int d\omega\, e^{-i\omega \, t} \,\omega^{n_l}$ which
give $t^{-\mu_l}$ with $\mu_l=n_l+1$.
The different cases above are actually summarized in the
parametric formula in Eq.~(15) of~\cite{Hod:2000fh}. However, the
parametrization there seems to miss the special case
$l'=l_0+1$ and $m=0$.  

The calculation at null infinity is similar to the one above. 
In this case different approximations for the Fourier transform $\tilde{G}_l(y,\theta,y',\theta',\omega)$ 
of Eq.~(\ref{eq:GreenFun}) have to be made but the remaining procedure is unchanged. In particular
using again Eq.~(\ref{eq:SlmexpC}) for the approximation of the
${}^sS_{lm}$ and that $\langle sim|\sin^2\theta'|sl'm\rangle=0$  
unless $i=l',l'\pm1,l'\pm2$ one finds the effective Green's function (cf.~\cite{Hod:2000fh}) of the $l$-mode

\begin{align}
  G^{C \, {\rm eff}}_l&= 
  \sum\limits_{k=l_0}^{\infty} 
  \tilde{K}_{ks} \, y'^{k+1} \, y^{-s}
 \int\limits_{0}^{-i \, \infty}
 d\omega \, e^{-i \, \omega \, (t-y)} \, \,\omega^{k-s+1} \, 
 C_{kl}  \,  (a\omega)^{|k-l|} \, Y_l(\theta)\\
 &\left\{\right. 
 \left[ 1 - b_1(y') \right] 
 C_{kl'} (a\omega)^{|k-l'|} Y_{l'}^*(\theta') \nonumber\\
 & - b_1(y') \left[\right. 
   C_{kl'-2} (a\omega)^{|k-(l'-2)|} Y_{l'-2}^*(\theta') 
   + C_{kl'-1} (a\omega)^{|k-(l'-1)|} Y_{l'-1}^*(\theta') \nonumber\\
   & + C_{kl'+1} (a\omega)^{|k-(l'+1)|} Y_{l'+1}^*(\theta')
   + C_{kl'+2} (a\omega)^{|k-(l'+2)|} Y_{l'+2}^*(\theta')\left.\right]\nonumber
\left.\right\} \;,
\end{align}
where $\tilde{K}_{ks}$ is a constant depending on $k$ and $s$. This Green's function
provides again all information to find the decay rates for any pair of $(l,l')$ by 
going through the sum over $k$ and picking the lowest power of $\omega$. While at $i^+$ 
the lowest power is always obtained by the $k=l_0$ term one finds that this is not 
true at $\scri^+$. Still in~\cite{Hod:2000fh} Hod found an appropriate way to parametrize
the resulting decay rates but again the parametrization contains an error for $l'=l_0+1$ 
with $m=0$. In this case the $C_{kl'-1} (a\omega)^{|k-(l'-1)|} Y_{l'-1}^*(\theta')$ term
in the curly brackets vanishes so that for $l=l_0$ Hod's prediction is an integer $1$ below 
the actual value. Higher $l$-modes of the $l'=l_0+1$, $m=0$ case are predicted correctly 
because in these cases the $k=l_0+1$ term gives the same power of $\omega$ without using the 
$C_{kl'-1} (a\omega)^{|k-(l'-1)|} Y_{l'-1}^*(\theta')$ term of the curly brackets.

\section*{References}

\end{document}